\documentclass[reqno,11pt,tikz]{article}
\usepackage{jheppub}
\usepackage{extarrows}
\usepackage{epsfig}
\usepackage{amssymb}
\usepackage{verbatim}
\usepackage{amsmath}
\usepackage{stmaryrd}
\usepackage{mathrsfs}
\usepackage{hyperref}
\usepackage{dsfont}
\usepackage{slashed}
\usepackage[dvipsnames]{xcolor}
\usepackage{epstopdf}
\usepackage{eucal}
\usepackage{graphicx}
\usepackage{color}
\usepackage{bbold}
\usepackage[version=3]{mhchem}
\usepackage{shuffle}

\usepackage[compat=1.1.0]{tikz-feynman}
\usepackage{textgreek}

\DeclareMathAlphabet{\mathsfit}{T1}{\sfdefault}{\mddefault}{\sldefault}
\SetMathAlphabet{\mathsfit}{bold}{T1}{\sfdefault}{\bfdefault}{\sldefault}

\newcommand*{\sbullet}{\raisebox{0.1ex}{\scalebox{0.6}{$\bullet$}}}

\def\RR{{\mathbb{R}}}
\def\CC{{\mathbb{C}}}

\def\ud{{\mathrm d}}
\def\ue{{\mathrm e}}
\def\ui{{\mathrm i}}

\def\Dsf{{\mathrm D}}
\def\Csf{{\mathrm C}}
\def\Ssf{{\mathrm S}}
\def\Gsf{{\mathrm G}}

\def\to{\rightarrow}

\newcommand{\coker}{\operatorname{coker}}
\newcommand{\im}{\operatorname{im}}

\newcommand{\SU}{\operatorname{SU}} 
\newcommand{\su}{\operatorname{\mathfrak{su}}}

\newcommand{\ba}{\begin{eqnarray}}
\newcommand{\ea}{\end{eqnarray}}

\let\originalleft\left
\let\originalright\right
\renewcommand{\left}{\mathopen{}\mathclose\bgroup\originalleft}
\renewcommand{\right}{\aftergroup\egroup\originalright}

\DeclareMathAlphabet{\mathbfit}{OML}{cmm}{b}{it}

\date{}
\allowdisplaybreaks
\numberwithin{equation}{section}

\title{The $L_{\infty}$ structure of gauge theories with matter}

\author[a,b]{Humberto Gomez}
\author[c]{Renann Lipinski Jusinskas}
\author[d]{Cristhiam Lopez-Arcos}
\author[d]{Alexander Quintero V\'elez}

\affiliation[a]{Niels Bohr International Academy and Discovery Center The Niels Bohr Institute, University of Copenhagen Blegdamsvej 17, DK-2100 Copenhagen, Denmark}
\affiliation[b]{Facultad de Ciencias, Basicas Universidad
Santiago de Cali, Calle 5 No 62-00 Barrio Pampalinda Cali, Valle, Colombia}
\affiliation[c]{Institute of Physics of the Czech Academy of Sciences \& CEICO Na Slovance 2, 182 21, Prague -- Czech Republic}
\affiliation[d]{Escuela de Matem\'{a}ticas, Universidad Nacional de Colombia Sede Medell\'{i}n, \\ \phantom{${}^{b}$}Carrera 65 $\#$ 59A--110, Medell\'{i}n, Colombia}

\emailAdd{humberto.gomez@nbi.ku.dk}
\emailAdd{renannlj@fzu.cz}
\emailAdd{cmlopeza@unal.edu.co}
\emailAdd{aquinte2@unal.edu.co}

\abstract{
In this work we present an algebraic approach to the dynamics and perturbation theory at tree-level for gauge theories coupled to matter. The field theories we will consider are: Chern-Simons-Matter, Quantum Chromodynamics, and scalar Quantum Chromodynamics. Starting with the construction of the master action in the classical Batalin-Vilkovisky forma\-lism, we will extract the $L_{\infty}$-algebra that allow us to recursively calculate the perturbiner expansion from its minimal model. The  Maurer-Cartan action obtained in this procedure will then motivate a generating function for all the tree-level scattering amplitudes. There are two interesting outcomes of this construction: a generator for fully-flavoured amplitudes via a localisation on Dyck words; and closed expressions for fermion and scalar lines attached to $n$-gluons with arbitrary polarisations.
}

\begin{document} 
\maketitle
\flushbottom

\section{Introduction} 

Quantum field theory is currently the best theoretical framework to
describe particle physics and most condensed matter systems. In fact,
gauge theories are at the core of our very understanding about the
fundamental forces of nature and how they intermediate matter interactions.

A central object of study in any quantum field theory is the S-matrix.
Whether computed perturbatively or not, the S-matrix encodes the main
dynamical data of a given theory. Therefore, studying its
structure and general properties is an essential step towards a more
complete physical and mathematical understanding of the problem at
hand. In this direction, recent results have considerably improved
and expanded the available tools for computing scattering amplitudes
in quantum field theory, establishing the area as one of the main
driving forces in contemporary theoretical physics research.

At this interface between physics and mathematics, lies the present
work. We will employ algebraic techniques inspired from String Field
Theory \cite{Stasheff1992,Lada-Markl1995,Kajiura:2003ax}  (see also \cite{Erler:2019loq} for a recent review) to obtain all the tree-level amplitudes
for $SU(N)$ gauge theories coupled to massive fermions or scalars.
We will focus our attention on Chern-Simons-Matter (CSM) theory, Quantum
Chromodynamics (QCD) and scalar QCD theory (sQCD).

$L_{\infty}$ algebras constitute a natural underlying structure in
the Batalin-Vilkovisky (BV) formalism (see, for example, \cite{Jurco:2018sby,Jurco:2019bvp} and references therein. See also \cite{Arvanitakis:2019ald} for a direct application to the S-matrix). From a physical point of view,
these algebras have a clear structure and help to organise the dynamics
of a given field theory in a chain complex that takes the form \cite{Hohm:2017pnh}:

$$
\begin{tikzpicture}[scale=1.0,baseline=-0.1cm, inner sep=1mm,>=stealth]
\node (0) at (0,0)  {$\dots$};
\node (1) at (2.1,0)  {$
  \bigl\{ \begin{smallmatrix} 
      \mathrm{gauge} &  \\ 
      \mathrm{parameters} 
   \end{smallmatrix} \bigr\}
$};
\node (2) at (4.6,0)  {$\bigl\{ \begin{smallmatrix} 
      \mathrm{fields} \end{smallmatrix} \bigr\}$};
\node (3) at (7.2,0)  {$\bigl\{ \begin{smallmatrix} 
      \mathrm{equations}\,\,\mathrm{of} &  \\ 
      \mathrm{motion} 
   \end{smallmatrix} \bigr\}$};
\node (4) at (9.4,0)  {$\dots$};
\tikzset{every node/.style={fill=white}} 
\draw[->, thick] (0) to node[midway,above=0.2pt] {} (1); 
\draw[->, thick] (1) to node[midway,above=0.2pt] {} (2); 
\draw[->, thick] (2) to node[midway,above=0.2pt] {} (3);
\draw[->, thick] (3) to node[midway,above=0.2pt] {} (4); 
\end{tikzpicture}
$$

As shown in \cite{Macrelli:2019afx}, single-particle solutions to
the free equations of motion of the field theory, given by plane waves
in flat space, are related to multiparticle solutions via a quasi-isomorphism
in the $L_{\infty}$ algebra, which are manifested through the Berends-Giele
recursion relations \cite{Berends:1987me} they satisfy.

The fundamental ingredient in our approach is the so-called perturbiner
expansion \cite{Rosly:1996vr,Rosly:1997ap,Selivanov:1997aq,Selivanov:1997ts,Rosly:1998vm,Selivanov:1998hn},
which is another multi-particle solution of the classical equations
of motion of the theory that can be directly used to compute tree-level
amplitudes, e.g. \cite{Mafra:2016ltu,Mizera:2018jbh,Mafra:2015gia,Mafra:2015vca,Lee:2015upy,Mafra:2016mcc,Garozzo:2018uzj,Bridges:2019siz}. 

The perturbiner construction has been formally justified only recently
by some of the authors \cite{Lopez-Arcos:2019hvg}, and it arises
in the context of $L_{\infty}$-algebras in the construction of their minimal models. In this work,
we will extend this construction to colour-dressed perturbiners, which
necessarily appear in theories involving fields that transform in
different representations of the gauge group (adjoint, fundamental
and anti-fundamental, for example)\footnote{Although structurally similar, we are not aware of such solutions with
mixed representations being explicitly presented in the literature.}. In addition, we will show that the Maurer-Cartan action built with
the perturbiner solutions can be turned into a generator of tree-level
scattering amplitudes that include several cases of interest depending
on the chosen \emph{boundary conditions}.

In order to define the Maurer-Cartan action, we need to introduce a suitable graded inner product $\langle\, ,\rangle$, guarateeing that the $L_{\infty}$-algebra$,\mathfrak{L}$, is cyclic.  We will denote by $l_k(a_1,\dots,a_k)$ the $L_\infty$ products associated to a given field theory,  with $a_i$ generically representing the elements of a graded vector space that includes the fields. The cyclicity of the algebra is then simply expressed as
\ba
\langle a_{k+1},l_k(a_1,\dots,a_k) \rangle = \pm \langle a_1,l_k(a_2,\dots,a_k,a_{k+1}) ,
\ea
where $\pm$ is determined by the grading of the elements $a_i$ involved in the product (see, for example, \cite{Kajiura:2003ax}).
The Maurer-Cartan action is then expressed as
\ba
S_{\mathrm{MC}}[a] =  \frac{1}{2} \langle a, l_{1}(a)\rangle + \sum_{n \geq 2} \frac{1}{(n+1)!} \langle a, l_{n}(a,\dots,a)\rangle,
\ea
where the $a$'s generically denote the classical field content of the theory and are called Maurer-Cartan elements. Observe that $l_{1}(a)$ is associated with the free field equations of motion while $l_{n}$ with $n\geq2$ corresponds to interaction contributions in the Lagrangian (vertices).
In this construction, the solutions to the linearised equations of motion are said to be in the cohomology of the algebra (free fields).
Interestingly, there is also a homotopy Maurer-Cartan action in the algebra cohomology, denoted by $S'_{\mathrm{MC}}[a]$, and they are related by a quasi-isomorphism\footnote{Any homotopy algebra is quasi-isomorphic to a minimal model, characterized by the vanishing of the $l_1$ product.}. As we will show here, $S'_{\mathrm{MC}}[a]$ can be seen as the generating function for all the tree-level amplitudes of the theory.

For our algebraic analysis, we will extract the $L_{\infty}$-algebra for the field theories from the master actions in the classical BV formalism. This type of algebraic construction for gauge theories with matter has been done from a more formal perspective in \cite{Zeitlin:2007fp}. Here we will use notation and conventions more familiar to the physics community. Once the $L_{\infty}$-algebra products are obtained, we will proceed to build the perturbiner expansion from the minimal model and then all the tree-level amplitudes. The procedure is detaily described in \cite{Lopez-Arcos:2019hvg}.

The first theory that we will study is Chern-Simons-Matter (CSM) with massive fermions.
This theory plays an important role in condensed matter physics, being
used to describe phenomena like the quantum Hall effect and anyonic
physics. The tree-level amplitudes for CSM were calculated for the
case of massless matter in \cite{Inbasekar:2017ieo} using the Britto-Cachazo-Feng-Witten
recursion relations \cite{Britto:2005fq}, which is an on-shell method.
Working with massive fermions is important if we want to go to the
condensed matter domain via a nonrelativistic limit.

We will then move on to QCD-like theories. The scattering amplitudes
for this case have been studied extensively using Feynman diagrams.
There is a result for all tree-level amplitudes with massless fermions
in \cite{Dixon:2010ik}, where they found a procedure to extract the
QCD amplitudes from the ones of $\mathcal{N}=4$ Super Yang-Mills.
Another interesting result for tree-level amplitudes is in \cite{Johansson:2015oia},
where a new colour decomposition was introduced together with evidence
that QCD obeys the colour-kinematics duality \cite{Bern:2008qj}. 

The last example that we will work in is scalar QCD. This theory models
the type of interactions between gluons and Higgs bosons in the Standard
Model, this being a topic of interest in particle physics. Another
interesting application of this theory is the problem of two bodies
in classical General Relativity, by the application of double-copy
\cite{Kawai:1985xq} in the scattering amplitudes and the infinite
mass limit for the scalars. In order to apply double-copy the amplitudes
must also satisfy the colour-kinematics duality, this procedure has
been studied in \cite{Plefka:2019wyg}.

This work is organised as follows. In section \ref{sec:actions} we will present the field theories and determine their BV master actions. Each theory will then be studied in a different section including the extraction of the $L_{\infty}$-algebra from the BV action, the construction of the perturbiner expansion and the generating function for the scattering amplitudes, finishing with a series of examples illustrating our method. CSM is described in section \ref{sec:CSM}, QCD in section \ref{sec:QCD} and scalar QCD in section \ref{sec:sQCD}. In section \ref{sec:concl} we present the final remarks.


\paragraph{Notation}

We will work in $d$-dimensional Minskowski spacetime $\mathbb{R}^{1,d-1}$, with coordinates $x^{\mu}$ and diagonal metric tensor $\eta_{\mu\nu}$ with the mostly minus signature $(+---\cdots)$, being $x^{0}$ the time direction. The standard volume form is
\ba
\ud^d x = \ud x^{0} \wedge \ud x^{1} \wedge \cdots \wedge \ud x^{d-1}.
\ea
The d'Alembertian operator is defined as $\square = \eta^{\mu\nu} \partial_\mu \partial_\nu$
where we use the shorthand notation $\partial_{\mu}$ for the partial derivative $\partial / \partial x^{\mu}$. 

For the gauge group, we take the Lie group $\SU(N)$ of all special unitary linear transformations in $\CC^{N}$ and, as is conventional, we take its Lie algebra $\su(N)$ to consist of complex $N \times N$ anti-Hermitian matrices with zero trace. Latin indices $a,b$, etc. from the beginning of the alphabet run over the $N^{2}-1$ generators of $\su(N)$. We let these generators be denoted by $T^{a}$, with structure constants $f^{ab}_{\phantom{ab}c}$ satisfying 
\begin{eqnarray}\label{eq:gaugealgebra}
[T_{a},T_{b}] & = & i\sqrt{2}f_{ab}^{\hphantom{ab}c}T_{c},\nonumber \\
 & \equiv & \tilde{f}_{ab}^{\hphantom{ab}c}T_{c},
\end{eqnarray}
We also normalise the Cartan-Killing form on $\su(N)$ in such a way that
\begin{equation}
\kappa_{ab}=tr(T_{a}T_{b})=\delta_{ab}.
\end{equation}
The indices $a,b$, and so on are raised and lowered by using $\kappa^{ab}$ and its inverse $\kappa_{ab}$, respectively. The fundamental representation of $\SU(N)$ we denote by $V = \CC^{N}$. Latin lowercase indices $i,j$, and so on run over the $N$ basis vectors of $V$. 

We will denote by $S$ the complexified spinor representation of $\textrm{Spin}(1,d-1)$. The Dirac matrices, $\gamma^{\mu}$, satisfy
\begin{equation}
\{\gamma^{\mu},\gamma^{\nu}\}=2\eta^{\mu\nu}.
\end{equation}
We also frequently make use of Feynman's ``slash'' notation: for any vector $v$ in $\RR^{1,d-1}$, we write $\slashed{v} = \eta_{\mu\nu}\gamma^{\mu} v^{\nu} = \gamma^{\mu} v_{\mu}$. The Dirac operator is then simply expressed as
$\slashed{\partial} = \gamma^{\mu} \partial_{\mu}$.
In our conventions, the product of two Grassmann numbers, $\xi$ and $\chi$, satisfies $(\xi\chi)^{\dagger}=\chi^{\dagger}\xi^{\dagger}$.


\section{Gauge theories with matter}\label{sec:actions}

Our focus in this work is on gauge theories coupled to matter fields in the fundamental representation of $\SU(N)$. In this section we briefly review their field content and action, and use the BV formalism to determine their master action and extract their $L_{\infty}$ structure through the corresponding BV transformations.

\subsection{Quick review of the BV formalism}
Let us first briefly recall the BV formalism for general gauge theories. An extended review can be found in \cite{Gomis:1994he}.

Consider a gauge theory with collective field content $\phi^{r}$, classical action $S_0[\phi]$ and gauge parameters $\varepsilon^{\alpha}$. For simplicity, we are assuming an irreducible gauge algebra, i.e. with linearly independent gauge transformations.

The starting point of the BV formalism is to promote the infinitesimal gauge parameters $\varepsilon^{\alpha}$ to ghost fields of opposite Grassmann character (bosonic gauge symmetries give origin to fermionic ghosts and vice-versa). Additionally, we assign a graduation to our fields, the ghost number, equal to $1$ for $\varepsilon^{\alpha}$ and $0$ for  $\phi^{r}$. The next step is to introduce an antifield for each field in the theory. We let $\Phi^{A}$ run over all the fields $\phi^{r}$ and $\varepsilon^{\alpha}$, and for each $\Phi^{A}$ we introduce an  antifield $\Phi^*_{A}$ with opposite statistics, and ghost number equal to $- \mathrm{gh}(\Phi^{A})-1$, where $\mathrm{gh}(\Phi^{A})$ is the ghost number of $\Phi^{A}$. Thus we can define an antibracket of two general functionals $F[\Phi,\Phi^*]$ and $G[\Phi,\Phi^*]$ by
\begin{equation}
(F,G) = F\left( \frac{\overleftarrow{\partial}}{\partial \Phi^{A}}\frac{\partial}{\partial \Phi^*_{A}} - \frac{\overleftarrow{\partial}}{\partial \Phi^*_{A}}\frac{\partial}{\partial \Phi^{A}} \right)G.
\end{equation}
From there we look for a BV action $S[\Phi,\Phi^*]$ of ghost number $0$ depending on $\Phi^{A}$ and $\Phi^*_{A}$ subjected to two requirements. The first requirement is that $S[\Phi,\Phi^*]$ reduces to the original action $S_{0}[\phi]$ when the antifields $\Phi^*_{A}$ are set to zero. The second requirement is that $S[\Phi,\Phi^*]$ satisfies what is called the classical master equation
\begin{equation}\label{eq:master-equation}
(S,S) = 0.
\end{equation}
The solution $S[\Phi,\Phi^*]$ exists as an expansion in powers of antifields, usually finite. Also, it follows from \eqref{eq:master-equation} that $S[\Phi,\Phi^*]$ is automatically invariant under the so-called BV transformations given by
\begin{align}\label{eq:BVtransf}
\begin{split}
\delta_{\mathrm{BV}} \Phi^{A} &= -(S,\Phi^{A} ) = \frac{\partial_r S}{\partial \Phi^*_{A}}, \\
 \delta_{\mathrm{BV}} \Phi^*_{A} &= -(S, \Phi^*_{A}) = -\frac{\partial_r S}{\partial \Phi^{A}}.
\end{split}
\end{align}
Differently put, the gauge symmetries of $S_0[\phi]$ are promoted to nilpotent global symmetries of the master action $S[\Phi,\Phi^*]$.

\subsection{Chern-Simons-Matter theory}\label{sec:2.2}

Chern-Simons-Matter theory has been highlighted in recent times in a number of papers \cite{Jain:2013gza,Giombi:2011kc,Halder:2019foo,Gaiotto:2007qi}. We refer to those references for more details.

The field content for CSM is given by a gauge field $A$, which we consider as a $1$-form on $\RR^{1,2}$ with values in $\su(N)$, a Dirac spinor $\psi$, which we consider as $0$-forms on $\RR^{1,2}$ with values in $S \otimes V$, and a conjugate Dirac spinor $\bar{\psi}$, which we consider as $0$-forms on $\RR^{1,2}$ with values $S \otimes \bar{V}$. More explicitly, $A$ is specified through the components of the $1$-forms $A^{a} =  A^{a}_{\mu} \ud x^{\mu}$, while $\psi$ and $\bar{\psi}$ can be described in terms of its components $\psi^{i}$ and $\bar{\psi}_{i}$.

The CSM action is
\begin{equation}\label{eq:CSMaction}
S_{0}[A,\psi,\bar{\psi}] =  \int_{\RR^{1,2}} \ud^3 x \,\left\{ -  \frac{\kappa}{4\pi}\varepsilon^{\mu\nu\rho} \left( A^{a}_{\mu}  \partial_{\nu} A_{\rho a}  -  \frac{\ui} {3} \tilde{f}_{abc} A^{a}_{\mu} A^{b}_{\nu} A^{c}_{\rho}\right) +  \bar{\psi}(\ui \slashed{D} - m) \psi \right\},
\end{equation}
where $\kappa$ is the Chern-Simons level, $m$ is the mass and $\varepsilon^{\mu\nu\rho}$ is the Levi-Civita symbol, with $\varepsilon^{012}=1$. The covariant derivative, $D_{\mu}$, is given by
\begin{align}\label{eq:covD-psi}
\begin{split}
(D_{\mu} \psi)^{i} &=\partial_{\mu} \psi^{i} - \ui A^{a}_{\mu} (T_a)^{i}_{\phantom{i}j} \psi^{j},  \\
(D_{\mu} \bar{\psi})_{i} &=\partial_{\mu} \bar{\psi}_{i} + \ui A^{a}_{\mu} \bar{\psi}_{j}  (T_a)^{j}_{\phantom{j}i}.
\end{split}
\end{align} 

This action is invariant under the following infinitesimal gauge transformations
\begin{align}\label{eq:inf-gauge}
\begin{split}
\delta A^{a}_{\mu} &= D_{\mu} c^{a}, \\
\delta \psi^{i} &= \ui c^{a} (T_a)^{i}_{\phantom{i}j} \psi^{j},\\
\delta \bar{\psi}_{i} &= -\ui c^{a} \bar{\psi}_{j}  (T_a)^{j}_{\phantom{j}i},    
\end{split}
\end{align}
where $c$ is the infinitesimal gauge parameter, which we interpret as a $0$-form on $\RR^{1,2}$ with values in $\su(N)$, and $D_{\mu} c^{a}= \partial_{\mu}c^{a} -\ui \tilde{f}_{bc}^{\phantom{bc}a}A_{\mu}^{b} c^{c}$ is its covariant derivative. 

The field strength $F$ is  a $2$-form on $\RR^{1,2}$ with values in $\su(N)$. In terms of components this is $F^{a} = \frac{1}{2} F^{a}_{\mu \nu} \ud x^{\mu} \wedge \ud x^{\nu}$ with
\begin{equation}\label{eq:field-strength}
F^{a}_{\mu \nu} = \partial_{\mu} A_{\nu}^{a} - \partial_{\nu} A_{\mu}^{a} - \ui \tilde{f}_{bc}^{\phantom{bc}a} A_{\mu}^{b} A_{\nu}^{c}. 
\end{equation}
In addition, there is a matter current $J$ associated with the spinors $\psi$ and $\bar{\psi}$, which can be thought of as the $1$-form on $\RR^{1,2}$ with values in $\su(N)$ having the components
\begin{equation}\label{eq:matter-current}
J^{a}_{\mu} = \bar{\psi}_{i} \gamma_{\mu} (T^{a})^{i}_{\phantom{i}j} \psi^{j}. 
\end{equation}

The equations of motion derived from the action above are simply
\begin{equation}
\begin{split}
\frac{\kappa}{4 \pi} \varepsilon^{\mu\nu \rho} F^{a}_{\mu\nu} &= J^{a \rho} ,  \\
\ui \gamma^{\mu}(D_{\mu}\psi)^{i} - m \psi^{i}&= 0, \\
\ui (D_{\mu}\bar{\psi})_i \gamma^{\mu} + m \bar{\psi}_i &= 0. 
\end{split}
\end{equation}
In particular, the Bianchi identity for $F$ is compatible with the covariant  conservation of $J^{\mu}_{a}$,
$D_{\mu} J^{\mu}_{a} = 0$.

We now recast the CSM theory in the BV formalism. First, we promote the $0$-form $c$ with values in $\su(N)$, appearing as the infinitesimal gauge parameter in \eqref{eq:inf-gauge}, to an anticommuting field of ghost number $1$. Next we introduce antifields $A^*$, $\bar{\psi}{}^*$, $\psi^*$ and $c^*$ for all the fields $A$, $\psi$, $\bar{\psi}$ and $c$, with $A^*$ regarded as a $1$-form on $\RR^{1,2}$ with values in $\su(N)$, $\bar{\psi}{}^*$ and $\psi^*$ as $0$-forms on $\RR^{1,2}$ with values in $S \otimes \bar{V}$ and $S \otimes V$, respectively, and $c^*$ as a $0$-form on $\RR^{1,2}$ with values in $\su(N)$. Therefore, $A^*$, $\bar{\psi}{}^*$ and $\psi^*$ have ghost number $-1$, while $c$ has ghost number $-2$. In this case, the solution of the master equation \eqref{eq:master-equation} is given by
\begin{align}
\begin{split}
S[A,A^*,&\psi,\bar{\psi}{}^*,\bar{\psi},\psi^*,c,c^*] = S_{0}[A,\psi,\bar{\psi}]   \\
& + \int_{\RR^{1,2}} \ud^3 x \,\left\{ D_{\mu}c^{a} A^{*\mu}_{a} + \frac{\ui}{2} \tilde{f}_{abc} c^{*a} c^{b} c^{c} + \ui c^{a} \bar{\psi}{}^*_{i} (T_{a})^{i}_{\phantom{i}j} \psi^{j} - \ui c^{a} \bar{\psi}_{i} (T_{a})^{i}_{\phantom{i}j} \psi^{*j}\right\}. 
\end{split}
\end{align}

The BV transformations defined in  \eqref{eq:BVtransf} are explicitly given by
\begin{align}\label{eq:BVtransf-CSM}
\begin{split}
\delta_{\mathrm{BV}} c^{a} &= \tfrac{\ui}{2} \tilde{f}_{bc}^{\phantom{bc}a} c^{b} c^{c},\\
\delta_{\mathrm{BV}} A^{a}_{\mu} &= D_{\mu}c^{a}, \\
\delta_{\mathrm{BV}} \psi^{i} &= \ui c^{a} (T_{a})^{i}_{\phantom{i}j} \psi^{j}, \\
\delta_{\mathrm{BV}} \bar{\psi}_{i} &= -\ui c^{a} \bar{\psi}_{j} (T_{a})^{j}_{\phantom{j}i}, \\
\delta_{\mathrm{BV}} A^{*a}_{\mu} &= \tfrac{\kappa}{4\pi} \varepsilon_{\mu \nu \rho} F^{a \nu\rho} - J^{a}_{\mu} - \ui \tilde{f}_{bc}^{\phantom{bc}a} c^{b} A^{*c}_{\mu}, \\
\delta_{\mathrm{BV}} \bar{\psi}{}^{*}_{i} &= \ui (D_{\mu}\bar{\psi})_{i} \gamma^{\mu} + m \bar{\psi}_{i} - \ui c^{a} \bar{\psi}{}^*_{j}(T_a)^{j}_{\phantom{j} i}, \\
\delta_{\mathrm{BV}} \psi^{*i}&= \ui \gamma^{\mu} (D_{\mu} \psi)^{i} - m \psi^{i} + \ui c^{a}(T_{a})^{i}_{\phantom{i}j} \psi^{*j},\\
\delta_{\mathrm{BV}} c^{*a} &= - D_{\mu} A^{*a \mu} - \ui \tilde{f}_{bc}^{\phantom{bc}a} c^{*b}c^{c} + \ui \bar{\psi}{}^*_{i}(T^{a})^{i}_{\phantom{i}j} \psi^{j} - \ui \bar{\psi}_{i} (T^{a})^{i}_{\phantom{i}j} \psi^{*j}, \\
\end{split}
\end{align}
and their nilpotency follows from the master equation.

\subsection{Quantum Chromodynamics}\label{sec:2.3}
Let us next turn to QCD. Standard references include \cite{Greiner:2002ui} and \cite{Pokorski:1987ed}. Our conventions are those of \cite{AlvarezGaume:2012zz}.

As in the previous section, the field content of QCD consists of a gauge field $A$, a Dirac spinor $\psi$ and a conjugate Dirac spinor $\bar{\psi}$. However, being a four-dimensional theory, we now consider $A$ as a $1$-form on $\RR^{1,3}$ with values in $\su(N)$, while $\psi$ and $\bar{\psi}$ are regarded as $0$-forms on $\RR^{1,3}$ with values in $S \otimes V$ and $S \otimes \bar{V}$, respectively. The field strength components $F^{a}_{\mu\nu}$ are defined by the same formula as \eqref{eq:field-strength}, and the covariant derivatives $D_{\mu} \psi$ and $D_{\mu}\bar{\psi}$, by the same formula as \eqref{eq:covD-psi}. The action for the theory reads
\begin{equation}\label{eq:QCDaction}
S_{0}[A,\psi,\bar{\psi}] =  \int_{\RR^{1,3}} \ud^4 x \,\left\{ -  \frac{1}{4}F^{a}_{\mu\nu} F^{\mu\nu}_a +  \bar{\psi}(\ui \slashed{D} - m) \psi \right\},
\end{equation}
where again $m$ is the mass of the matter fields $\psi$ and $\bar{\psi}$. This action is invariant under infinitesimal gauge transformations as in \eqref{eq:inf-gauge}, with the infinitesimal gauge parameter $c$ being now a $0$-form on $\RR^{1,3}$ with values in $\su(N)$. The equations of motion derived from \eqref{eq:QCDaction} are
\begin{align}
\begin{split}
 D_{\nu} F^{\mu\nu}_{a} &= J_{a}^{\mu}, \\
\ui \gamma^{\mu}(D_{\mu}\psi)^{i} - m \psi^{i}&= 0, \\
\ui (D_{\mu}\bar{\psi})_i \gamma^{\mu} + m \bar{\psi}_i &= 0. 
\end{split}
\end{align}
Here $D_{\nu} F^{\mu\nu}_{a} = \partial_{\nu} F^{\mu\nu}_{a} - \ui \tilde{f}_{\phantom{bc}a}^{bc} A_{b \nu} F_{c}^{\mu\nu}$ is the covariant derivative appropriate to the field strength $F$, and $J^{\mu}_{a}$ are the components of the covariantly conserved matter current defined by the same formula as that appearing in \eqref{eq:matter-current}.

We next consider the BV formulation of QCD.  Let us therefore promote the infinitesimal gauge parameter $c$ to an anticommuting field of ghost number $1$. Following the CSM case, we denote the antifields as $A^*$, $\bar{\psi}{}^*$, $\psi^*$ and $c^*$, with respective ghost numbers $-1$, $-1$, $-1$ and $-2$. We think of $A^{*}$ as a $1$-form on $\RR^{1,3}$ with values in $\su(N)$, $\bar{\psi}{}^*$ and $\psi^*$ as $0$-forms on $\RR^{1,3}$ with values in $S \otimes \bar{V}$ and $S \otimes V$, respectively, and $c^*$ as a $0$-form on $\RR^{1,3}$ with values in $\su(N)$. The BV action is
\begin{align}
\begin{split}
S[A,A^*,&\psi,\bar{\psi}{}^*,\bar{\psi},\psi^*,c,c^*]  = S_{0}[A,\psi,\bar{\psi}]\\
& + \int_{\RR^{1,3}} \ud^4 x \,\left\{ D_{\mu}c^{a} A^{*\mu}_{a} + \frac{\ui}{2} \tilde{f}_{abc} c^{*a} c^{b} c^{c} + \ui c^{a} \bar{\psi}{}^*_{i} (T_{a})^{i}_{\phantom{i}j} \psi^{j} - \ui c^{a} \bar{\psi}_{i} (T_{a})^{i}_{\phantom{i}j} \psi^{*j}\right\},
\end{split}
\end{align}
such that
\begin{align}\label{eq:BVtransf-QCD}
\begin{split}
\delta_{\mathrm{BV}} c^{a} &= \tfrac{\ui}{2} \tilde{f}_{bc}^{\phantom{bc}a} c^{b} c^{c},\\
\delta_{\mathrm{BV}} A^{a}_{\mu} &= D_{\mu}c^{a}, \\
\delta_{\mathrm{BV}} \psi^{i} &= \ui c^{a} (T_{a})^{i}_{\phantom{i}j} \psi^{j}, \\
\delta_{\mathrm{BV}} \bar{\psi}_{i} &= -\ui c^{a} \bar{\psi}_{j} (T_{a})^{j}_{\phantom{j}i}, \\
\delta_{\mathrm{BV}} A^{*a}_{\mu} &= D^{\nu}F^{a}_{\mu\nu} - J^{a}_{\mu} - \ui \tilde{f}_{bc}^{\phantom{bc}a} c^{b} A^{*c}_{\mu}, \\
\delta_{\mathrm{BV}} \bar{\psi}{}^{*}_{i} &= \ui (D_{\mu}\bar{\psi})_{i} \gamma^{\mu} + m \bar{\psi}_{i} - \ui c^{a} \bar{\psi}{}^*_{j}(T_a)^{j}_{\phantom{j} i}, \\
\delta_{\mathrm{BV}} \psi^{*i}&= \ui \gamma^{\mu} (D_{\mu} \psi)^{i} - m \psi^{i} + \ui c^{a}(T_{a})^{i}_{\phantom{i}j} \psi^{*j},\\
\delta_{\mathrm{BV}} c^{*a} &= - D_{\mu} A^{*a \mu} - \ui \tilde{f}_{bc}^{\phantom{bc}a} c^{*b}c^{c} + \ui \bar{\psi}{}^*_{i}(T^{a})^{i}_{\phantom{i}j} \psi^{j} - \ui \bar{\psi}_{i} (T^{a})^{i}_{\phantom{i}j} \psi^{*j}. \\
\end{split}
\end{align}
constitute the nilpotent BV transformations of the fieds and antifields.

\subsection{Scalar Quantum Chromodynamics}\label{sec:2.4}
Lastly, let us describe scalar QCD. For some recent references see \cite{Plefka:2018zwm} and \cite{Plefka:2019wyg}. 

The fields of interest here are a gauge field $A$ with field strength $F$, together with a scalar field $\phi$ and a conjugate scalar field $\bar{\phi}$, which we consider as $0$-forms on $\RR^{1,3}$ with values in $V$ and $\bar{V}$, respectively. In terms of components, $\phi$ and $\bar{\phi}$ are represented by functions $\phi^{i}$ and $\bar{\phi}_{i}$.

The action for the theory is given by
\begin{equation}\label{eq:sQCDaction}
S_{0}[A,\phi,\bar{\phi}] =  \int_{\RR^{1,3}} \ud^4 x \,\left\{ -  \frac{1}{4}F^{a}_{\mu\nu} F^{\mu\nu}_a +  D_{\mu} \bar{\phi} D^{\mu} \phi - m^2 \bar{\phi} \phi \right\},
\end{equation}
with $m$ being the mass of $\phi$ and $\bar{\phi}$.  The covariant derivatives are defined in the standard way as
\begin{align}
\begin{split}
(D_{\mu} \phi)^{i} &= \partial_{\mu} \phi^{i} - \ui A^{a}_{\mu} (T_a)^{i}_{\phantom{i}j} \phi^{j}, \\
(D_{\mu} \bar{\phi})_{i} &= \partial_{\mu}\bar{\phi}_{i} + \ui A^{a}_{\mu}  \bar{\phi}_{j}  (T_a)^{j}_{\phantom{j}i}.
\end{split}
\end{align}

The gauge transformations can be cast as
\begin{align} \label{eq:2.19}
\begin{split}
\delta A^{a}_{\mu} &= D_{\mu} c^{a}, \\
\delta \phi^{i} &= \ui c^{a} (T_a)^{i}_{\phantom{i}j} \phi^{j},\\
\delta \bar{\phi}_{i} &= -\ui c^{a} \bar{\phi}_{j} (T_a)^{j}_{\phantom{j}i},    
\end{split}
\end{align}
where the infinitesimal gauge parameter $c$ is as in the QCD case. We can also define a matter current $J$ associated with the scalar fields $\phi$ and $\bar{\phi}$, which we view as the $1$-form on $\RR^{1,3}$ with values in $\su(N)$ with components
\begin{equation} 
J^{a}_{\mu} = (D_{\mu}\bar{\phi})_{i} (T^{a})^{i}_{\phantom{i}j} \phi^{j} - \bar{\phi}_{i} (T^{a})^{i}_{\phantom{i}j} (D_{\mu}\phi)^{i}. 
\end{equation}
The equations of motion derived from \eqref{eq:sQCDaction} are given by
\begin{align}
\begin{split}
 D_{\nu} F^{\mu\nu}_{a} &= J_{a}^{\mu}, \\
D^{\mu}D_{\mu}\phi^{i} + m^2 \phi^{i} &= 0, \\
D^{\mu}D_{\mu}\bar{\phi}_{i} + m^2 \bar{\phi}_{i} &= 0,
\end{split}
\end{align}
and, again, imply the covariant conservation of the matter current, $D_{\mu} J^{\mu}_a = 0$. 

As in the previous cases, the BV algorithm can be reproduced if we simply promote the infinitesimal gauge parameter $c$ in \eqref{eq:2.19} to an anticommuting field of ghost number $1$, and introduce antifields $A^*$, $\bar{\phi}^{*}$, $\phi^*$ and $c^*$ conjugate to $A$, $\phi$, $\bar{\phi}$ and $c$, with ghost numbers $-1$, $-1$, $-1$ and $-2$. Here $A^*$ and $c^*$ are interpreted as in the QCD case, and $\bar{\phi}^{*}$ and $\phi^*$ as $0$-forms on $\RR^{1,3}$ with values in $\bar{V}$ and $V$, respectively. The solution to the master equation is
\begin{align}
\begin{split}
S[A,A^*,&\phi,\bar{\phi}^{*},\bar{\phi},\phi^*,c,c^*] = S_{0}[A,\phi,\bar{\phi}]\\
& + \int_{\RR^{1,3}} \ud^4 x \,\left\{ D_{\mu}c^{a} A^{*\mu}_{a} + \frac{\ui}{2} \tilde{f}_{abc} c^{*a} c^{b} c^{c} + \ui c^{a}\bar{\phi}^{*}_{i} (T_{a})^{i}_{\phantom{i}j} \phi^{j} - \ui c^{a} \bar{\phi}_{i} (T_{a})^{i}_{\phantom{i}j} \phi^{*j}\right\}, 
\end{split}
\end{align}
which directly lead to the nilpotent BV transformations:
\begin{align}\label{eq:BVtransf-scalarQCD}
\begin{split}
\delta_{\mathrm{BV}} c^{a} &= \tfrac{\ui}{2} \tilde{f}_{bc}^{\phantom{bc}a} c^{b} c^{c},\\
\delta_{\mathrm{BV}} A^{a}_{\mu} &= D_{\mu}c^{a}, \\
\delta_{\mathrm{BV}} \phi^{i} &= \ui c^{a} (T_{a})^{i}_{\phantom{i}j} \phi^{j}, \\
\delta_{\mathrm{BV}} \bar{\phi}_{i} &= -\ui c^{a} \bar{\phi}_{j} (T_{a})^{j}_{\phantom{j}i}, \\
\delta_{\mathrm{BV}} A^{*a}_{\mu} &= D^{\nu}F^{a}_{\mu\nu} - J^{a}_{\mu} - \ui \tilde{f}_{bc}^{\phantom{bc}a} c^{b} A^{*c}_{\mu}, \\
\delta_{\mathrm{BV}} \bar{\phi}^{*}_{i} &= (D^{\mu} D_{\mu}\bar{\phi})_{i}  + m^2 \bar{\phi}_{i} - \ui c^{a} \bar{\phi}^{*}_{j}(T_a)^{j}_{\phantom{j} i}, \\
\delta_{\mathrm{BV}} \phi^{*i}&= (D^{\mu} D_{\mu}\phi)^{i} + m^2 \phi^{i} + \ui c^{a}(T_{a})^{i}_{\phantom{i}j} \phi^{*j},\\
\delta_{\mathrm{BV}} c^{*a} &= - D_{\mu} A^{*a \mu} - \ui \tilde{f}_{bc}^{\phantom{bc}a} c^{*b}c^{c} + \ui \bar{\phi}^{*}_{i} (T^{a})^{i}_{\phantom{i}j} \phi^{j} - \ui \bar{\phi}_{i} (T^{a})^{i}_{\phantom{i}j} \phi^{*j}. \\
\end{split}
\end{align}

As we will see in the next sections, the BV transformations in \eqref{eq:BVtransf-CSM}, \eqref{eq:BVtransf-QCD} and \eqref{eq:BVtransf-scalarQCD} are the basic structures needed to extract the $L_{\infty}$ algebra of their respective theories.


\section{$L_{\infty}$-structure for CSM theory}\label{sec:CSM}
In this section we shall present the $L_{\infty}$-structure underlying CSM theory and use it to determine recursion relations for its tree-level scattering amplitudes. The key ingredient in this construction is the perturbiner expansion, which can be formally obtained in the transition to the minimal model of its corresponding $L_{\infty}$ algebra \cite{Lopez-Arcos:2019hvg}. We will also describe how to derive exact expressions for the tree-level scattering amplitudes in CSM theory by analysing the respective homotopy Maurer-Cartan action in the cohomology.

\subsection{The CSM $L_\infty$ algebra}

We will first describe the $L_{\infty}$-algebra associated to the CSM theory. For more information on the theory of $L_{\infty}$-algebras we refer the reader to \cite{Jurco:2018sby} (see also \cite{Jurco:2019bvp}). 

Denote by $\Omega^{r}(\mathbb{R}^{1,2},\mathfrak{su}(N))$ the space of $r$-forms on $\mathbb{R}^{1,2}$ with values in $\mathfrak{su}(N)$, and by $\Omega^{0}(\mathbb{R}^{1,2}, S \otimes V)$ the space of $0$-forms on $\mathbb{R}^{1,2}$ with values in $S \otimes V$. We let ${\rm d}$ be the exterior differential, $\ast$ the Hodge star operator induced by the Minkowski metric, and $\delta= \ast {\rm d} \ast$ the corresponding codifferential. Then, the graded vector space $L$ underlying the CSM $L_{\infty}$-algebra is 
\begin{align}
L^{0} &= \Omega^{0}(\RR^{1,2},\su(N)), \nonumber\\
L^{1} &= \Omega^{1}(\RR^{1,2},\su(N)) \oplus \Omega^{0}(\RR^{1,2}, S \otimes V) \oplus \Omega^{0}(\RR^{1,2}, S \otimes \bar{V}), \nonumber \\
L^{2} &= \Omega^{1}(\RR^{1,2},\su(N)) \oplus \Omega^{0}(\RR^{1,2}, S \otimes \bar{V}) \oplus \Omega^{0}(\RR^{1,2}, S \otimes V), \nonumber \\
L^{3} &= \Omega^{0}(\RR^{1,2},\su(N)). \nonumber
\end{align}
In the terminology of subsection \ref{sec:2.2}, an element $c \in L^{0}$ is related to the gauge parameters, an element $A + \psi + \bar{\psi} \in L^{1}$ is a triple consisting of a gauge field, a Dirac spinor and a conjugate Dirac spinor, an element $A^* + \bar{\psi}^*+\psi^* \in L^{2} $ is the triple of antifields conjugate to $A$, $\psi$ and $\bar{\psi}$, and an element $c^* \in  L^{3}$ correponds to the antifields of $c$. The higher order brackets on $L$ are conveniently encoded in the BV transformations displayed in equation \eqref{eq:BVtransf-CSM}. As explained in the subsection 4.3 of \cite{Jurco:2018sby}, they can be cast as
\begin{align}\label{eq:BV-Linfinity}
\begin{split}
\delta_{\mathrm{BV}} c^{a} &= - \tfrac{1}{2} l_2(c,c)^{a}, \\
\delta_{\mathrm{BV}} A^{a}_{\mu} &=  l_1(c)^{a}_{\mu} + l_2(A,c)^{a}_{\mu} + \tfrac{1}{2} l_3(A,A,c)^{a}_{\mu} + \tfrac{1}{2} l_3(\psi + \bar{\psi},\psi + \bar{\psi},c)^{a}_{\mu} + \tfrac{1}{2} l_3(c,c,A^*)^{a}_{\mu}, \\
\delta_{\mathrm{BV}} \psi^{i} &=l_2(\psi ,c)^{i} + l_3(A,\psi ,c)^{i} + l_3(c,c, \psi^*)^{i}, \\ 
\delta_{\mathrm{BV}} \bar{\psi}_{i} &= l_2(\bar{\psi},c)_{i} + l_3(A,\bar{\psi},c)_{i} + l_3(c,c,\bar{\psi}^*)_{i},\\
\delta_{\mathrm{BV}} A^{*a}_{\mu} &= - l_1(A)^{a}_{\mu} - \tfrac{1}{2} l_2(A,A)^{a}_{\mu} - \tfrac{1}{2} l_2(\psi + \bar{\psi},\psi + \bar{\psi})^{a}_{\mu}- l_2(c,A^*)^{a}_{\mu}  + \cdots, \\
\delta_{\mathrm{BV}} \bar{\psi}^*_i &= -l_1(\psi + \bar{\psi})_{i} - l_2(A,\bar{\psi})_{i}  - l_2(c,\bar{\psi}^*)_{i} + \cdots,\\
\delta_{\mathrm{BV}} \psi^{*i} &= -l_1(\psi + \bar{\psi})^{i} - l_2(A,\psi )^{i}  - l_2(c, \psi^*)^{i} + \cdots,\\
\delta_{\mathrm{BV}} c^{*a} &= l_1(A^*)^{a} + l_2(A,A^*)^{a} - l_2(c,c^*)^{a} + l_2(\psi + \bar{\psi}, \bar{\psi}^* + \psi^*)^{a} +  \cdots ,  
\end{split}
\end{align}
where for the antifields we have explicitly written only those terms which are needed for our discussion. The non-vanishing higher order brackets $l_n$ can be determined by a direct comparisson between the above expressions with \eqref{eq:BVtransf-CSM}. For $l_{1}$ we have
\begin{equation}\label{eq:CSMl1}
\begin{array}{rc}
l_{1}(c)_{\mu a}=-\partial_{\mu}c_{a} & \in L^{1},\\
l_{1}(A)_{\mu}^{a}=-\frac{k}{2\pi}\epsilon^{\mu\nu\rho}\partial_{\nu}A_{\rho}^{a} & \in L^{2},\\
l_1(\psi + \bar{\psi})^{i} = -\ui \gamma^{\mu} \partial_{\mu} \psi^{i} + m \psi^{i} & \in L^{2},\\
l_1(\psi + \bar{\psi})_{i} = -\ui  \partial_{\mu} \bar{\psi}_i \gamma^{\mu} - m \bar{\psi}_i & \in L^{2},\\
l_{1}(A^{*})_{a}=-\partial^{\mu}A_{\mu a}^{*} & \in L^{3},\\
\end{array}
\end{equation}
satisfying the $L_{\infty}$ relation
\begin{equation}
l_{1}(l_{1}(x))=0.\label{eq:l1-relation}
\end{equation}
For $l_{2}$ we have
\begin{equation}\label{eq:CSMl2}
\begin{array}{rc}
l_2(c_1,c_2)^{a} = \ui \tilde{f}_{bc}^{\phantom{bc}a} c_1^{b} c_2^{c} & \in L^{0},\\
l_2(c_1,A_2)^{a}_{\mu} =  \ui \tilde{f}_{bc}^{\phantom{bc}a}   c_1^{b} A_{2\mu}^{c} & \in L^{1},\\
l_2(c_1,A_2^*)^{a}_{\mu} = \ui \tilde{f}_{bc}^{\phantom{bc}a} c_1^{b} A_{2\mu}^{*c} & \in L^{2},\\
l_2(c_1,c_2^*)^{a}  =- \ui \tilde{f}_{bc}^{\phantom{bc} a}  c_1^{b} c_2^{*c} & \in L^{3},\\
l_2(A_1,A_2)^{a}_{\mu} = \frac{\ui \kappa}{2\pi} \varepsilon_{\mu\nu\rho} \tilde{f}_{bc}^{\phantom{bc}a}  A^{b\nu}_{1} A^{c\rho}_2 & \in L^{2}\\
l_2(A_1,A_2^*)^{a} = \ui f_{bc}^{\phantom{bc}a} A_{1\mu}^{b} A_2^{*c\mu} & \in L^{3}\\
l_2(c_1,\psi_2)^{i} = -\ui c_1^{a} (T_{a})^{i}_{\phantom{i}j} \psi_2^{j} & \in L^{1},\\
l_2(c_1,\bar{\psi}_2)_{i} = \ui c_1^{a} \bar{\psi}_{2j} (T_{a})^{j}_{\phantom{j}i} & \in L^{1},\\
l_2(c_1,\psi^*_2)^{i} = - \ui c_1^{a} (T_{a})^{i}_{\phantom{i}j} \psi_2^{*j} & \in L^{2},\\
l_2(c_1,\bar{\psi}^*_2)_{i} =  \ui c_1^{a} \bar{\psi}_{2j}(T_{a})^{j}_{\phantom{j}i} & \in L^{2},\\
l_2(A_1,\psi_2)^{i} = \ui \gamma^{\mu} A_{1\mu}^{a} (T_{a})^{i}_{\phantom{i}j}\psi_2^{j} & \in L^{2},\\
l_2(A_1,\bar{\psi}_2)_{i} =- \ui  A_{1\mu}^{a} \bar{\psi}_{2j}(T_{a})^{j}_{\phantom{j}i} \gamma^{\mu} & \in L^{2},\\
l_2 (\psi_1 + \bar{\psi}_1,\psi_2 + \bar{\psi}_2)^{a}_{\mu}= (T^{a})^{i}_{\phantom{i}j}[(\bar{\psi}_{1 i} \gamma_{\mu} \psi^{j}_2) + (\bar{\psi}_{2 i} \gamma_{\mu}\psi^{j}_1)] & \in L^{2},\\
l_2 (\psi_1 + \bar{\psi}_1,\bar{\psi}^*_2 + \psi^*_2)^{a} = \ui  (T^{a})^{i}_{\phantom{i}j}(\bar{\psi}{}^*_{2i} \psi_1^{j} - \bar{\psi}{}_{1i}\psi_2^{*j}) & \in L^{3},\\
\end{array}
\end{equation}
such that
\begin{equation}
l_{1}\left(l_{2}\left(x_{1},x_{2}\right)\right)=l_{2}\left(l_{1}\left(x_{1}\right),x_{2}\right)+\left(-1\right)^{|x_{1}|}l_{2}\left(x_{1},l_{1}\left(x_{2}\right)\right),\label{eq:l2-relation}
\end{equation}
with $|x_{1}|$ denoting the grade of $x_{1}$. All the higher products
$l_{n}$, with $n\geq3$, vanish.

We can now identify the underlying cochain complex of this algebra:
$$
\begin{tikzpicture}[scale=1.0,baseline=-0.1cm, inner sep=1mm,>=stealth]
\node (0) at (-0.9,0)  {$\Omega^{0}(\RR^{1,2},\su(N))$};
\node (1) at (2.68,0)  {$\Omega^{1}(\RR^{1,2},\su(N))$};
\node (2) at (8.32,0)  {$\Omega^{1}(\RR^{1,2},\su(N))$};
\node (3) at (11.9,0)  {$\Omega^{0}(\RR^{1,2},\su(N)).$};
\node (4) at (2.68,-0.6) {$\oplus$};
\node (5) at (8.32,-0.6) {$\oplus$};
\node (6) at (2.68,-1.2) {$\Omega^{0}(\RR^{1,2},S \otimes V)$};
\node (7) at (8.32,-1.2) {$\Omega^{0}(\RR^{1,2},S \otimes \bar{V})$};
\node (8) at (2.68,-1.8) {$\oplus$};
\node (9) at (8.32,-1.8) {$\oplus$};
\node (10) at (3.95,-1.8) {};
\node (11) at (7.05,-1.8) {};
\node (12) at (2.68,-2.4) {$\Omega^{0}(\RR^{1,2},S \otimes \bar{V})$};
\node (13) at (8.32,-2.4) {$\Omega^{0}(\RR^{1,2},S \otimes V)$};
\tikzset{every node/.style={fill=white}} 
\draw[->, thick] (0) to node[midway,above=0.2pt] {\scriptsize$\ud$} (1); 
\draw[->, thick] (1) to node[midway,above=0.2pt] {\scriptsize$\ast\ud$} (2); 
\draw[->, thick] (2) to node[midway,above=0.2pt] {\scriptsize$\delta$} (3); 
\draw[->, thick] (10) to node[midway,above=0.2pt] {\scriptsize$\begin{pmatrix}
0 & \ui \slashed{\partial} + m  \\
\ui \slashed{\partial} - m & 0 
\end{pmatrix}$} (11);
\end{tikzpicture}
$$

This $L_{\infty}$-algebra can be made cyclic by setting the non-vanishing  inner products between the different elemets of the algebra to be
\begin{align}\label{eq:inner-CSM}
\begin{split}
\langle c , c^*  \rangle &=  \int_{\RR^{1,2}} \ud^3 x \,  c^*_{a} c^{a}, \\
\langle A , A^*  \rangle &=  \int_{\RR^{1,2}} \ud^3 x \, A^{*\mu}_{a} A_{\mu}^{a} , \\
\langle \psi + \bar{\psi} , \bar{\psi}^* + \psi^*  \rangle &= \int_{\RR^{1,2}} \ud^3 x \, (  \psi^{*i} \bar{\psi}_{i} +  \bar{\psi}{}^{*}_{i} \psi^{i}  ), 
\end{split}
\end{align}
where $c \in L^{0}$, $c^* \in L^{3}$, $A+\psi + \bar{\psi}\in L^{1}$ and $A^* +  \bar{\psi}^* + \psi^* \in L^{2}$. 

Using the inner products above, it is straightforward to verify that  the homotopy Maurer-Cartan action for the $L_{\infty}$-algebra $\mathfrak{L}_{\mathrm{CSM}}$, given by
\begin{equation}\label{eq:MCaction}
S_{\mathrm{MC}}[A + \psi + \bar{\psi}] = \sum_{n=0}^{\infty} \frac{1}{(n+1)!} \langle A + \psi + \bar{\psi}, l_n(A + \psi + \bar{\psi},\ldots,A + \psi + \bar{\psi}) \rangle,
\end{equation}
indeed coincides with the CSM action \eqref{eq:CSMaction},
\begin{equation}
S_{\mathrm{MC}}[A + \psi + \bar{\psi}] =  \int_{\RR^{1,2}} \ud^3 x \,\left\{ -  \frac{\kappa}{4\pi}\varepsilon^{\mu\nu\rho} \left( A^{a}_{\mu}  \partial_{\nu} A_{\rho a}  -  \frac{\ui} {3} \tilde{f}_{abc} A^{a}_{\mu} A^{b}_{\nu} A^{c}_{\rho}\right) +  \bar{\psi}(\ui \slashed{D} - m) \psi \right\}. 
\end{equation}

Since our goal is to deal with multiparticle solutions (perturbiner expansion), we have to slighty modify the definition of the $L_{\infty}$-algebra $L$. To that end, let $(k_p)_{p \geq 1}$ be an infinite set of momentum vectors in $\RR^{1,2}$. Let also $\mathcal{OW}_n$ be the set of words $P = p_1 p_2 \cdots p_n$ of length $n$ with $p_1 < p_2 < \cdots < p_n$. If $P = p_1 p_2 \cdots p_n$ is one such word, we put $k_P = k_{p_1} + k_{p_2} + \cdots + k_{p_n}$. We denote by $\mathscr{E}^{0}(\RR^{1,2})$ the space of formal series of the form
\begin{equation}\label{eq:3.4}
h(x) = \sum_{n \geq 1} \sum_{P \in \mathcal{OW}_n}  h_{P} \,  \ue^{-\ui k_{P} \cdot x},
\end{equation}
by $\mathscr{E}^{r}(\RR^{1,2},\su(N))$ the space of $r$-forms on $\RR^{1,2}$ with values in $\su(N)$ and coefficients on $\mathscr{E}^{0}(\RR^{1,2})$, by $\mathscr{E}^{0}(\RR^{1,2},S \otimes V)$ the space of $0$-forms on $\RR^{1,2}$ with values in $S \otimes V$ and coefficients on $\mathscr{E}^{0}(\RR^{1,2})$, and by $\mathscr{E}^{0}(\RR^{1,2},S \otimes \bar{V})$ the space of $0$-forms on $\RR^{1,2}$ with values in $S \otimes \bar{V}$ and coefficients on $\mathscr{E}^{0}(\RR^{1,2})$. We further extend the definition of the exterior differential $\ud$, the Hodge star operator $\ast$, the codifferential $\delta$ and the Dirac operator $\slashed{\partial}$ to the spaces $\mathscr{E}^{\sbullet}(\RR^{1,2},\su(N))$, $\mathscr{E}^{0}(\RR^{1,2},S \otimes V)$ and $\mathscr{E}^{0}(\RR^{1,2},S \otimes \bar{V})$ in the obvious way. 

In terms of these new spaces, the graded vector space underlying the cyclic $L_{\infty}$-algebra $L$ controlling the perturbiner expansion for the CSM theory is 
\begin{align*}
L^{0} &= \mathscr{E}^{0}(\RR^{1,2},\su(N)), \\
L^{1} &= \mathscr{E}^{1}(\RR^{1,2},\su(N)) \oplus \mathscr{E}^{0}(\RR^{1,2}, S \otimes V) \oplus \mathscr{E}^{0}(\RR^{1,2}, S \otimes \bar{V}), \\
L^{2} &= \mathscr{E}^{1}(\RR^{1,2},\su(N)) \oplus \mathscr{E}^{0}(\RR^{1,2}, S \otimes \bar{V}) \oplus \mathscr{E}^{0}(\RR^{1,2}, S \otimes V), \\
L^{3} &= \mathscr{E}^{0}(\RR^{1,2},\su(N)). 
\end{align*}
The higher order brackets and the cyclic inner product are determined by the same formulas as in \eqref{eq:CSMl1}, \eqref{eq:CSMl2} and \eqref{eq:inner-CSM}.

\subsection{The perturbiner expansion for CSM}\label{sec:3.2}
Now we turn our attention to the perturbiner expansion for the CSM theory. As explained in \cite{Lopez-Arcos:2019hvg}, this can be naturally obtained from the minimal model for the $L_{\infty}$-algebra $L$. 

Let us start by describing the minimal $L_{\infty}$-structure on the cohomology\footnote{The cohomology is given by the plane-wave states annihilated by $l_1$. For the gauge bosons, this is empty as Chern-Simons fields do not have dynamical degrees of freedom. For the spinors, the cohomology is simply given by states satisfying the massive Dirac equation.} $H^{\sbullet}(L)$ of $L$. For this purpose, we need to define a projection $p \colon L \to H^{\sbullet}(L)$ and a contracting homotopy $h \colon L \to L$. The first thing to notice is that the cochain complex underlying $H^{\sbullet}(L)$ is
$$
\begin{tikzpicture}[scale=1.0,baseline=-0.1cm, inner sep=1mm,>=stealth]
\node (0) at (-0.7,0)  {$\ker (\ud)$};
\node (1) at (2,0)  {$0$};
\node (2) at (8,0)  {$0$};
\node (3) at (11,0)  {$\coker(\delta).$};
\node (4) at (2,-0.6) {$\oplus$};
\node (5) at (8,-0.6) {$\oplus$};
\node (6) at (2,-1.55) {$\ker \begin{pmatrix}
0 & \ui \slashed{\partial} + m  \\
\ui \slashed{\partial} - m & 0 
\end{pmatrix}$};
\node (7) at (8,-1.55) {$\coker \begin{pmatrix}
0 & \ui \slashed{\partial} + m  \\
\ui \slashed{\partial} - m & 0 
\end{pmatrix}$};
\tikzset{every node/.style={fill=white}} 
\draw[->, thick] (0) to node[midway,above=0.2pt] {\scriptsize$0$} (1); 
\draw[->, thick] (1) to node[midway,above=0.2pt] {\scriptsize$0$} (2); 
\draw[->, thick] (2) to node[midway,above=0.2pt] {\scriptsize$0$} (3); 
\draw[->, thick] (6) to node[midway,above=0.2pt] {\scriptsize$0$} (7);
\end{tikzpicture}
$$ 
The first two components of the projection $p^{(0)}$ and $p^{(1)}$ are thus chosen to be the natural projections induced by the Hodge-Kodaira decomposition\footnote{See  \cite{Kajiura:2003ax, Macrelli:2019afx} for more details. For physyicsts, the Hodge-Kodaira decomposition might be better understood as follows. In a given BRST quantized theory, any set of states can be decomposed into three parts: spurious (BRST-exact), physical and unphysical. Here, $l_1$ plays the role of the nilpotent BRST charge.}, while the last two components $p^{(2)}$ and $p^{(3)}$ are chosen to be the ones induced by the quotient maps.  As for the contracting homotopy, it is necessary to consider the massless Feynman propagator $\Dsf^{\mathrm{F}}$, the Chern-Simons Feynman propagator $\Csf^{\mathrm{F}}$ and the spinor Feynman propagator $\Ssf^{\mathrm{F}}$, defined on the spaces of $0$-forms on $\RR^{1,2}$, $1$-forms on $\RR^{1,2}$, and $0$-forms on $\RR^{1,2}$ with values in $(S \otimes V) \oplus (S \otimes \bar{V})$, respectively. Their explicit expressions are
\begin{align}\label{eq:3.5}
\begin{split}
\Dsf^{\mathrm{F}} =  \frac{1}{\square}, \quad  \Csf^{\mathrm{F}} = \frac{2 \pi}{\kappa} \varepsilon_{\mu\nu\rho} \frac{\partial^{\mu}}{\square},\quad \Ssf^{\mathrm{F}} = -\frac{1}{\square + m^2}\begin{pmatrix}
0 & \ui \slashed{\partial} + m  \\
\ui \slashed{\partial} - m & 0 
\end{pmatrix},
\end{split}
\end{align}
or, when acting on plane waves of the form $\ue^{- \ui k \cdot x}$,
\begin{align}\label{eq:3.6}
\begin{split}
\Dsf^{\mathrm{F}} = - \frac{1}{k^2}, \quad  \Csf^{\mathrm{F}} =  \frac{2 \pi}{\kappa} \varepsilon_{\mu\nu\rho} \frac{\ui k^{\mu}}{k^2},\quad \Ssf^{\mathrm{F}} = \frac{1}{k^2 - m^2}\begin{pmatrix}
0 & \slashed{k} + m  \\
 \slashed{k} - m & 0 
\end{pmatrix}. 
\end{split}
\end{align}
We extend all three propagators $\Dsf^{\mathrm{F}}$, $\Csf^{\mathrm{F}}$ and $\Ssf^{\mathrm{F}}$ in such a way as to yield linear operators $\Dsf^{\mathrm{F}} \colon \mathscr{E}^{0}(\RR^{1,2},\su(N)) \to \mathscr{E}^{0}(\RR^{1,2},\su(N))$, $\Csf^{\mathrm{F}} \colon \mathscr{E}^{1}(\RR^{1,2},\su(N)) \to \mathscr{E}^{1}(\RR^{1,2},\su(N))$ and $\Ssf^{\mathrm{F}} \colon \mathscr{E}^{0}(\RR^{1,2},S \otimes \bar{V}) \oplus \mathscr{E}^{0}(\RR^{1,2},S \otimes V) \to  \mathscr{E}^{0}(\RR^{1,2},S \otimes V) \oplus \mathscr{E}^{0}(\RR^{1,2},S \otimes \bar{V})$. The three non-zero components of the contracting homotopy $h$ then read as
\begin{align}\label{eq:3.7}
\begin{split}
h^{(1)} = \left(\begin{array}{cc} \Dsf^{\mathrm{F}}\circ \delta &\, 0 \end{array} \right) &\colon L^{1} \longrightarrow L^{0}, \\ 
h^{(2)} = \left(\begin{array}{cc} \Csf^{\mathrm{F}} & 0 \\ 0 & \Ssf^{\mathrm{F}} \end{array} \right) &\colon L^{2} \longrightarrow L^{1}, \\
h^{(3)} = \left(\begin{array}{c} \ud \circ \Dsf^{\mathrm{F}}  \\ 0 \end{array} \right) &\colon L^{3}  \longrightarrow L^{2},
\end{split}
\end{align}
or, more explicitly,
\begin{align}\label{eq:3.8}
\begin{split}
h^{(1)}(A)^{a} &= - \frac{\partial_{\mu}}{\square} A^{a \mu},  \\
h^{(2)}(A^*)^{a}_{\mu} &= \frac{2 \pi}{\kappa} \varepsilon_{\mu\nu\rho} \frac{\partial^{\nu}}{\square} A^{*a\rho}, \\
 h^{(2)}(\bar{\psi}^* + \psi^*)^{i} &= -\frac{\ui \slashed{\partial} + m }{\square + m^2} \psi^{* i}, \\
  h^{(2)}(\bar{\psi}^* + \psi^*)_{i} &= -\frac{\ui \slashed{\partial} - m }{\square + m^2} \bar{\psi}^{*}_{i}, \\
h^{(3)}(c^{*})^{a}_{\mu} &=  - \frac{\partial_{\mu}}{\square} c^{* a},
\end{split}
\end{align}
where $A \in L^{1}$, $A^* + \bar{\psi}^* + \psi^* \in L^{2}$ and $c^* \in L^{3}$. With these prescriptions, the quasi-isomorphism between $H^{\sbullet}(L)$ and $L$ is determined by maps $f_n \colon H^{\sbullet}(L)^{\otimes n} \to L$ which are constructed recursively using $h$, whereas the higher order brackets $l'_n \colon H^{\sbullet}(L)^{\otimes n} \to H^{\sbullet}(L)$ are constructed recursively using $p$. We shall not reproduce here the explicit expressions, but instead refer the reader to the Appendix A of \cite{Macrelli:2019afx} (see also \cite{Jurco:2018sby}). It is important to note, though, that such expressions are derived under the assumption that $h^{(1)}(A + \psi + \bar{\psi}) = 0$ for any $A + \psi + \bar{\psi} \in L^{1}$. By the definition above, this implies that the first summands $A \in \mathscr{E}^{1}(\RR^{1,2},\su(N))$ should satisfy the Lorenz gauge condition $\partial_{\mu} A^{a\mu} = 0$, since this corresponds to our chosen representatives of the cohomology. Different gauge conditions are of course possible but they would have to be accompanied by a corresponding change in the contracting homotopy (i.e. the propagators in the field theory).

We shall now obtain the perturbiner expansion for the CSM theory by using the minimal $L_{\infty}$-structure on $H^{\sbullet}(L)$. For this, we consider a Maurer-Cartan element $\psi' + \bar{\psi}' \in H^1(L) = \ker \left( \begin{smallmatrix}
0 & \ui \slashed{\partial} + m  \\
\ui \slashed{\partial} - m & 0 
\end{smallmatrix} \right)$, in which each summand is written in the form
\begin{align}\label{eq:coh-elem-CSM}
\begin{split}
\psi' &= \sum_{p \geq 1} \Psi_{p}\, \ue^{-\ui k_p \cdot x}, \\ 
\bar{\psi}' &= \sum_{p \geq 1} \bar{\Psi}_{p} \, \ue^{-\ui k_p \cdot x},
\end{split}
\end{align}
which are the simplest multi-particle solutions of the free Dirac's equation. 

Then we define the perturbiner expansion as the Maurer-Cartan element $A + \psi + \bar{\psi}$ in $L$ given by the formula
\begin{equation}
A + \psi + \bar{\psi} = \sum_{n \geq 1} \frac{1}{n!} f_{n} (\psi' + \bar{\psi}',\dots, \psi' + \bar{\psi}'). 
\end{equation}
To help understand this definition, we work out the components $A \in \mathscr{E}^{1}(\RR^{1,2},\su(N))$, $\psi \in \mathscr{E}^{0}(\RR^{1,2},S \otimes V)$ and $\bar{\psi} \in \mathscr{E}^{0}(\RR^{1,2},S \otimes \bar{V})$ separately. Here we may borrow from the analysis carried out in the subsection 3.2 of \cite{Lopez-Arcos:2019hvg}. To start off, we notice that, on general grounds,
\begin{align}
\begin{split}
&f_{n}(\psi' + \bar{\psi}',\dots,\psi'  + \bar{\psi}') \\
 &\qquad = -\tfrac{1}{2} \sum_{k=1}^{n-1} \binom{n}{k} (h^{(2)} \circ l_2) (f_{k}(\psi' + \bar{\psi}',\dots,\psi' +  \bar{\psi}'),f_{n-k}(\psi' + \bar{\psi}',\dots,\psi' + \bar{\psi}')). 
\end{split}
\end{align}
The higher order brackets in \eqref{eq:CSMl2} and the definition of $h^{(2)}$ in \eqref{eq:3.7} then tells us that
\begin{align}\label{eq:3.12}
\begin{split}
A^{a \mu} &= \sum_{n \geq 1} \frac{1}{(2n)!} f_{2n} (\psi' + \bar{\psi}',\dots, \psi' + \bar{\psi}')^{\mu a}, \\ 
\psi^{i} &= \sum_{n \geq 1} \frac{1}{(2n+1)!} f_{2n+1} (\psi' + \bar{\psi}',\dots, \psi' + \bar{\psi}')^{i}, \\
\bar{\psi}_{i} &= \sum_{n \geq 1} \frac{1}{(2n+1)!} f_{2n+1} (\psi' + \bar{\psi}',\dots, \psi' + \bar{\psi}')_{i},
\end{split}
\end{align}
where now
\begin{align}
\begin{split}
&f_{2n}(\psi' + \bar{\psi}',\dots,\psi' + \bar{\psi}')^{a \mu} = -\tfrac{1}{2} \sum_{k=1}^{2n-1} \binom{2n}{k}  (\Csf^{\mathrm{F}} \circ l_2) (f_{k}(\psi' + \bar{\psi}',\dots,\psi' + \bar{\psi}'), \\
&\qquad \qquad \qquad \qquad\qquad \qquad\qquad \qquad\qquad \qquad\qquad \quad f_{2n-k}(\psi' + \bar{\psi}',\dots,\psi' + \bar{\psi}'))^{a\mu}, \\ 
&f_{2n+1}(\psi' + \bar{\psi}',\dots,\psi' + \bar{\psi}')^{i}  = -\tfrac{1}{2} \sum_{k=1}^{2n} \binom{2n+1}{k} (\Ssf^{\mathrm{F}} \circ l_2) (f_{k}(\psi' + \bar{\psi}',\dots,\psi' + \bar{\psi}'), \\
&\qquad \qquad \qquad \qquad\qquad \qquad\qquad \qquad\qquad \qquad\qquad \quad f_{2n+1 -k}(\psi' + \bar{\psi}',\dots,\psi' + \bar{\psi}'))^{i}, \\
&f_{2n+1}(\psi' + \bar{\psi}',\dots,\psi' + \bar{\psi}')_{i}  = -\tfrac{1}{2} \sum_{k=1}^{2n} \binom{2n+1}{k} (\Ssf^{\mathrm{F}} \circ l_2) (f_{k}(\psi' + \bar{\psi}',\dots,\psi' + \bar{\psi}'), \\
&\qquad \qquad \qquad \qquad\qquad \qquad\qquad \qquad\qquad \qquad\qquad \quad f_{2n+1 -k}(\psi' + \bar{\psi}',\dots,\psi' + \bar{\psi}'))_{i}.
\end{split}
\end{align}
Using mathematical induction and taking note of \eqref{eq:3.6}, we reduce the above to
\begin{align}\label{eq:3.15}
\begin{split}
f_{2n}(\psi' + \bar{\psi}',\dots,\psi' + \bar{\psi}')^{a \mu} &= (2n)! \sum_{P \in \mathcal{OW}_{2n}} \ \mathcal{A}^{a \mu}_{P} \,\ue^{-\ui k_{P} \cdot x}, \\ 
f_{2n+1}(\psi' + \bar{\psi}',\dots,\psi' + \bar{\psi}')^{i} &=(2n + 1)! \sum_{P \in \mathcal{OW}_{2n + 1}} \Psi_{P}^{i} \, \ue^{-\ui k_{P} \cdot x},  \\
f_{2n+1}(\psi' + \bar{\psi}',\dots,\psi' + \bar{\psi}')_{i} &=(2n + 1)! \sum_{P \in \mathcal{OW}_{2n + 1}} \bar{\Psi}_{P i} \,\ue^{-\ui k_{P} \cdot x}, 
\end{split}
\end{align}
where the coefficients $\mathcal{A}^{a\mu}_{P}$, $\Psi^{i}_{P}$ and $\bar{\Psi}_{Pi}$ are determined from the recursion relations
\begin{align}\label{eq:pert-CSM}
\begin{split}
\mathcal{A}^{\mu a}_{P} &= \frac{\mathcal{J}^{\mu a}_{P}}{s_P} -\frac{k_{P\lambda}}{s_{P}} \sum_{P = Q \cup R}  \varepsilon^{\lambda \mu \nu} \varepsilon_{\nu \rho \sigma} \tilde{f}^{\phantom{bc}a}_{bc}  \mathcal{J}^{\rho b}_{Q} \mathcal{J}^{\sigma c}_{R}, \\
\mathcal{J}^{\mu a}_{P} &=  \frac{2\pi \ui }{\kappa } \, k_{P\lambda} \sum_{P = Q \cup R} \varepsilon^{\lambda \mu \nu} \bar{\Psi}_{Qi} \gamma_{\nu} (T^{a})^i_{\,j} \Psi_R^j, \\
\Psi_P^i &= - \frac{\slashed{k}_P + m}{s_{P} - m^2} \sum_{P = Q \cup R} \slashed{\mathcal{A}}_{Q}^{a} (T_{a})^i_{\, j} \Psi^j_R, \\
\bar{\Psi}_{Pi} &= \sum_{P = Q \cup R} \bar{\Psi}_{Rj} (T_{a})^j_{\, i} \slashed{\mathcal{A}}_{Q}^{a} \frac{\slashed{k}_P - m}{s_{P} - m^2}.
\end{split}
\end{align}
Here, $P = Q \cup R$ denotes the distribution of the letters of the ordered words $P$ into non-empty ordered words $Q$ and $R$. In line with the terminology used \cite{Mafra:2015vca,Mafra:2016mcc,Mafra:2016ltu,Mizera:2018jbh}, we refer to the coefficients $\mathcal{A}^{a\mu}_{P}$, $\Psi^{i}_{P}$ and $\bar{\Psi}_{Pi}$ as the Berends-Giele currents for $A$, $\psi$ and $\bar{\psi}$, respectively. Observe that the auxiliary quantities $\mathcal{J}^{\mu a}$ are the perturbiner coefficients of the topological current associated to the matter current \eqref{eq:matter-current}. While the latter is covariantly conserved, the former is divergenceless ($k_{P\mu}\mathcal{J}^{\mu a}_{P}=0$). The topological current is the one that generates the field $A^{\mu a}$, as can be seen in the equations. By plugging equation \eqref{eq:3.15} back in \eqref{eq:3.12}, we obtain
\begin{align}
\begin{split}
A^{a \mu} &= \sum_{n \geq 1} \sum_{P \in \mathcal{OW}_{2n}} \mathcal{A}^{a \mu}_{P} \, \ue^{-\ui k_P \cdot x} = \sum_{p < q}  \mathcal{A}^{a \mu}_{pq} \, \ue^{-\ui k_{pq} \cdot x} + \sum_{p < q < r < s} \mathcal{A}^{a \mu}_{pqrs} \, \ue^{-\ui k_{pqrs} \cdot x} + \cdots, \\
\psi^{i} &=  \sum_{n \geq 1} \sum_{P \in \mathcal{OW}_{2n+1}} \Psi^{i}_{P} \, \ue^{-\ui k_P \cdot x} = \sum_{p} \Psi^{i}_p \, \ue^{-\ui k_p \cdot x} + \sum_{p < q} \Psi^{i}_{pq} \, \ue^{-\ui k_{pq} \cdot x} + \cdots \\
\bar{\psi}_{i} &=  \sum_{n \geq 1} \sum_{P \in \mathcal{OW}_{2n+1}} \bar{\Psi}_{P i} \, \ue^{-\ui k_P \cdot x} = \sum_{p} \bar{\Psi}_{pi} \, \ue^{-\ui k_{p} \cdot x} + \sum_{p < q} \bar{\Psi}_{pq i} \, \ue^{-\ui k_{pq} \cdot x} + \cdots.
\end{split}
\end{align}

These expressions are called color-dressed pertubiners \cite{Mizera:2018jbh}, since they yield color-dressed amplitudes when applying the Berends-Giele formula. In conclusion, as advertised above, the recursion relations for the perturbiner coefficients are encoded in the recursion relations for the $L_{\infty}$-quasi-isomorphism from $H^{\sbullet}(L)$ onto $L$. 

Before continuing, we would like to add some physical input. The multi-particle
expansions above may be slightly misleading because there does not seem to be any restriction on repeated particle labels, e.g. $\mathcal{A}^{a \mu}_{112}$. In addition, a given label cannot simultaneously be used to describe different polarizations. In order to remediate this, we will formally
assume that the polarisations $A_{p\mu}^{a}$, $\psi_{p}^{i}$ and
$\bar{\psi}_{pi}$ have an internal structure given by
\begin{subequations}\label{eq:grassmann-polarization}
\begin{eqnarray}
A_{p\mu}^{a} & = & \tilde{A}_{p\mu}^{a}\theta_{p},\\
\psi_{p}^{i} & = & \tilde{\psi}_{p}^{i}\theta_{p},\\
\bar{\psi}_{pi} & = & \tilde{\bar{\psi}}_{pi}\theta_{p},
\end{eqnarray}
\end{subequations}where the tilded polarisations have the opposite
statistics to the original ones and $\theta_{p}$ is a Grasmannian
variable satisfying $\theta_{p} \theta_{q}=-\theta_{q} \theta_{p}$. Therefore, the single-particle
labels will never appear twice in a given ordered word and the computed multiparticle expansions are well defined. In spite of this construction, the final results
will always be written in terms of the original polarisations. This also applies to the following perturbiners in the present work.

Now we turn to the analysis of the tree-level scattering amplitudes for the CSM theory.

\subsection{Tree-level scattering amplitudes for CSM}\label{sec:3.3}

In this section, we will see  how  the homotopy Maurer-Cartan action  for $H^{\sbullet}(\mathfrak{L}_{\mathrm{CSM}})$ can be seen as a generator for tree level scattering amplitudes.
The starting point is to insert the multi-particle solution \eqref{eq:coh-elem-CSM} into the homotopy Maurer-Cartan action
\begin{equation}\label{eq:MC-ferm}
S'_{\mathrm{MC}}[\psi' + \bar{\psi}'] = \sum_{n \geq 2} \frac{1}{(n+1)!} \langle \psi' + \bar{\psi}', l'_{n}(\psi' + \bar{\psi}',\dots,\psi' + \bar{\psi}') \rangle. 
\end{equation}
In order to compute the sum, first notice that
\begin{equation}
 l'_{n}(\psi' + \bar{\psi}',\dots,\psi' + \bar{\psi}') = \tfrac{1}{2} \sum_{i=1}^{n-1} \binom{n}{i} (p^{(2)} \circ l_2) (f_{i}(\psi' + \bar{\psi}',\dots,\psi' + \bar{\psi}'),f_{n-i}(\psi' + \bar{\psi}',\dots,\psi' + \bar{\psi}')).
\end{equation}
Thus, using the $L_{\infty}$ relations \eqref{eq:CSMl2} and inner products displayed in \eqref{eq:inner-CSM}, the action \eqref{eq:MC-ferm} can be cast as
\begin{equation}\label{eqn:3.18}
S'_{\mathrm{MC}}[\psi' + \bar{\psi}',\dots,\psi' + \bar{\psi}'] = \sum_{n \geq 1} \frac{1}{(2n+2)!} \langle \psi' + \bar{\psi}', l'_{2n+1}(\psi' + \bar{\psi}',\dots,\psi' + \bar{\psi}') \rangle. 
\end{equation}
such that
\begin{eqnarray}
l'_{2n+1}(\psi' + \bar{\psi}',\dots,\psi' + \bar{\psi}')^{i} &=&  (2n+1)! \sum_{\substack{P \in \mathcal{OW}_{2n+1} \\ P = Q \cup R}}\slashed{\mathcal{A}}_{Q}^{a} (T_{a})^{i}_{\phantom{i}j} \Psi^{j}_R \ue^{-\ui k_{P} \cdot x},\\
l'_{2n+1}(\psi' + \bar{\psi}',\dots,\psi' + \bar{\psi}')_{i} &=& -(2n+1)! \sum_{\substack{P \in \mathcal{OW}_{2n+1} \\ P = Q \cup R}} \bar{\Psi}_{Rj} (T_{a})^{j}_{\phantom{j}i} \slashed{\mathcal{A}}^{a}_{Q} \ue^{-\ui k_{P} \cdot x}.
\end{eqnarray}
These results are straightforward to obtain using mathematical induction.

Now, using
\begin{equation}
\int_{\RR^{1,2}} \ud^3 x \, \ue^{-\ui k \cdot x} = (2\pi)^3 \delta(k),
\end{equation}
we finally obtain 
\begin{equation}\label{eq:MC-CSM}
S'_{\mathrm{MC}}= -(2 \pi)^3 \sum_{n \geq 1}\frac{1}{(2n+2)} \sum_{p \geq 1} \sum_{\substack{P \in \mathcal{OW}_{2n+1} \\ \hspace{-11pt}P = Q \cup R}} \delta(k_{pP}) \big\{\bar{\Psi}_{Rj} (T_{a})^{j}_{\phantom{j}i} \slashed{\mathcal{A}}^{a}_{Q}\Psi^j_{p} + \bar{\Psi}_{pi}\slashed{\mathcal{A}}_{Q}^{a} (T_{a})^{i}_{\phantom{i}j} \Psi^{j}_R \big\}.
\end{equation}
This action contains all the possible tree-level scattering amplitudes with arbitrary labels, i.e. redundant information for practical purposes.  In order to improve the output, we can constrain the multiple sums in several ways.  First, we can restrict the sum to a ``diagonal'' form by taking $p=2n+2$ in the summand. This takes care of the permutation factor $\frac{1}{(2n+2)}$. Additionally, we can remove the global constant factor and leave the momentum conservation delta function implicit.  We are then left with
\begin{equation}\label{eq:MC-CSM-diag}
S'^{\mathrm{CSM}}_{\mathrm{MC}} =  \sum_{n \geq 1}\sum_{\substack{P \in \mathcal{OW}_{2n+1} \\ \hspace{-11pt}P = Q \cup R}}\big\{\bar{\Psi}_{Rj} (T_{a})^{j}_{\phantom{j}i} \slashed{\mathcal{A}}^{a}_{Q}\Psi^j_{(2n+2)} + \bar{\Psi}_{(2n+2)i}\slashed{\mathcal{A}}_{Q}^{a} (T_{a})^{i}_{\phantom{i}j} \Psi^{j}_R \big\}.
\end{equation}

Since this expression is multilinear in the coefficients $\Psi_p^i$ and $\bar{\Psi}_{qi}$, we can apply an operator that exchanges a given coefficient by its boundary value, i.e. the actual polarisations we want to compute. Such operators can be defined as:
\ba\label{eq:ferm-oper} 
\tilde{\bar{\Psi}}_{pi}\frac{\partial}{\partial\bar{\Psi}_{pi}},\quad\quad\tilde{\Psi}_{q}^i\frac{\partial}{\partial\Psi_q^i}.
\ea
Note, in particular, that we do not have to worry about extra signs being introduced in the amplitude computation since they have bosonic statistic. The boundary conditions are usually expressed as $\tilde{\bar{\Psi}}_{pi}=\bar{v}_p\delta_i^{i_p}$ and  $\tilde{\Psi}_q^i=u_q\delta^i_{i_q}$, where $\bar{v}$ and $u$ satisfy
\ba\label{eq:fer-boundry} 
\bar{v}_p(\slashed{k}_p+m)=0,\quad\quad(\slashed{k}_q-m)u_q=0.
\ea

In Chern-Simons Matter theories, the amplitudes always have an even number of particles since the only bosonic field, the gauge vector, does not have any propagating degrees of freedom. Therefore, $2m$-point amplitudes of $m$ fermion-antifermion pairs are given by
\ba\label{eq:CMS-general-amps}
\mathscr{A}_{2m} = \prod_{q\in \mathcal{S}^1_{m}}\left(\tilde{\bar{\Psi}}_{qi}\frac{\partial}{\partial\bar{\Psi}_{qi}}\right)\prod_{r\in \mathcal{S}^2_{m}}\left(\tilde{\Psi}_{r}^i\frac{\partial}{\partial\Psi_r^i}\right)S'^{\mathrm{CSM}}_{\mathrm{MC}} \Bigg|_{\Psi, \bar{\Psi}=0},
\ea 
where the $\mathcal{S}^i_m$ are non-intersecting particle label sets of size $m$ which are determined by the explicit process to be computed. For example, the choice of two three-element sets picked from $(1,2,3,4,5,6)$ for the $\mathscr{A}_{6}$ amplitude.

It is worth pointing out that these amplitudes can also be obtained using the Berends-Giele formula on the currents with the respective boundary conditions. The $L_\infty$ construction offers an underlying algebraic structure, including a formalisation of the perturbiner expansion.

\subsection{Dyck words and flavoured amplitudes}\label{DyckLocalization}

One particular kind of amplitudes we are interested in are the ones with distinct flavors for the fermions, these can be extracted from the Maurer-Cartan action \eqref{eq:MC-CSM-diag} by restricting the set of words we take in the sum to Dyck words, as we will now show. 

Ampltidues as the ones above, involving only fermion-antifermion pairs, were studied in \cite{Melia:2013bta,Melia:2013epa} for QCD, where a new basis for amplitudes with matter was introduced employing Dyck words. This is known as the Melia basis. A conjecture for the generalisation of Melia's result was done in \cite{Johansson:2015oia} and it is known as the Johansson-Ochirov basis. This conjecture was later proved in \cite{Melia:2015ika}.  

Dyck words are regular words decorated with a string of parentheses ``$($'' and ``$)$'' which are \textit{properly} closed. These words help to keep track of the fermion-antifermion pairs in the amplitude with the decoration $\overline{p}$ (``$($'') for the antifermions and $\underline{q}$ (``$)$'') for the fermions.\footnote{Since we took solutions with $\ue^{-\ui k\cdot x}$ our conventions for ``$($'' and ``$)$'' are in the inverse order from \cite{Johansson:2015oia} .} For example,$$\overline{1}\overline{2}\underline{3}\underline{4}=(1(23)4)$$is a Dyck word, representing two antifermions, $1$ and $2$, colour paired with two fermions, respectively $4$ and $3$. On the other hand,$$\underline{1}\overline{2}\underline{3}\overline{4}=1)(23)(4$$ is not a Dyck word.

In our construction, we also have the ordering restriction of the words, which furhter simplifies the amplitude generator.

The last ingredient concerns the \textit{boundary conditions} of the problem. We will fix the particle in $\bar{\Psi}_1$ to be an antifermion and the $(2n+2)$ particle to be a fermion, so the second term on the right-hand side of \eqref{eq:MC-CSM-diag} drops out. This choice also means that the first and last particle have the same flavour. With all the previous considerations, we propose the following generator:
\begin{equation}
\label{eq:flav-CSM-gener}
\mathbf{G}_{CSM}=\sum_{n \geq 1} \sum_{\substack{Q \in \mathcal{OW}_{2n} \\ \hspace{-8pt}Q\in \mathrm{Dyck}}}\Big\{ \bar{\Psi}_{1i}\slashed{\mathcal{A}}_{Q}^{a} (T_{a})^{i}_{\phantom{i}j} \Psi_{2n+2}^j + \sum_{\substack{Q = R \cup S \\ \hspace{-8pt}R,S\in \mathrm{Dyck}}} \bar{\Psi}_{1i}(T_b)^{i}_{\phantom{i}j}\slashed{\mathcal{A}}_{R}^{b}\left(\frac{\slashed{k}_{1R} - m}{s_{1R} - m^2}\right)\slashed{\mathcal{A}}_{S}^{a} (T_{a})^{j}_{\phantom{j}k} \Psi_{2n+2}^k  \Big\}.
\end{equation}

In order to make sense of this generator, the perturbiner expansions have to be analogously localised. We will call a word even or odd according to its length. By construction, the (anti)fermion currents in \eqref{eq:pert-CSM} have only odd words $P$. In this case, due to the ordering condition, the label that corrupts the Dyck ``character'' of a word  will always be located in one of the extremes, such that $P=pQ$ and $Q\in\mathrm{Dyck}$. Thus, our perturbiner coefficients for odd words will be cast as 
\begin{subequations}\label{odd-localized}
\begin{multline}
\Psi^i_{pQ}= -\left(\frac{\slashed{k}_{pQ} + m}{s_{pQ} - m^2}\right)\slashed{\mathcal{A}}_{Q}^{a} (T_{a})^{i}_{\phantom{i}j}\Psi^j_p \\
+\left(\frac{\slashed{k}_{pQ} + m}{s_{pQ} - m^2}\right) \sum_{\substack{\hspace{-5pt}Q = R \cup S \\ \hspace{-2pt}R,S\in \mathrm{Dyck}}} \slashed{\mathcal{A}}_{R}^{a} (T_{a})^{i}_{\phantom{i}j}\left(\frac{\slashed{k}_{pS} + m}{s_{pS} - m^2}\right)(T_{b})^{j}_{\phantom{j}k} \slashed{\mathcal{A}}_{S}^{b}\Psi^k_p,
\end{multline}
\begin{multline}
\bar{\Psi}_{pQi}= \bar{\Psi}_{pj}\slashed{\mathcal{A}}_{Q}^{a} (T_{a})^{k}_{\phantom{k}i} \left(\frac{\slashed{k}_{pQ} - m}{s_{pQ} - m^2}\right) \\
 +\sum_{\substack{\hspace{-5pt}Q = R \cup S \\ \hspace{-2pt}R,S\in \mathrm{Dyck}}}\bar{\Psi}_{pj}(T_b)^{j}_{\phantom{j}k}\slashed{\mathcal{A}}_{R}^{b}\left(\frac{\slashed{k}_{pR} - m}{s_{pR} - m^2}\right)\slashed{\mathcal{A}}_{S}^{a} (T_{a})^{k}_{\phantom{k}i} \left(\frac{\slashed{k}_{pQ} - m}{s_{pQ} - m^2}\right).
\end{multline}
\end{subequations}

Analogously, the even part of the perturbiners in \eqref{eq:pert-CSM} shall be written instead as
\begin{subequations}\label{even-localized}
\begin{equation}
\mathcal{A}^{\mu a}_{P} = \frac{\mathcal{J}^{\mu a}_{P}}{s_P} -\frac{k_{P\lambda}}{s_{P}} \sum_{\substack{P = Q \cup R \\ Q \cup R\in \mathrm{Dyck}}}  \varepsilon^{\lambda \mu \nu} \varepsilon_{\nu \rho \sigma} \tilde{f}^{\phantom{bc}a}_{bc}  \mathcal{J}^{\rho b}_{Q} \mathcal{J}^{\sigma c}_{R},
\end{equation}
\begin{equation}
\mathcal{J}^{\mu a}_{P} =  \frac{2\pi \ui}{\kappa}\, k_{P\lambda} \sum_{\substack{P = Q \cup R \\ Q \cup R\in \mathrm{Dyck}}} \varepsilon^{\lambda \mu \nu} \bar{\Psi}_{Qi} \gamma_{\nu} (T^{a})^i_{\, j} \Psi^j_R.
\end{equation} 
\end{subequations}

In this construction, even words that are not Dyck are automatically projected out of the generator \eqref{eq:flav-CSM-gener}, i.e. they do not contribute to the fully-flavoured amplitudes. Once the boundary conditions are set, the $(2n+2)$-point fully-flavoured scattering amplitude for CSM can be finally expressed as
\ba\label{eq:CSM-amps}
\mathscr{A}^{\rm Flav.}_{2m} = \Big(\tilde{\bar{\Psi}}_{1i}\frac{\partial}{\partial\bar{\Psi}_{1i}}\Big) 
\Big(\tilde{\Psi}_{2m}^j\frac{\partial}{\partial\Psi^j_{2m}}\Big)
 \prod_{\substack{q=2}}^{m}\Big(\tilde{\bar{\Psi}}_{2q-2,i}\frac{\partial}{\partial\bar{\Psi}_{2q-2,i}}\Big)\Big(\tilde{\Psi}_{2q-1}^j\frac{\partial}{\partial\Psi^j_{2q-1}}\Big)\mathbf{G}_{CSM}\Bigg|_{\Psi,\bar{\Psi}=0}. \qquad
\ea 

We are now ready to work out some examples using the previous construction.

\subsection{Examples}

We will now present some examples, focusing exclusively on fully-flavoured amplitudes. 

First, the case of the four-point scattering amplitude. The only possible Dyck word in \eqref{eq:flav-CSM-gener} is $Q=\overline{2}\underline{3}$. From \eqref{eq:CSM-amps}, it follows that
\ba\label{eq:CSM-4-p-flav}
\mathscr{A}^{\rm Flav.}_{4} &=& (T_{a})^{i_{1}}_{\phantom{i_{1}}i_{4}}\bar{v}_1\left(\tilde{\bar{\Psi}}_{2i}\frac{\partial}{\partial\bar{\Psi}_{2i}}\tilde{\Psi}_{3}^i\frac{\partial}{\partial\Psi^i_3}\slashed{\mathcal{A}}_{23}^{a}\right)u_4\nonumber\\
&=& \frac{2\pi}{\kappa}(T_{a})^{i_{1}}_{\phantom{i_{1}}i_{4}}(T^{a})^{i_{2}}_{\phantom{i_{2}}i_{3}}\bar{v}_1\gamma_{\mu}u_4\left(\frac{\ui k_{23\lambda}\varepsilon^{\lambda \mu \nu}}{s_{23}}\right)\bar{v}_2\gamma_{\nu}u_3,
\ea
which can be diagramatically expressed \footnote{In this section we present the Feynman diagrams just as a pictorial representation of the amplitude. We do not present the Feynman rules for the graphs, since the procedure with those is usually the other way around, i.e. graphs$\longrightarrow$Feynman rules$\longrightarrow$amplitude. In the preceding sections they will be cast out in order to generalise some results.} as
\begin{equation*}
\mathscr{A}^{\rm Flav.}_{4}=
\begin{gathered}
\feynmandiagram [layered layout, large, horizontal=a to b] {
i1 [particle=\(\overline 2\)]
-- [fermion] b1
-- [fermion] b2 [particle=\(\underline 3\)],
i3 [particle=\(\overline 1\)]
-- [fermion] a1
-- [fermion] a2 [particle=\(\underline 4\)],
b1 -- [gluon] a1,
};
\end{gathered}
\end{equation*}

For the six-point case, we will choose the labels to be $(\overline{1},\overline{2},\underline{3},\overline{4},\underline{5},\underline{6})$. Because of the ordering restriction, there are very few possible Dyck words and the only contributions in \eqref{eq:CSM-amps} are
\ba 
(R,S)=(\overline{2}\underline{3},\overline{4}\underline{5}),\quad(R,S)=(\overline{4}\underline{5},\overline{2}\underline{3}).
\ea
Therefore, the six-point amplitude reads
\begin{multline}\label{eq:CSM-6-p-prev} 
\mathscr{A}^{\rm Flav.}_{6} = \bar{v}_1\tilde{\bar{\Psi}}_{2h}\frac{\partial}{\partial\bar{\Psi}_{2h}}\tilde{\Psi}_{3}^i\frac{\partial}{\partial\Psi^i_3}\tilde{\bar{\Psi}}_{4k}\frac{\partial}{\partial\bar{\Psi}_{4k}}\tilde{\Psi}_{5}^l\frac{\partial}{\partial\Psi^l_5}\Bigg\{(T_{a})^{i_{1}}_{\phantom{i_{1}}j}\slashed{\mathcal{A}}_{23}^{a}\left(\frac{\slashed{k}_{123} - m}{s_{123} - m^2}\right)\slashed{\mathcal{A}}_{45}^{b} (T_{b})^{j}_{\phantom{j}i_{6}} \\ + (T_{a})^{i_{1}}_{\phantom{i_{1}}j}\slashed{\mathcal{A}}_{45}^{a}\left(\frac{\slashed{k}_{145} - m}{s_{145} - m^2}\right)\slashed{\mathcal{A}}_{23}^{b} (T_{b})^{j}_{\phantom{j}i_{6}} + (T_{a})^{i_{1}}_{\phantom{i_{1}}i_{6}}\slashed{\mathcal{A}}_{2345}^{a}\Bigg\}u_6.
\end{multline}

The first two terms inside the curly brackets are straightforward to compute. For the third term, we have to take into account the deconcatenations of the word $\overline{2}\underline{3}\overline{4}\underline{5}$ in the Dyck localisation. The allowed deconcatenations for it are
\ba 
(R,S)=(\overline{2},\underline{3}\overline{4}\underline{5}),\,(\overline{4},\overline{2}\underline{3}\underline{5}), \,(\overline{2}\overline{4}\underline{5},\underline{3}), \,(\overline{2}\underline{3}\overline{4},\underline{5}),\,(\overline{2}\underline{3},\overline{4}\underline{5}),\,(\overline{4}\underline{5},\overline{2}\underline{3}).
\ea
There is also the deconcatenation $\overline{2}\overline{4}\,\,\underline{3}\underline{5}$, but we can see that they correspond to non-Dyck words and they do not contribute to the calculation. In the odd parts of the deconcatenations, it is easy to identify a letter in one of the extremes that breaks the Dyck character of the word. A similar computation is presented in more details in the appendix \ref{Comp-Psi}.

After applying the boundary condition derivatives and using the momentum conservation of the scattering, the fully-flavoured $6$-point amplitude can be presented as
\ba\label{eq:CSM-6-p-flav}
\mathscr{A}^{\rm Flav.}_{6}= \left(\frac{4\pi^2}{\kappa^2}\right) \sum_{r=1}^{7} \frac{c_r n_r}{D_r},
\ea
where the $c_i$'s are the colour factors that give information about the gauge group representation, the $n_i$'s are the kinematic numerators which depend only in kinematical variables (momenta and polarisations), and the  $D_i$'s that are the propagators. The colour factors are given by
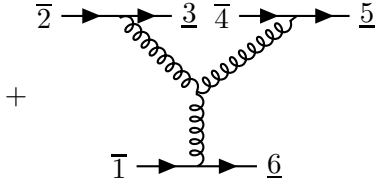
\begin{equation}
\begin{array}{lc}
c_{1}=(T_{a})^{i_{1}}_{\phantom{i_{1}}j}(T_{b})^{j}_{\phantom{j}i_{6}}(T^{a})^{i_{2}}_{\phantom{i_{2}}i_{3}}(T^{b})^{i_{4}}_{\phantom{i_{4}}i_{5}},\\
c_{2}=(T_{b})^{i_{1}}_{\phantom{i_{1}}j}(T_{a})^{j}_{\phantom{j}i_{6}}(T^{b})^{i_{2}}_{\phantom{i_{2}}i_{3}}(T^{a})^{i_{4}}_{\phantom{i_{4}}i_{5}},\\
c_{3}=(T_{a})^{i_{2}}_{\phantom{i_{2}}j}(T_{b})^{j}_{\phantom{j}i_{3}}(T^{a})^{i_{1}}_{\phantom{i_{1}}i_{6}}(T^{b})^{i_{4}}_{\phantom{i_{4}}i_{5}},\\
c_{4}=(T_{b})^{i_{2}}_{\phantom{i_{2}}j}(T_{a})^{j}_{\phantom{j}i_{3}}(T^{b})^{i_{1}}_{\phantom{i_{1}}i_{6}}(T^{a})^{i_{4}}_{\phantom{i_{4}}i_{5}},\\
c_{5}=(T_{a})^{i_{4}}_{\phantom{i_{4}}j}(T_{b})^{j}_{\phantom{j}i_{5}}(T^{a})^{i_{1}}_{\phantom{i_{1}}i_{6}}(T^{b})^{i_{2}}_{\phantom{i_{2}}i_{3}},\\
c_{6}=(T_{b})^{i_{4}}_{\phantom{i_{4}}j}(T_{a})^{j}_{\phantom{j}i_{5}}(T^{b})^{i_{1}}_{\phantom{i_{1}}i_{6}}(T^{a})^{i_{2}}_{\phantom{i_{2}}i_{3}},\\
c_{7}=\tilde{f}_{abc}(T^{a})^{i_{1}}_{\phantom{i_{1}}i_{6}}(T^{b})^{i_{2}}_{\phantom{i_{2}}i_{3}}(T^{c})^{i_{4}}_{\phantom{i_{4}}i_{5}}.
\end{array}
\end{equation}
while the kinematic numerators and propagators can be expressed as
\begin{equation}
\begin{array}{lc}
n_{1}=k_{23\lambda}\varepsilon^{\lambda \mu \nu}k_{45\rho}\varepsilon^{\rho \sigma \tau} (\bar{v}_1\gamma_{\mu}(\slashed{k}_{123} - m)\gamma_{\sigma}u_6)(\bar{v}_{2} \gamma_{\nu}u_{3})(\bar{v}_{4} \gamma_{\tau}u_{5}),\\
n_{2}=k_{23\lambda}\varepsilon^{\lambda \mu \nu}k_{45\rho}\varepsilon^{\rho \sigma \tau} (\bar{v}_1\gamma_{\sigma}(\slashed{k}_{145} - m)\gamma_{\mu}u_6)(\bar{v}_{2} \gamma_{\nu}u_{3})(\bar{v}_{4} \gamma_{\tau}u_{5}),\\
n_{3}=k_{16\lambda}\varepsilon^{\lambda \mu \nu}k_{45\rho}\varepsilon^{\rho \sigma \tau} (\bar{v}_{2}\gamma_{\nu}(\slashed{k}_{345} + m)\gamma_{\sigma}u_{3})(\bar{v}_{1}\gamma_{\mu}u_{6})(\bar{v}_{4} \gamma_{\tau}u_{5}),\\
n_{4}=-k_{16\lambda}\varepsilon^{\lambda \mu \nu}k_{45\rho}\varepsilon^{\rho \sigma \tau} (\bar{v}_{2}\gamma_{\sigma}(\slashed{k}_{245} - m)\gamma_{\nu}u_{3})(\bar{v}_{1}\gamma_{\mu}u_{6})(\bar{v}_{4} \gamma_{\tau}u_{5}),\\
n_{5}=k_{16\lambda}\varepsilon^{\lambda \mu \nu}k_{23\rho}\varepsilon^{\rho \sigma \tau} (\bar{v}_{4}\gamma_{\nu}(\slashed{k}_{235} + m)\gamma_{\sigma}u_{5})(\bar{v}_{1}\gamma_{\mu}u_{6})(\bar{v}_{2} \gamma_{\tau}u_{3}),\\
n_{6}=-k_{16\lambda}\varepsilon^{\lambda \mu \nu}k_{23\rho}\varepsilon^{\rho \sigma \tau} (\bar{v}_{4}\gamma_{\sigma}(\slashed{k}_{234} - m)\gamma_{\nu}u_{5})(\bar{v}_{1}\gamma_{\mu}u_{6})(\bar{v}_{2} \gamma_{\tau}u_{3}),\\
n_{7}=-2\varepsilon_{\nu\rho\sigma}\tilde{\mathcal{J}}^{\nu}_{16}\tilde{\mathcal{J}}^{\rho}_{23}\tilde{\mathcal{J}}^{\sigma}_{45},
\end{array}
\quad
\begin{array}{lc}
D_1=(s_{123} - m^2)s_{23}s_{45},\\
D_2=(s_{145} - m^2)s_{23}s_{45},\\
D_3=(s_{345} - m^2)s_{45}s_{16},\\
D_4=(s_{245} - m^2)s_{45}s_{16},\\
D_5=(s_{235} - m^2)s_{23}s_{16},\\
D_6=(s_{234} - m^2)s_{23}s_{16},\\
D_7=s_{23}s_{45}s_{16},
\end{array}
\end{equation}
where we wrote the colour-stripped topological currents $\tilde{\mathcal{J}}^{\mu}_{pq}=k_{pq\lambda}\varepsilon^{\lambda\mu\nu}(\bar{v}_p\gamma_{\nu}u_q)$.

Like previously, this amplitude can be diagramatically expressed as
\begin{eqnarray*}
\mathscr{A}^{\rm Flav.}_{6}&=&
\begin{gathered}
\feynmandiagram [layered layout, large, horizontal=a to b] {
i2 -- [opacity=0] q1 -- [opacity=0] q2 
[particle=\(\overline 4\)]
-- [fermion] c1 
-- [fermion] c2 [particle=\(\underline 5\)],
i1 [particle=\(\overline 2\)]
-- [fermion] b1
-- [fermion] b2 [particle=\(\underline 3\)],
i3 [particle=\(\overline 1\)]
-- [fermion] a1
-- [opacity=0] p2
-- [opacity=0] a2
-- [fermion] a3 [particle=\(\underline 6\)],
a1 -- [gluon] b1,
a2 -- [gluon] c1,
a1 -- [fermion] a2
};
\end{gathered}
+
\begin{gathered}
\feynmandiagram [layered layout, large, horizontal=a to b] {
i2 -- [opacity=0] q1 -- [opacity=0] q2 
[particle=\(\overline 2\)]
-- [fermion] c1 
-- [fermion] c2 [particle=\(\underline 3\)],
i1 [particle=\(\overline 4\)]
-- [fermion] b1
-- [fermion] b2 [particle=\(\underline 5\)],
i3 [particle=\(\overline 1\)]
-- [fermion] a1
-- [opacity=0] p2
-- [opacity=0] a2
-- [fermion] a3 [particle=\(\underline 6\)],
a1 -- [gluon] b1,
a2 -- [gluon] c1,
a1 -- [fermion] a2
};
\end{gathered}
+
\begin{gathered}
\feynmandiagram [layered layout, large, horizontal=a to b] {
i3-- [opacity=0] q1 [particle=\(\overline 4\)]
-- [fermion] c1 
-- [fermion] c2 [particle=\(\underline 5\)],
i2 [particle=\(\overline 2\)]
-- [fermion] b1
-- [fermion] b2
-- [fermion] b3 [particle=\(\underline 3\)],
i1 [particle=\(\overline 1\)]
-- [fermion] a1
-- [opacity=0] p1
-- [opacity=0] a3 [particle=\(\underline 6\)],
b2 -- [gluon] c1,
a1 -- [gluon] b1,
a1-- [fermion] a3
};
\end{gathered}\\
&&+
\begin{gathered}
\feynmandiagram [layered layout, large, horizontal=a to b] {
i3[particle=\(\overline 4\)]
-- [fermion] c1 
-- [fermion] c2 [particle=\(\underline 5\)],
i2 [particle=\(\overline 2\)]
-- [fermion] b1
-- [fermion] b2
-- [fermion] b3 [particle=\(\underline 3\)],
i1 [particle=\(\overline 1\)]
-- [opacity=0] p1
-- [opacity=0] a1
-- [fermion] a3 [particle=\(\underline 6\)],
b1 -- [gluon] c1,
a1 -- [gluon] b2,
i1-- [fermion] a1
};
\end{gathered}
+
\begin{gathered}
\feynmandiagram [layered layout, large, horizontal=a to b] {
i3-- [opacity=0] q1 [particle=\(\overline 2\)]
-- [fermion] c1 
-- [fermion] c2 [particle=\(\underline 3\)],
i2 [particle=\(\overline 4\)]
-- [fermion] b1
-- [fermion] b2
-- [fermion] b3 [particle=\(\underline 5\)],
i1 [particle=\(\overline 1\)]
-- [fermion] a1
-- [opacity=0] p1
-- [opacity=0] a3 [particle=\(\underline 6\)],
b2 -- [gluon] c1,
a1 -- [gluon] b1,
a1-- [fermion] a3
};
\end{gathered}
+
\begin{gathered}
\feynmandiagram [layered layout, large, horizontal=a to b] {
i3[particle=\(\overline 2\)]
-- [fermion] c1 
-- [fermion] c2 [particle=\(\underline 3\)],
i2 [particle=\(\overline 4\)]
-- [fermion] b1
-- [fermion] b2
-- [fermion] b3 [particle=\(\underline 5\)],
i1 [particle=\(\overline 1\)]
-- [opacity=0] p1
-- [opacity=0] a1
-- [fermion] a3 [particle=\(\underline 6\)],
b1 -- [gluon] c1,
a1 -- [gluon] b2,
i1-- [fermion] a1
};
\end{gathered}\\
&&+
\begin{gathered}
\begin{tikzpicture}
\begin{feynman}[large]
\vertex (a1) {\(\overline 2\)};
\vertex[right=1cm of a1] (a2);
\vertex[right=.7cm of a2] (a3) {\(\underline 3\)};
\vertex[right=.1cm of a3] (a4);
\vertex[right=.1cm of a4] (a5) {\(\overline 4\)};
\vertex[right=1cm of a5] (a6);
\vertex[right=.7cm of a6] (a7) {\(\underline 5\)};
\vertex[below=1cm of a4] (b1);
\vertex[below=1cm of b1] (c1);
\vertex[left=.8cm of c1] (c2) {\(\overline 1\)};
\vertex[right=.8cm of c1] (c3) {\(\underline 6\)};
\diagram* {
(a1) --[fermion] (a2),
(a2) --[fermion] (a3),
(a5) --[fermion] (a6),
(a6) --[fermion] (a7),
(c2) --[fermion] (c1),
(c1) --[fermion] (c3),
(c1) --[gluon] (b1),
(b1) --[gluon] (a2),
(b1) --[gluon] (a6),
};
\end{feynman}
\end{tikzpicture}
\end{gathered}
\end{eqnarray*}

It is interesting that the Dyck localisation in the cohomology Maurer-Cartan action take us to an expansion that perfectly encodes the fully-flavoured diagrams in a compact way. For example, in \eqref{eq:CSM-6-p-prev} we can see how the first two terms are the possible ways of attach to the base fermion line two flavoured pairs directly via two internal vectors, and the third one gives all the possible ways to organise the two flavoured lines with only one internal vector attached to the base fermion line. Going to a higher number of points the number of diagrams will increase exponentially, but the number of words for our first expression in the cohomology Maurer-Cartan action will be considerably smaller due to the localisation.     
 
In addition, we would like to point out that amplitudes with fermions of the same flavour can be obtained from the fully-flavoured ones just by adding label permutations. Since different flavours usually have different masses, in section \ref{sec:concl} we give a prescription to take this into account.
  

\section{$L_{\infty}$-structure for QCD}\label{sec:QCD}
We will now use the underlying $L_{\infty}$-structure of QCD to derive recursion relations for the tree-level scattering amplitudes. These are obtained by carrying out the same procedure described in the previous section with slightly more involved steps.

\subsection{The QCD $L_\infty$ algebra}
Our notation here will also be similar: $\Omega^r(\RR^{1,3},\su(N))$ is the space of $r$-forms on $\RR^{1,3}$ with values in $\su(N)$, $\ud$ is the exterior differential, $\delta$ the codifferential induced by the Minkowski metric, and  $\Omega^{0}(\RR^{1,3},S \otimes V)$ the space of $0$-forms on $\RR^{1,3}$ with values in $S \otimes V$. As a graded vector space, the QCD $L_{\infty}$-algebra $L$ is  
\begin{align*}
L^{0} &= \Omega^{0}(\RR^{1,3},\su(N)), \\
L^{1} &= \Omega^{1}(\RR^{1,3},\su(N)) \oplus \Omega^{0}(\RR^{1,3}, S \otimes V) \oplus \Omega^{0}(\RR^{1,3}, S \otimes \bar{V}), \\
L^{2} &= \Omega^{1}(\RR^{1,3},\su(N)) \oplus \Omega^{0}(\RR^{1,3}, S \otimes \bar{V}) \oplus \Omega^{0}(\RR^{1,3}, S \otimes V), \\
L^{3} &= \Omega^{0}(\RR^{1,3},\su(N)). 
\end{align*} 
Adopting the terminology of subsection \ref{sec:2.3}, we think of the elements $c \in L^{0}$ as coming from infinitesimal gauge parameters, elements $A + \psi + \bar{\psi} \in L^{1}$ as triples containing a gauge field, a Dirac spinor and a conjugate Dirac spinor, elements $A^* + \bar{\psi}^* + \psi^* \in L^{2}$ as triples of antifields conjugate to $A$, $\psi$ and $\bar{\psi}$, and elements $c^* \in L^{3}$ as antifields conjugate to $c$. The link between the BV transformations \eqref{eq:BVtransf-QCD} and the higher order brackets on $L$ is provided by formulas identical to those of \eqref{eq:BV-Linfinity}, except that for the BV transformation of the antifield $A^{*a}_{\mu}$ we will have the extra term $\frac{1}{3!} l_3(A,A,A)^{a}_{\mu}$. The non-vanishing products for $l_1$ are
\begin{equation}\label{eq:QCDl1}
\begin{array}{rc}
l_{1}(c)_{\mu a}=-\partial_{\mu}c_{a} & \in L^{1},\\
l_{1}(A)_{\mu}^{a}=-\partial^{\nu}(\partial_{\mu}A_{\nu}^{a}-\partial_{\nu}A_{\mu}^{a}) & \in L^{2},\\
l_1(\psi + \bar{\psi})^{i} = -\ui \gamma^{\mu} \partial_{\mu} \psi^{i} + m \psi^{i} & \in L^{2},\\
l_1(\psi + \bar{\psi})_{i} = -\ui  \partial_{\mu} \bar{\psi}_{i} \gamma^{\mu} - m \bar{\psi}_{i} & \in L^{2},\\
l_{1}(A^{*})_{a}=-\partial^{\mu}A_{\mu a}^{*} & \in L^{3},
\end{array}
\end{equation}
satisfying the relation \eqref{eq:l1-relation}. For $l_{2}$ we have
\begin{equation}\label{eq:QCDl2}
\begin{array}{rc}
l_2(c_1,c_2)^{a} = \ui \tilde{f}_{bc}^{\phantom{bc}a} c_1^{b} c_2^{c} & \in L^{0},\\
l_2(c_1,A_2)^{a}_{\mu} =  \ui \tilde{f}_{bc}^{\phantom{bc}a}   c_1^{b} A_{2\mu}^{c} & \in L^{1},\\
l_2(c_1,A_2^*)^{a}_{\mu} = \ui \tilde{f}_{bc}^{\phantom{bc}a} c_1^{b} A_{2\mu}^{*c} & \in L^{2},\\
l_2(c_1,c_2^*)^{a}  =- \ui \tilde{f}_{bc}^{\phantom{bc} a}  c_1^{b} c_2^{*c} & \in L^{3},\\
l_{2}(A^{1},A^{2})_{\mu a}=\tilde{f}_{abc}\partial^{\nu}(A_{\mu b}^{1}A_{\nu c}^{2})-\tilde{f}_{abc}(\partial_{\mu}A_{\nu b}^{1}-\partial_{\nu}A_{\mu b}^{1})A_{c}^{2\nu}+(1\leftrightarrow2) & \in L^{2}\\
l_2(A_1,A_2^*)^{a} = \ui f_{bc}^{\phantom{bc}a} A_{1\mu}^{b} A_2^{*c\mu} & \in L^{3}\\
l_2(c_1,\psi_2)^{i} = -\ui c_1^{a} (T_{a})^{i}_{\phantom{i}j} \psi_2^{j} & \in L^{1},\\
l_2(c_1,\bar{\psi}_2)_{i} = \ui c_1^{a} \bar{\psi}_{2j} (T_{a})^{j}_{\phantom{j}i} & \in L^{1},\\
l_2(c_1,\psi^*_2)^{i} = - \ui c_1^{a} (T_{a})^{i}_{\phantom{i}j} \psi_2^{*j} & \in L^{2},\\
l_2(c_1,\bar{\psi}^*_2)_{i} =  \ui c_1^{a} \bar{\psi}_{2j}(T_{a})^{j}_{\phantom{j}i} & \in L^{2},\\
l_2(A_1,\psi_2)^{i} = \ui \gamma^{\mu} A_{1\mu}^{a} (T_{a})^{i}_{\phantom{i}j}\psi_2^{j} & \in L^{2},\\
l_2(A_1,\bar{\psi}_2)_{i} =- \ui  A_{1\mu}^{a} \bar{\psi}_{2j}(T_{a})^{j}_{\phantom{j}i} \gamma^{\mu} & \in L^{2},\\
l_2 (\psi_1 + \bar{\psi}_1,\psi_2 + \bar{\psi}_2)^{a}_{\mu}= (T^{a})^{i}_{\phantom{i}j}[(\bar{\psi}_{1 i} \gamma_{\mu} \psi^{j}_2) + (\bar{\psi}_{2 i} \gamma_{\mu}\psi^{j}_1)] & \in L^{2},\\
l_2 (\psi_1 + \bar{\psi}_1,\bar{\psi}^*_2 + \psi^*_2)^{a} = \ui  (T^{a})^{i}_{\phantom{i}j}(\bar{\psi}{}^*_{2i} \psi_1^{j} - \bar{\psi}{}_{1i}\psi_2^{*j}) & \in L^{3},\\
\end{array}
\end{equation}
satisfying the relation \eqref{eq:l2-relation}. And for $l_{3}$
we have:
\begin{equation}\label{eq:QCDl3}
\begin{array}{rc}
l_{3}(A^{1},A^{2},A^{3})_{\mu a}=-\tilde{f}_{abe}\tilde{f}_{cde}(A_{\nu b}^{1}A_{c}^{2\nu}A_{\mu d}^{3}+\mathrm{sym}(1,2,3)) & \in L^{2}.\end{array}
\end{equation}
satisfying the relation
\begin{multline}
l_{2}\left(l_{2}\left(x_{1},x_{2}\right),x_{3}\right)+\left(-1\right)^{|x_{1}(x_{2}+x_{3})|}l_{2}\left(l_{2}\left(x_{2},x_{3}\right),x_{1}\right)+\left(-1\right)^{|x_{3}(x_{1}+x_{2})|}l_{2}\left(l_{2}\left(x_{3},x_{1}\right),x_{2}\right)\\
+l_{3}\left(l_{1}\left(x_{1}\right),x_{2},x_{3}\right)+\left(-1\right)^{|x_{1}|}l_{3}\left(x_{1},l_{1}\left(x_{2}\right),x_{3}\right)\\
+\left(-1\right)^{|x_{1}+x_{2}|}l_{3}\left(x_{1},x_{2},l_{1}\left(x_{3}\right)\right)+l_{1}\left(l_{3}\left(x_{1},x_{2},x_{3}\right)\right)=0.\label{eq:l3-relation}
\end{multline}

The underlying cochain complex is
$$
\begin{tikzpicture}[scale=1.0,baseline=-0.1cm, inner sep=1mm,>=stealth]
\node (0) at (-0.9,0)  {$\Omega^{0}(\RR^{1,3},\su(N))$};
\node (1) at (2.68,0)  {$\Omega^{1}(\RR^{1,3},\su(N))$};
\node (2) at (8.32,0)  {$\Omega^{1}(\RR^{1,3},\su(N))$};
\node (3) at (11.9,0)  {$\Omega^{0}(\RR^{1,3},\su(N)).$};
\node (4) at (2.68,-0.6) {$\oplus$};
\node (5) at (8.32,-0.6) {$\oplus$};
\node (6) at (2.68,-1.2) {$\Omega^{0}(\RR^{1,3},S \otimes V)$};
\node (7) at (8.32,-1.2) {$\Omega^{0}(\RR^{1,3},S \otimes \bar{V})$};
\node (8) at (2.68,-1.8) {$\oplus$};
\node (9) at (8.32,-1.8) {$\oplus$};
\node (10) at (3.95,-1.8) {};
\node (11) at (7.05,-1.8) {};
\node (12) at (2.68,-2.4) {$\Omega^{0}(\RR^{1,3},S \otimes \bar{V})$};
\node (13) at (8.32,-2.4) {$\Omega^{0}(\RR^{1,3},S \otimes V)$};
\tikzset{every node/.style={fill=white}} 
\draw[->, thick] (0) to node[midway,above=0.2pt] {\scriptsize$\ud$} (1); 
\draw[->, thick] (1) to node[midway,above=0.2pt] {\scriptsize$\delta \ud$} (2); 
\draw[->, thick] (2) to node[midway,above=0.2pt] {\scriptsize$\delta$} (3); 
\draw[->, thick] (10) to node[midway,above=0.2pt] {\scriptsize$\begin{pmatrix}
0 & \ui \slashed{\partial} + m  \\
\ui \slashed{\partial} - m & 0 
\end{pmatrix}$} (11);
\end{tikzpicture}
$$

This $L_{\infty}$-algebra is endowed with a cyclic structure defined by
\begin{align}\label{eq:inner-QCD}
\begin{split}
\langle c , c^*  \rangle &=  \int_{\RR^{1,3}} \ud^4 x \,  c^*_{a} c^{a}, \\
\langle A , A^*  \rangle &=  \int_{\RR^{1,3}} \ud^4 x \, A^{*\mu}_{a} A_{\mu}^{a} , \\
\langle \psi + \bar{\psi} , \bar{\psi}^* + \psi^*  \rangle &= \int_{\RR^{1,3}} \ud^4 x \, (  \psi^{*i} \bar{\psi}_{i} +  \bar{\psi}{}^{*}_{i} \psi^{i}  ), 
\end{split}
\end{align}
where $c \in L^{0}$, $c^* \in L^{3}$, $A+\psi + \bar{\psi}\in L^{1}$ and $A^* +  \bar{\psi}^* + \psi^* \in L^{2}$. 

The Maurer-Cartan action has the same form as in \eqref{eq:MCaction} and can be computed to be
\begin{equation*} 
S_{\mathrm{MC}}[A + \psi + \bar{\psi}] =  \int_{\RR^{1,3}} \ud^4 x \,\left\{ -  \frac{1}{4}F^{a}_{\mu\nu} F^{\mu\nu}_a +  \bar{\psi}(\ui \slashed{D} - m) \psi \right\},
\end{equation*}
reproducing the QCD action \eqref{eq:QCDaction}.  

Next we must modify the definition of the $L_{\infty}$-algebra $L$ to work with perturbiner expansions. For this, we will again consider an infinite set $(k_p)_{p \geq 1}$ of momentum vectors, now in $\RR^{1,3}$. As before, we denote by $\mathscr{E}^{0}(\RR^{1,3})$ the space of formal series of the form \eqref{eq:3.4}, by $\mathscr{E}^{r}(\RR^{1,3},\su(N))$ the space of $r$-forms on $\RR^{1,3}$ with values in $\su(N)$ and coefficients on $\mathscr{E}^{0}(\RR^{1,3})$, and by $\mathscr{E}^{r}(\RR^{1,3},S \otimes V)$ the space of $0$-forms on $\RR^{1,3}$ with values in $S \otimes V$ and coefficients on $\mathscr{E}^{0}(\RR^{1,3})$. We also extend the exterior differential $\ud$, the codifferential $\delta$ and the Dirac operator $\slashed{\partial}$ to act on the spaces $\mathscr{E}^{\sbullet}(\RR^{1,3},\su(N))$ and $\mathscr{E}^{0}(\RR^{1,3},S \otimes V)$. 

With this modification, the graded vector space underlying the cyclic $L_{\infty}$-algebra $L$ that encodes the perturbiner expansion for QCD is
\begin{align*}
L^{0} &= \mathscr{E}^{0}(\RR^{1,3},\su(N)), \\
L^{1} &= \mathscr{E}^{1}(\RR^{1,3},\su(N)) \oplus \mathscr{E}^{0}(\RR^{1,3}, S \otimes V) \oplus \mathscr{E}^{0}(\RR^{1,3}, S \otimes \bar{V}), \\
L^{2} &= \mathscr{E}^{1}(\RR^{1,3},\su(N)) \oplus \mathscr{E}^{0}(\RR^{1,3}, S \otimes \bar{V}) \oplus \mathscr{E}^{0}(\RR^{1,3}, S \otimes V), \\
L^{3} &= \mathscr{E}^{0}(\RR^{1,3},\su(N)). 
\end{align*} 
The higher order brackets and the cyclic inner product are given by exactly the same formulas presented above.

\subsection{The perturbiner expansion for QCD}\label{sec:4.2}
Let us now turn to the perturbiner expansion for QCD. Again, we will that it is encoded in the minimal model for the $L_{\infty}$-algebra $L$ introduced above. 

In order to describe the minimal $L_{\infty}$-structure on the cohomology $H^{\sbullet}(L)$ of $L$, we need to specify a projection $p \colon L \to  H^{\sbullet}(L)$ and a contracting homotopy $h \colon L \to L$. First notice that, as a consequence of the abstract Hodge-Kodaira decomposition, the cochain complex underlying $H^{\sbullet}(L)$ is
$$
\begin{tikzpicture}[scale=1.0,baseline=-0.1cm, inner sep=1mm,>=stealth]
\node (0) at (-1,0)  {$\ker (\ud)$};
\node (1) at (2,0)  {$\ker(\delta \ud)/\im(\ud)$};
\node (2) at (8,0)  {$\ker(\delta \ud)/\im(\ud)$};
\node (3) at (11,0)  {$\ker(\ud).$};
\node (4) at (2,-0.6) {$\oplus$};
\node (5) at (8,-0.6) {$\oplus$};
\node (6) at (2,-1.55) {$\ker \begin{pmatrix}
0 & \ui \slashed{\partial} + m  \\
\ui \slashed{\partial} - m & 0 
\end{pmatrix}$};
\node (7) at (8,-1.55) {$\coker \begin{pmatrix}
0 & \ui \slashed{\partial} + m  \\
\ui \slashed{\partial} - m & 0 
\end{pmatrix}$};
\tikzset{every node/.style={fill=white}} 
\draw[->, thick] (0) to node[midway,above=0.2pt] {\scriptsize$0$} (1); 
\draw[->, thick] (1) to node[midway,above=0.2pt] {\scriptsize$0$} (2); 
\draw[->, thick] (2) to node[midway,above=0.2pt] {\scriptsize$0$} (3); 
\draw[->, thick] (6) to node[midway,above=0.2pt] {\scriptsize$0$} (7);
\end{tikzpicture}
$$ 
With this in mind, we choose the four components of the projection $p^{(0)}$, $p^{(1)}$, $p^{(2)}$ and $p^{(3)}$ to be the natural projections induced by such decomposition. On the other hand, to define the contracting homotopy $h$, we consider the massless Feynman propagator $\Dsf^{\mathrm{F}}$ and the spinor Feynman propagator $\Ssf^{\mathrm{F}}$, given by the same expressions as those in \eqref{eq:3.5}, along with their extension to linear operators $\Dsf^{\mathrm{F}} \colon \mathscr{E}^{r}(\RR^{1,3},\su(N)) \to  \mathscr{E}^{r}(\RR^{1,3},\su(N))$ and $\Ssf^{\mathrm{F}}\colon \mathscr{E}^{0}(\RR^{1,3},S \otimes \bar{V}) \oplus \mathscr{E}^{0}(\RR^{1,3},S \otimes V) \to  \mathscr{E}^{0}(\RR^{1,3},S \otimes V) \oplus \mathscr{E}^{0}(\RR^{1,3},S \otimes \bar{V})$. In terms of these, the three non-zero components of the contracting homotopy $h$ read as
\begin{align}
\begin{split}
h^{(1)} = \left(\begin{array}{cc} \Dsf^{\mathrm{F}}\circ \delta &\, 0 \end{array} \right) &\colon L^{1} \longrightarrow L^{0}, \\
h^{(2)} = \left(\begin{array}{cc} \Dsf^{\mathrm{F}} \circ P_{\mathrm{e}} & 0 \\ 0 & \Ssf^{\mathrm{F}} \end{array} \right) &\colon L^{2} \longrightarrow L^{1}, \\
h^{(3)} = \left(\begin{array}{c} \ud \circ \Dsf^{\mathrm{F}}  \\ 0 \end{array} \right) &\colon L^{3} \longrightarrow L^{2},
\end{split}
\end{align}
where $P_{\mathrm{e}} \colon  \mathscr{E}^{r}(\RR^{1,3},\su(N)) \to  \mathscr{E}^{r}(\RR^{1,3},\su(N))$ denotes the projector onto the image of $\delta \ud$. More explicitly, 
\begin{align}\label{eq:QCDhomotopy}
\begin{split}
h^{(1)}(A)^{a} &= - \frac{\partial_{\mu}}{\square} A^{a \mu},  \\
h^{(2)}(A^*)^{a}_{\mu} &= \frac{1}{\square}\left( \eta_{\mu\nu} - \frac{\partial_{\mu} \partial_{\nu}}{\square}\right) A^{*a \nu}, \\
 h^{(2)}(\bar{\psi}^* + \psi^*)^{i} &= -\frac{\ui \slashed{\partial} + m }{\square + m^2} \psi^{* i}, \\
  h^{(2)}(\bar{\psi}^* + \psi^*)_{i} &= -\frac{\ui \slashed{\partial} - m }{\square + m^2} \bar{\psi}^{*}_{i}, \\
h^{(3)}(c^{*})^{a}_{\mu} &=  - \frac{\partial_{\mu}}{\square} c^{* a},
\end{split}
\end{align}
where $A \in L^{1}$, $A^* + \bar{\psi}^* + \psi^* \in L^{2}$ and $c^* \in L^{3}$. Once more, it should be stressed that the formulas for the $L_{\infty}$-quasi-isomorphism and the higher-order brackets for the minimal $L_{\infty}$-structure on $H^{\sbullet}(L)$ are derived under the assumption that $h^{(1)}(A + \psi + \bar{\psi}) = 0$ for any $A + \psi + \bar{\psi} \in L^{1}$. In view of \eqref{eq:QCDhomotopy}, the first summands $A \in \mathscr{E}^{0}(\RR^{1,3},\su(N))$ are then bound to satisfy the Lorenz gauge condition $\partial_{\mu} A^{a\mu} = 0$.

Let us next examine how the perturbiner expansion for QCD can be extracted from the minimal $L_{\infty}$-structure on $H^{\sbullet}(L)$. In analogy with the CSM theory case, we start with a Maurer-Cartan element $A ' + \psi' + \bar{\psi}' \in H^{1}(L) = \left( \ker(\delta \ud)/\im(\ud)\right) \oplus \ker \left( \begin{smallmatrix}
0 & \ui \slashed{\partial} + m  \\
\ui \slashed{\partial} - m & 0 
\end{smallmatrix} \right)$,
for which we have
\begin{align}\label{eqn:QCDpw}
\begin{split}
A'^{a \mu} &= \sum_{p \geq 1} \mathcal{A}_{p}^{a\mu} \, \ue^{-\ui k_p \cdot x}, \\
\psi'^{i} &= \sum_{p \geq 1} \Psi^{i}_{p} \, \ue^{-\ui k_p \cdot x}, \\
\bar{\psi}'_{i} &= \sum_{p \geq 1} \bar{\Psi}_{pi} \, \ue^{-\ui k_p \cdot x}.
\end{split}
\end{align}
Again, we are automatically assuming the physical input described around equation \eqref{eq:grassmann-polarization}.

We then define the perturbiner expansion to be the Maurer-Cartan element $A + \psi + \bar{\psi}$ in $L$ given by the expression
\begin{equation}
A + \psi + \bar{\psi} = \sum_{n \geq 1} \frac{1}{n!} f_{n}(A' + \psi' + \bar{\psi}', \dots, A' + \psi' + \bar{\psi}').
\end{equation}  
Let us make this construction clearer and work out the separate components $A \in \mathscr{E}^{1}(\RR^{1,3},\su(N))$, $\psi \in \mathscr{E}^{0}(\RR^{1,3},S \otimes V)$ and $\bar{\psi} \in \mathscr{E}^{0}(\RR^{1,3},S \otimes \bar{V})$. The first remark is that, since $f_n$ is graded symmetric, we have
\begin{equation}\label{eq:QCDfnexpansion}
f_{n}(A' + \psi' + \bar{\psi}', \dots, A' + \psi' + \bar{\psi}') = \sum_{p + q = n} \binom{n}{q} f_{n}(A' , \dots, A', \psi' + \bar{\psi}',\dots, \psi' + \bar{\psi}'),
\end{equation}
where, on the right-hand side, there are $p$ copies of $A'$ and $q$ copies of $\psi' + \bar{\psi}'$. In addition, for each decomposition of $n$ into $p + q$, the general prescription for writing down the maps $f_n$ gives
\begin{align}
\begin{split}
& f_{n}(A' , \dots, A', \psi' + \bar{\psi}', \dots, \psi' + \bar{\psi}') \\
&\, = -\tfrac{1}{2!} \sum_{i + j = n} \binom{n}{i} (h^{(2)} \circ l_2)\left(f_{i}(A' , \dots, A', \psi' + \bar{\psi}', \dots, \psi' + \bar{\psi}'), \right. \\
& \phantom{a = -\tfrac{1}{2!} \sum_{i + j = n} \binom{n}{i} (h^{(2)} \circ l_2)} \left. f_{j}(A' , \dots, A', \psi' + \bar{\psi}', \dots, \psi' + \bar{\psi}')\right) \\
&\phantom{\, =\,} -\tfrac{1}{3!} \sum_{i + j + k = n} \binom{n}{i} \binom{n - i}{j} (h^{(2)} \circ l_3) \left(f_{i}(A' , \dots, A', \psi' + \bar{\psi}', \dots, \psi' + \bar{\psi}'), \right.\\
& \phantom{\, = -\tfrac{1}{3!} \sum_{i + j + k = n} \binom{n}{i} \binom{n - i}{j} (h^{(2)} \circ l_3) a\:} f_{j}(A' , \dots, A', \psi' + \bar{\psi}', \dots, \psi' + \bar{\psi}'), \\
&\phantom{\, = -\tfrac{1}{3!} \sum_{i + j + k = n} \binom{n}{i} \binom{n - i}{j} (h^{(2)} \circ l_3) a\:}  \left. f_{k}(A' , \dots, A', \psi' + \bar{\psi}', \dots, \psi' + \bar{\psi}') \right). 
\end{split}
\end{align}
Thus, using the higher order brackets in \eqref{eq:QCDl2} and \eqref{eq:QCDl3},  and the definition of $h^{(2)}$ in \eqref{eq:QCDhomotopy}, we can show that
\begin{align}\label{eq:QCDperturbiner-quasi}
\begin{split}
A^{a \mu} &= \sum_{n \geq 1} \sum_{p + 2q = n}  \frac{1}{p! (2q)!}  f_{n}(A' , \dots, A', \psi' + \bar{\psi}', \dots, \psi' + \bar{\psi}')^{a \mu}, \\
\psi^{i} & = \sum_{n \geq 1} \sum_{p + 2q +1 = n}  \frac{1}{p! (2q + 1)!}  f_{n}(A' , \dots, A', \psi' + \bar{\psi}', \dots, \psi' + \bar{\psi}')^{i}, \\
\bar{\psi}_{i} & = \sum_{n \geq 1} \sum_{p + 2q +1 = n}  \frac{1}{p! (2q + 1)!}  f_{n}(A' , \dots, A', \psi' + \bar{\psi}', \dots, \psi' + \bar{\psi}')_{i}.
\end{split}
\end{align}
Here, for each decomposition of $n$ into $p + 2q$, 
\begin{align}\label{eq:QCDfnvector}
\begin{split}
& f_{n}(A' , \dots, A', \psi' + \bar{\psi}', \dots, \psi' + \bar{\psi}')^{a \mu} \\
  &\qquad = -\tfrac{1}{2!} \sum_{l+m = n} \binom{n}{l} (\Dsf^{\mathrm{F}} \circ l_2)\left(f_{l}(A' , \dots, A', \psi' + \bar{\psi}', \dots, \psi' + \bar{\psi}'), \right.\\
  &\qquad\phantom{= -\tfrac{1}{2!} \sum_{l+m = n} \binom{n}{m} (\Dsf^{\mathrm{F}} \circ l_2)a} \left. f_{m}(A' , \dots, A', \psi' + \bar{\psi}', \dots, \psi' + \bar{\psi}')\right)^{a \mu},
\end{split}
\end{align}
where we bear in mind that in the case when $q = 0$, 
\begin{align}\label{eqn:4.11}
\begin{split}
 f_{n}(A' , \dots, A')^{a \mu} &= -\tfrac{1}{2!} \sum_{l+m= n} \binom{n}{l} (\Dsf^{\mathrm{F}} \circ l_2)\left(f_{l}(A' , \dots, A'), f_{m}(A' , \dots, A')\right)^{a \mu} \\
&\phantom{ =}\, -\tfrac{1}{3!} \sum_{k+l+m= n} \binom{n}{k} \binom{n - k}{l} (\Dsf^{\mathrm{F}} \circ l_3) \left(f_{k}(A' , \dots, A'), f_{l}(A' , \dots, A'), \right. \\
&\phantom{ = -\tfrac{1}{3!} \sum_{k+l+m= n} \binom{n}{k} \binom{n - k}{l} (\Dsf^{\mathrm{F}} \circ l_3) a}  \left. f_{m}(A' , \dots, A') \right)^{a \mu},
\end{split}
\end{align}
whilst for each decomposition of $n$ into $p + 2q + 1$,
\begin{align}\label{eq:QCDfnfundamental}
\begin{split}
& f_{n}(A' , \dots, A', \psi' + \bar{\psi}', \dots, \psi' + \bar{\psi}')^{i} \\
  &\qquad = -\tfrac{1}{2!} \sum_{l+ m = n} \binom{n}{l} (\Ssf^{\mathrm{F}} \circ l_2)\left(f_{l}(A' , \dots, A', \psi' + \bar{\psi}', \dots, \psi' + \bar{\psi}'), \right.\\
  &\qquad\phantom{= -\tfrac{1}{2!} \sum_{l+m = n} \binom{n}{l} (\Ssf^{\mathrm{F}} \circ l_2)a} \left. f_{m}(A' , \dots, A', \psi' + \bar{\psi}', \dots, \psi' + \bar{\psi}')\right)^{i},
\end{split}
\end{align}
and 
\begin{align}\label{eq:QCDfnantifundamental}
\begin{split}
& f_{n}(A' , \dots, A', \psi' + \bar{\psi}', \dots, \psi' + \bar{\psi}')_{i} \\
  &\qquad = -\tfrac{1}{2!} \sum_{l+m = n} \binom{n}{l} (\Ssf^{\mathrm{F}} \circ l_2)\left(f_{l}(A' , \dots, A', \psi' + \bar{\psi}', \dots, \psi' + \bar{\psi}'), \right.\\
  &\qquad\phantom{= -\tfrac{1}{2!} \sum_{l+m = n} \binom{n}{l} (\Ssf^{\mathrm{F}} \circ l_2)a} \left. f_{m}(A' , \dots, A', \psi' + \bar{\psi}', \dots, \psi' + \bar{\psi}')\right)_{i}.
\end{split}
\end{align}
The recursion relations \eqref{eq:QCDfnvector}, \eqref{eq:QCDfnfundamental} and \eqref{eq:QCDfnantifundamental} can be solved by induction. Next, in a long but straightforward procedure, we substitute the resulting expressions into \eqref{eq:QCDperturbiner-quasi} to obtain
\begin{align}\label{eq:QCDperturbiner}
\begin{split}
A^{a \mu} &= \sum_{n \geq 1} \sum_{P \in \mathcal{OW}_{n}} \mathcal{A}^{a \mu}_{P}\, \ue^{-\ui k_{P} \cdot x} = \sum_{p} \mathcal{A}^{a \mu}_{p} \,\ue^{-\ui k_{p} \cdot x} + \sum_{p < q} \mathcal{A}^{a \mu}_{pq} \, \ue^{-\ui k_{pq} \cdot x} +  \cdots, \\
\psi^{i} &= \sum_{n \geq 1} \sum_{P \in \mathcal{OW}_{n}} \Psi^{i}_{P} \ue^{-\ui k_{P} \cdot x} = \sum_{p} \Psi^{i}_{p} \, \ue^{-\ui k_{p} \cdot x} + \sum_{p < q} \Psi^{i}_{pq}\, \ue^{-\ui k_{pq} \cdot x}  + \cdots, \\
\bar{\psi}_{i} &= \sum_{n \geq 1} \sum_{P \in \mathcal{OW}_{n}} \bar{\Psi}_{Pi} \ue^{-\ui k_{P} \cdot x} = \sum_{p} \bar{\Psi}_{p i} \, \ue^{-\ui k_{p} \cdot x} + \sum_{p < q} \bar{\Psi}_{pq i} \, \ue^{-\ui k_{pq} \cdot x}  + \cdots,
\end{split}
\end{align}
where the Berends-Giele currents $\mathcal{A}^{a \mu}_{P}$, $\Psi^{i}_{P}$ and $\bar{\Psi}_{Pi}$ are determined from the recursion relations
\begin{align}\label{eq:QCDperturbinercoefficients}
\begin{split}
\mathcal{A}^{a \mu}_{P} &=  \frac{1}{s_{P}}  \mathcal{J}^{a \mu}_{P} +   \frac{\ui}{s_{P}} \sum_{P = Q \cup R} \big\{ -\ui\tilde{f}^{\phantom{bc}a}_{bc} (k_{Q} \cdot \mathcal{A}^{b}_{R}) \mathcal{A}^{c \mu}_{Q} + \mathcal{A}^{b}_{Q \nu} \mathcal{F}^{c \nu \mu}_{R}\big\} , \\
\mathcal{J}^{a \mu}_{P} &=  \sum_{P = Q \cup R} \bar{\Psi}_{Q i} \gamma^{\mu} (T^{a})^{i}_{\phantom{i}j} \Psi^{j}_{R}, \\
\mathcal{F}^{a \mu \nu}_{P} &=\ui k^{\mu}_{P} \mathcal{A}^{a \nu}_{P} -\ui k^{\nu}_{P} \mathcal{A}^{a \mu}_{P} +  \ui \tilde{f}^{\phantom{bc}a}_{bc} \sum_{P = Q \cup R}  \mathcal{A}^{b \mu}_{Q} \mathcal{A}^{c \nu}_{R}, \\
\Psi^{i}_P &= - \left(\frac{\slashed{k}_P + m}{s_{P} - m^2}\right) \sum_{P = Q \cup R} \slashed{\mathcal{A}}_{Q}^{a} (T_{a})^{i}_{\phantom{i}j} \Psi^{j}_R, \\
\bar{\Psi}_{P i} &= - \sum_{P = Q \cup R} \bar{\Psi}_{R j} (T_{a})^{j}_{\phantom{j}i} \slashed{\mathcal{A}}_{Q}^{a} \left(\frac{\slashed{k}_P - m}{s_{P} - m^2}\right).
\end{split} 
\end{align}

Observe that these expansions are once again color-dressed versions of perturbiners. Also, it is important to realise that the auxiliary quantities $\mathcal{J}^{a \mu}_{P}$ and $\mathcal{F}^{a \mu \nu}_{P}$ represent the expansion coefficients of the matter current $J^{a\mu}$ and field strength $F^{a \mu \nu}$, respectively. The requirement of the Lorenz gauge condition $\partial_{\mu}A^{a \mu} = 0$ translates then into the divergenceless condition $k_{P\mu} \mathcal{A}^{a \mu}_{P} = 0$. Thus, as promised, we see that all the information contained in the perturbiner coefficients can be encoded in the recursion relations for the $L_{\infty}$-quasi-isomorphism between $H^{\sbullet}(L)$ and $L$.

Next, we will present the generator for tree-level scattering amplitudes for QCD

\subsection{Tree-level scattering amplitudes for QCD}\label{sec:4.3}

As we saw in the subsection \ref{sec:3.3}, the starting point to build the amplitude generator is to insert the plane wave superposition in \eqref{eqn:QCDpw} into the homotopy Maurer-Cartan action in the cohomology, which is simply
\begin{equation}\label{eqn:4.17}
S'_{\mathrm{MC}}[A' + \psi' + \bar{\psi}'] = \sum_{n \geq 2} \frac{1}{(n+1)!} \langle A' + \psi' + \bar{\psi}' , l'_{n}(A' + \psi' + \bar{\psi}',\dots,A' + \psi' + \bar{\psi}')\rangle.
\end{equation}

Analogously to \eqref{eq:QCDfnexpansion},  we can expand $l'_{n}(A' + \psi' + \bar{\psi}', \dots, A' + \psi' + \bar{\psi}')$ as
\begin{equation}
l'_{n}(A' + \psi' + \bar{\psi}', \dots, A' + \psi' + \bar{\psi}') = \sum_{p + q = n} \binom{n}{q} l'_{n} (A' , \dots, A', \psi' + \bar{\psi}',\dots, \psi' + \bar{\psi}'),
\end{equation}
where, on the right-hand side, there are $p$ copies of $A'$ and $q$ copies of $\psi'+ \bar{\psi}'$. Next, for each decomposition of $n$ into $p + q$, we use the general expression that gives us the higher order brackets on $H^{\sbullet}(\mathfrak{L}_{\mathrm{QCD}})$ to write
\begin{align}\label{eqn:4.19}
\begin{split}
& l'_{n}(A' , \dots, A', \psi' + \bar{\psi}', \dots, \psi' + \bar{\psi}') \\
&\, = -\tfrac{1}{2!} \sum_{i + j = n} \binom{n}{i} (p^{(2)} \circ l_2)\left(f_{i}(A' , \dots, A', \psi' + \bar{\psi}', \dots, \psi' + \bar{\psi}'), \right. \\
& \phantom{a = -\tfrac{1}{2!} \sum_{i + j = n} \binom{n}{i} (p^{(2)} \circ l_2)} \left. f_{j}(A' , \dots, A', \psi' + \bar{\psi}', \dots, \psi' + \bar{\psi}')\right) \\
&\phantom{\, =\,} -\tfrac{1}{3!} \sum_{i + j + k = n} \binom{n}{i} \binom{n - i}{j} (p^{(2)} \circ l_3) \left(f_{i}(A' , \dots, A', \psi' + \bar{\psi}', \dots, \psi' + \bar{\psi}'), \right.\\
& \phantom{\, = -\tfrac{1}{3!} \sum_{i + j + k = n} \binom{n}{i} \binom{n - i}{j} (h^{(2)} \circ l_3) a\:} f_{j}(A' , \dots, A', \psi' + \bar{\psi}', \dots, \psi' + \bar{\psi}'), \\
&\phantom{\, = -\tfrac{1}{3!} \sum_{i + j + k = n} \binom{n}{i} \binom{n - i}{j} (h^{(2)} \circ l_3) a\:}  \left. f_{k}(A' , \dots, A', \psi' + \bar{\psi}', \dots, \psi' + \bar{\psi}') \right). 
\end{split}
\end{align}

Now we can follow the same procedure that took us to the Maurer-Cartan action in the cohomology for CSM in \eqref{eq:MC-CSM}, to obtain the following result for QCD
\begin{multline}\label{eq:MC-QCD}
S'_{\mathrm{MC}} =(2 \pi)^4 \sum_{n \geq 1}\frac{1}{(n+1)}\sum_{p \geq 1} \sum_{\substack{P \in \mathcal{OW}_{n} \\ \hspace{-1pt}P = Q \cup R}} \delta(k_{pP}) \big\{ \tilde{f}_{abc} \big( (k_{Q} \cdot \mathcal{A}^{b}_{R})(\mathcal{A}^{c}_{Q}\cdot\mathcal{A}_{p}^{a}) + \mathcal{A}^{b}_{Q \nu} \mathcal{A}_{p\mu}^{a}\mathcal{F}^{\mu \nu c}_{R}\big) \\
+\bar{\Psi}_{Qi} \slashed{\mathcal{A}}_{p}^{a} (T^{a})^i_{\phantom{i}j}\Psi^j_{R} -\bar{\Psi}_{Ri}\slashed{\mathcal{A}}_{Q}^{a}(T_{a})^i_{\phantom{i}j} \Psi^j_{p} - \bar{\Psi}_{pi}\slashed{\mathcal{A}}_{Q}^{a} (T_{a})^i_{\phantom{i}j}\Psi^j_{R} \big\},
\end{multline}
which contains all the possible QCD tree-level scattering amplitudes for any given set of labels for the particles.

Without loss of generality, we can present this generator in a more compact form. As in the CSM case, we will leave momentum conservation implicit and restrict the sum by fixing one particle index (diagonal sum), removing the symmetry factor $\frac{1}{n+1}$. Therefore, 
\ba\label{eq:MC-QCD-diag}
S'^{\mathrm{QCD}}_{\mathrm{MC}} &=& \sum_{n \geq 1} \sum_{\substack{P \in \mathcal{OW}_{n} \\ \hspace{-1pt}P = Q \cup R}} \big\{ \tilde{f}_{abc} \big( (k_{Q} \cdot \mathcal{A}^{b}_{R})(\mathcal{A}^{c}_{Q}\cdot\mathcal{A}_{p}^{a}) + \mathcal{A}^{b}_{Q \nu} \mathcal{A}_{p\mu}^{a}\mathcal{F}^{\mu \nu c}_{R}\big)\nonumber \\
&&\hspace{1.7cm}+\bar{\Psi}_{Qi} \slashed{\mathcal{A}}_{p}^{a} (T^{a})^i_{\phantom{i}j}\Psi^j_{R} -\bar{\Psi}_{Ri}\slashed{\mathcal{A}}_{Q}^{a}(T_{a})^i_{\phantom{i}j} \Psi^j_{p} - \bar{\Psi}_{pi}\slashed{\mathcal{A}}_{Q}^{a} (T_{a})^i_{\phantom{i}j}\Psi^j_{R} \big\}
\ea
will be our generator. 

The amplitudes can then be extracted with the application of the boundary condition operators similar to \eqref{eq:ferm-oper} together with an additional one for the gluons,
\ba\label{eq:glu-ferm-oper} 
\tilde{\mathcal{A}}_{p}^{\mu a}\frac{\partial}{\partial\mathcal{A}_{p}^{\mu a}}.
\ea
with $\tilde{\mathcal{A}}_{p}^{\mu a}=\epsilon_p\delta^a_{a_p}$, with $k_p\cdot\epsilon_p=0$. A general $m$-point amplitude with $m-2k$ gluons and $k$ quark-antiquark pairs is given by 
\ba\label{eq:QCD-amps}
\mathscr{A}_{m,k} = \prod_{p\in \mathcal{S}^1_{m-2k}}\left(\tilde{\mathcal{A}}_{p}^{\mu a}\frac{\partial}{\partial\mathcal{A}_{p}^{\mu a}}\right)\prod_{q\in \mathcal{S}^2_{k}}\left(\tilde{\bar{\Psi}}_{qi}\frac{\partial}{\partial\bar{\Psi}_{qi}}\right)\prod_{r\in \mathcal{S}^3_{k}}\left(\tilde{\Psi}_{r}^j\frac{\partial}{\partial\Psi_r^j}\right)S'^{\mathrm{diag}}_{\mathrm{MC}} \Bigg|_{\mathcal{A},\Psi, \bar{\Psi}=0}
\ea 
with $\mathcal{S}^i_k$ denoting non-intersecting particle label sets of size $k$, which are assigned by the given amplitude to be computed (e.g. $(\overline{1},2,3,4,\underline{5})$ for $\mathscr{A}_{5,1}^{\mathrm{tree}}$).

When $k=0$, the expression above generates amplitudes containing only gluons, which are widely available in the literature. When $m=2k$, the amplitudes in \eqref{eq:QCD-amps} have no external gluons. Apart from dimensions and some kinematic factors, they already appeared in the previous section for CSM.

\paragraph{Flavoured amplitudes} Using the Dyck decomposition presented in the previous section, we can discuss fully-flavoured amplitudes in QCD, possibly with external gluons. Basically, the labels that represent gluons do not come with any kind of parentheses and can be placed anywhere without affecting the closure condition of a Dyck word. For example, 
\ba  
\overline{1} \, 2 \, \overline{3} \,\,4 \, \underline{5}  \underline{6}\, \rightarrow \,  (1\,2\,(3\,4\,5)6)  ,
\ea
is a Dyck word.

Here we define the generator for fully-flavoured amplitudes for QCD in the same way we did in the previous chapter:
\begin{equation}
\label{eq:flav-QCD-gener}
\mathbf{G}_{QCD}=\sum_{n \geq 1} \sum_{\substack{P \in \mathcal{OW}_{n} \\ \hspace{-8pt}P\in \mathrm{Dyck}}}\Big\{ \bar{\Psi}_{1i}\slashed{\mathcal{A}}_{P}^{a} (T_{a})^{i}_{\phantom{i}j} \Psi_{n+2}^j + \sum_{\substack{P = Q \cup R \\ \hspace{-8pt}Q,R\in \mathrm{Dyck}}} \bar{\Psi}_{1i}(T_b)^{i}_{\phantom{i}j}\slashed{\mathcal{A}}_{Q}^{b}\left(\frac{\slashed{k}_{1Q} - m}{s_{1Q} - m^2}\right)\slashed{\mathcal{A}}_{R}^{a} (T_{a})^{j}_{\phantom{j}k} \Psi_{n+2}^k  \Big\}.
\end{equation}

For simplicity we will take the sets for the particle labels with the quark-antiquark pairs first and the external gluons later in the ordering $(\overline{1},\overline{2},\underline{3},\dots,m-2,m-1,\underline{m})$. Therefore, the $m$-point fully-flavoured amplitude reads
\begin{multline}\label{eq:QCD-flav-amps}
\mathscr{A}^{\rm Flav.}_{m,k} =  \Big(\tilde{\bar{\Psi}}_{1i}\frac{\partial}{\partial\bar{\Psi}_{1i}}\Big)
\Big(\tilde{\Psi}_{m}^j\frac{\partial}{\partial\Psi^j_m}\Big) \times \\
\times \prod_{\substack{q=2 }}^{k} \Big(\tilde{\bar{\Psi}}_{2q-2,i}\frac{\partial}{\partial\bar{\Psi}_{2q-2,i}}\Big)\Big(\tilde{\Psi}_{2q-1}^j\frac{\partial}{\partial\Psi_{2q-1}^j}\Big)
\prod_{p = 2k}^{m-1}\Big(\tilde{\mathcal{A}}_{p}^{\mu a}\frac{\partial}{\partial\mathcal{A}_{p}^{\mu a}}\Big)
\mathbf{G}_{QCD} \Bigg|_{\mathcal{A},\Psi,\bar{\Psi}=0}.
\end{multline}
Let us now work out some examples and check that the amplitude generator is indeed a solid tool.

\subsection{Examples}\label{ExampleFermion}

In this section, we are going to present several cases of interest. First, we consider the scattering of one quark-antiquark pair with two and three-gluons. It will then be simple to obtain a generalisation to $n-2$ gluons, since the gluonic current, that we will denote as $a_M^{\mu }$, may be written in a  compact way. 

The second class of examples concerns scattering amplitudes involving only fermions, both with and without flavour.

In what follows we chose a slightly different convention for the perturbiner currents. In the perturbiner expansions \eqref{eq:QCDperturbiner}-\eqref{eq:QCDperturbinercoefficients} we will replace the particle momenta $k_p$ by $-k_p$. Equivalently, the fermion polarisations will by exchanged as $u \leftrightarrow v$. The reason is that we want the outcome to be readily comparable with other results in the literature (e.g. \cite{Johansson:2015oia}). 

\subsubsection{Four-point ampltidue with two gluons}

In this case we will choose the boundary conditions to be
\begin{eqnarray}
&& \tilde{\bar{\Psi}}_{1\,i}=\bar{u}_1 \, \delta^{i_1}_i, \qquad  \tilde{{\Psi}}_{4}^{i}= v_4 \, \delta^{i}_{i_4} , \nonumber \\ 
&& \tilde{\cal A}^\mu_{2\,a}= a_2^\mu \, \delta_{a}^{a_2}, \qquad \tilde{\cal A}^\mu_{3\,a}= a_3^\mu \, \delta_{a}^{a_3}  ,
\end{eqnarray}
where $\bar{u}_1$ and $v_4$ satisfy the equation of motion
\begin{equation}\label{EOMspinors}
\bar{u}_1  (\slashed{k}_{1} - m) =0 , \qquad  (\slashed{k}_{4} + m) v_4=0.
\end{equation}

Following \eqref{eq:QCD-amps}, the total amplitude is expressed as
\begin{eqnarray}
\mathscr{A}_{4,1}
&=&  
(T^{a_2} T^{a_3}  )^{i_{1}}_{\phantom{i_{1}}i_4} \frac{  \bar{u}_{1}   \slashed{a}_{2}   (\slashed{k_{12}} +m ) \slashed{a}_{3} v_4 }{ s_{12} - m^2} + \tilde f_{a_2a_3c} (T^c)^{i_{1}}_{\phantom{i_{1}}i_4}   \bar{u}_1 \slashed{a}_{23} v_4 \nonumber \\
& &+
 (T^{a_3} T^{a_2}  )^{i_{1}}_{\phantom{i_{1}}i_4} \frac{  \bar{u}_{1}   \slashed{a}_{3}   (\slashed{k_{13}} +m ) \slashed{a}_{2} v_4  }{ s_{13} - m^2},
\end{eqnarray}
with the slight change of notation
\begin{equation}\label{eq:amudef1}
\begin{array}{rclcrcl}
n_p^\mu & \equiv& a_{p}^\mu, & & n^\mu_{[pq]} &=& n_{p}^\mu  (k_{p} \cdot n_{q})  -  F_{p}^{\mu \nu}n_{\nu q}  -  (p \,  \leftrightarrow \, q), \\
a^\mu_{pq} &\equiv& \tfrac{n_{[pq]}^\mu  }{s_{pq}}, & & F_p^{\mu \nu} &=&  k_{p}^\mu \, n_{p}^\nu  -   k_{p}^\nu \, n_{p}^\mu.
\end{array}
\end{equation}

Now, if we use the gauge group algebra in \eqref{eq:gaugealgebra}, $\mathscr{A}_{4,1}$ can be finally rewritten in a colour-stripped form as
\begin{eqnarray}
\mathscr{A}_{4,1} = (T^{a_2} T^{a_3} )^{i_{1}}_{\phantom{i_{1}}i_4} A ( \underline{1},2, 3 , \overline{4} ) + (T^{a_3} T^{a_2} )^{i_{1}}_{\phantom{i_{1}}i_4} A ( \underline{1},3, 2 , \overline{4} ),
\end{eqnarray}
where the partial amplitudes are given by
\begin{align}\label{4P-qqg}
 A ( \underline{1},2, 3 , \overline{4} )=
\frac{\bar{u}_1 \slashed{a}_{2} ( \slashed{k}_{12}+m) \slashed{a}_{3}  v_4 }{ (s_{12}-m^2)  }  +
 \bar{u}_1 \slashed{a}_{23}   v_4  .
\end{align}

\subsubsection{Five-point ampltidue with three gluons}

For this example, the chosen boundary conditions are
\begin{eqnarray}
&& \tilde{\bar{\Psi}}_{1\,i}=\bar{u}_1 \, \delta^{i_1}_i, \qquad \tilde{{\Psi}}_{5}^{i}= v_5 \, \delta^{i}_{i_5} , \nonumber \\ 
&& \tilde{\cal A}^\mu_{p\,a}= a_p^\mu \, \delta_{a}^{a_n}, \qquad p=2,3,4 ,
\end{eqnarray}
where $\bar{u}_1$ and $v_5$ satisfy the equation of motions in \eqref{EOMspinors}.

Again, the total amplitude can be read from the general construction in \eqref{eq:QCD-amps}. It is straightforward to check the building blocks supported on the boundary conditions above:

\begin{subequations}
\begin{eqnarray}
{\cal A}_{pq\, a}^\mu &=&   \tilde f_{a_pa_qa} \, a^\mu_{pq}, \\
\bar{\Psi}_{1p\, j} &=& - (T^{a_p})^{i_{1}}_{\phantom{i_{1}}j} \,
\frac{ \bar{u}_{1} \slashed{a}_p  ( \slashed{k}_{1p} + m )  }{s_{1p}-m^2}, \\
{\cal A}_{234\, a}^\mu &=& \frac{1}{s_{234}} \Big\{ \tilde f_{a_2 a_3c } \tilde f_{c a_4 a}   \frac{ n^\mu_{[[23]4]} }{s_{23}}  +   \tilde f_{a_3 a_4c} \tilde f_{ca_2a}  \frac{ n^\mu_{[[34]2]} }{s_{34}} +  \tilde f_{ a_4 a_2 c } \tilde f_{ca_3a}  \frac{ n^\mu_{[[42]3] } }{s_{42}}\Big\}, \\
\bar{\Psi}_{1pq\, j} &=&
(T^{a_p} T^{a_q})^{i_{1}}_{\phantom{i_{1}}j} \frac{[   \bar{u}_{1} \slashed{a}_p  ( \slashed{k}_{1p} + m )   \slashed{a}_q  ( \slashed{k}_{1pq} + m )  ]}{ (s_{1p} - m^2)(s_{1pq}-m^2) }
+
\tilde f_{a_pa_q a}(T^a)^{i_{1}}_{\phantom{i_{1}}j} \frac{ [- \bar{u}_1 \slashed{a}_{pq} ( \slashed{k}_{1pq} + m )  ]}{(s_{1pq}-m^2)} 
 \nonumber \\
& &
+
(T^{a_q} T^{a_p})^{i_{1}}_{\phantom{i_{1}}j} \frac{[  \bar{u}_{1} \slashed{a}_q  ( \slashed{k}_{1q} + m )  \slashed{a}_p  ( \slashed{k}_{1qp} + m )  ]}{ (s_{1q} - m^2) (s_{1pq}-m^2) }
\end{eqnarray}
\end{subequations}
where we have generalised the definitions in \eqref{eq:amudef1} to
\begin{subequations}
\begin{eqnarray}
n^\mu_{[PQ]} &\equiv&  n_{P}^\mu (k_{P} \cdot n_{Q})  -   F_{P}^{\mu \nu} n_{\nu Q}  -  (P \leftrightarrow Q), \\
F^{\mu \nu}_{[PQ]} &\equiv&  k_{PQ}^\mu n_{[PQ]}^\nu + s_{PQ} n_P^\mu n_Q^\nu  - (\mu  \leftrightarrow \nu),
\end{eqnarray}
\end{subequations}
and
\begin{equation}\label{amudef2}
a^\mu_{p_1p_2\ldots p_r} =
\frac{ n^\mu_{ [[\ldots [ p_1 p_2] p_3] \ldots p_r]  }   }{ s_{p_1p_2}  s_{p_1p_2 p_3} \cdots s_{p_1p_2\ldots p_r} } +   (\text{all possible nested brackets}).
\end{equation}
For example, $a^\mu_{234} $ and $a^\mu_{2345} $ are given by
\begin{align}
&
a^\mu_{234} =
\frac{ n^\mu_{ [ [ 23] 4]}   }{ s_{23}  s_{234}  } + \frac{ n^\mu_{ [ 2[3 4]]   }   }{ s_{234}  s_{34}  } , \nonumber \\
&
a^\mu_{2345} =
\frac{ n^\mu_{ [ [ 23] 4] 5]}   }{ s_{23}  s_{234}  s_{2345} } + \frac{ n^\mu_{ [[  2 [3 4]] 5]}   }{ s_{34}  s_{234}  s_{2345} } +\frac{ n^\mu_{ [ 2 [[3 4] 5]]}   }{ s_{34}  s_{345}  s_{2345} } 
+\frac{ n^\mu_{ [ 2 [3 [4 5]]]}   }{ s_{45}  s_{345}  s_{2345} } +\frac{ n^\mu_{ [ [2 3] [4 5]]}   }{ s_{23}  s_{45}  s_{2345} }.
\end{align}

Again, using the gauge group algebra we can rewrite the total amplitude $\mathscr{A}_{5,1}$ in terms of the partial amplitudes as
\begin{eqnarray}
\mathscr{A}_{5,1} = \sum_{\rho\in S_3} (T^{a_{\rho(2)}} T^{a_{\rho(3)}} T^{a_{\rho(4)}} )^{i_1}_{\phantom{i_1}i_5} \, A ( \underline{1},\rho(2), \rho(3), \rho(4) , \overline{5} ),
\end{eqnarray}
with
\begin{eqnarray}
A ( \underline{1},2, 3,4 , \overline{5} ) &=& \bar{u}_1 \slashed{a}_{234} v_5  + 
\frac{\bar{u}_1 \slashed{a}_{2} ( \slashed{k}_{12}+m) \slashed{a}_{3} ( \slashed{k}_{123}+m) \slashed{a}_{4}  v_5 }{ (s_{12}-m^2)  (s_{123}-m^2) } \nonumber \\
& & +
\frac{\bar{u}_1 \slashed{a}_{23} ( \slashed{k}_{123}+m) \slashed{a}_{4}  v_5 }{  (s_{123}-m^2) } 
+
\frac{\bar{u}_1 \slashed{a}_{2} ( \slashed{k}_{12}+m) \slashed{a}_{34}  v_5 }{ (s_{12}-m^2) }.
\end{eqnarray}

\subsubsection{Generalisation $n$-point amplitudes with $(n-2)$ gluons}

In order to generalise the color-ordered amplitude, $A ( \underline{1},2, \ldots , \overline{n} )$,  
it is useful to built intuition from Feynman diagrams. The color-stripped Feynman rules from the QCD action are
\begin{align}\label{FeynmanSpinors}
\begin{gathered}
\feynmandiagram [layered layout, large, horizontal=a to b] {
i2-- [opacity=0] b1 [particle=\( \mu \)]
-- [opacity=0] b2 [particle=\( \)],
i1 [particle=\( \)]
-- [fermion] a1
-- [fermion] a2 [particle=\(  \)],
a1 -- [gluon] b1,
};
\end{gathered}
=\gamma^\mu \, , 
\qquad\,\,\,
\begin{gathered}
\feynmandiagram [layered layout, large, horizontal=a to b] {
i1 [particle=\(k_P\)]
-- [opacity=1] a1
-- [fermion] a2
-- [opacity=1] a3 [particle=\((-k_P)\)],
};
\end{gathered}
= \frac{( \slashed{k}_{P}+m) }{k_P^2-m^2} \, ,
\end{align}
where $P$ is a multiparticle label.

Thus, in terms of diagrams, the partial amplitude $A (\underline{1} ,2, 3, \overline{4} )$ can be simply represented as
\begin{equation}\label{A1234Q}
A (\underline{1} ,2, 3, \overline{4} ) =
\begin{gathered}
\feynmandiagram [layered layout, large, horizontal=a to b] {
i2-- [opacity=0] b1 [particle=\( a_{23}\)]
-- [opacity=0] b2 [particle=\( \)],
i1 [particle=\(\underline 1\)]
-- [fermion] a1
-- [fermion] a2 [particle=\(\overline 4\)],
a1 -- [gluon] b1,
};
\end{gathered}
+
\begin{gathered}
\feynmandiagram [layered layout, large, horizontal=a to b] {
i2-- [opacity=0] b1 [particle=\( a_2\)]
-- [opacity=0] b2 [particle=\( a_3\)],
i1 [particle=\(\underline 1\)]
-- [fermion] a1
-- [fermion] a2
-- [fermion] a3 [particle=\(\overline 4\)],
a1 -- [gluon] b1,
a2 -- [gluon] b2,
};
\end{gathered}\,\,\,.
\end{equation}

Similarly, $A (\underline{1} ,2, 3,4, \overline{5} ) $  may be cast as
\begin{eqnarray}
A (\underline{1} ,2, 3, 4, \overline{5} ) &=& 
\begin{gathered}
\feynmandiagram [layered layout, large, horizontal=a to b] {
i2-- [opacity=0] b1 [particle=\( a_{234}\)]
-- [opacity=0] b2 [particle=\( \)],
i1 [particle=\(\underline 1\)]
-- [fermion] a1
-- [fermion] a2 [particle=\(\overline 5\)],
a1 -- [gluon] b1,
};
\end{gathered}\nonumber
+
\begin{gathered}
\feynmandiagram [layered layout, large, horizontal=a to b] {
i2-- [opacity=0] b1 [particle=\( a_2\)]
-- [opacity=0] b2 [particle=\( a_3\)]
-- [opacity=0] b3 [particle=\( a_4\)],
i1 [particle=\(\underline 1\)]
-- [fermion] a1
-- [fermion] a2
-- [fermion] a3 
-- [fermion] a4
[particle=\(\overline 5\)],
a1 -- [gluon] b1,
a2 -- [gluon] b2,
a3 -- [gluon] b3,
};
\end{gathered}+\nonumber \\
& &
+
\begin{gathered}
\feynmandiagram [layered layout, large, horizontal=a to b] {
i2-- [opacity=0] b1 [particle=\( a_{23}\)]
-- [opacity=0] b2 [particle=\( a_4\)],
i1 [particle=\(\underline 1\)]
-- [fermion] a1
-- [fermion] a2
-- [fermion] a3 [particle=\(\overline 5\)],
a1 -- [gluon] b1,
a2 -- [gluon] b2,
};
\end{gathered}
+
\begin{gathered}
\feynmandiagram [layered layout, large, horizontal=a to b] {
i2-- [opacity=0] b1 [particle=\( a_{2}\)]
-- [opacity=0] b2 [particle=\( a_{34}\)],
i1 [particle=\(\underline 1\)]
-- [fermion] a1
-- [fermion] a2
-- [fermion] a3 [particle=\(\overline 5\)],
a1 -- [gluon] b1,
a2 -- [gluon] b2,
};
\end{gathered}
\,\,\,.
\end{eqnarray}

Therefore, following this pattern, we propose the following compact formula for the $n$-point partial amplitude, $A (\underline{1} ,2, \ldots,n-1, \overline{n} )$,
\begin{eqnarray}
A (\underline{1} ,2, \ldots,n-1, \overline{n} ) &=&
\begin{gathered}
\feynmandiagram [layered layout, large, horizontal=a to b] {
i2-- [opacity=0] b1 [particle=\( a_{23...n-1}\)]
-- [opacity=0] b2 [particle=\( \)],
i1 [particle=\(\underline 1\)]
-- [fermion] a1
-- [fermion] a2 [particle=\(\overline n\)],
a1 -- [gluon] b1,
};
\end{gathered}+
\begin{gathered}
\feynmandiagram [layered layout, large, horizontal=a to b] {
i2-- [opacity=0] b1 [particle=\( a_2\)]
-- [opacity=0] b2 [particle=\( a_3\)]
-- [opacity=0] b3 [particle=\( a_{n-1}\)],
i1 [particle=\(\underline 1\)]
-- [fermion] a1
-- [fermion] a2
-- [ghost] a3 
-- [fermion] a4
[particle=\(\overline n\)],
a1 -- [gluon] b1,
a2 -- [gluon] b2,
a3 -- [gluon] b3,
};
\end{gathered}+\\
&&+
\begin{gathered}
\feynmandiagram [layered layout, large, horizontal=a to b] {
i2-- [opacity=0] b1 [particle=\( a_{23}\)]
-- [opacity=0] b2 [particle=\( a_{n-1}\)],
i1 [particle=\(\underline 1\)]
-- [fermion] a1
-- [ghost] a2
-- [fermion] a3 [particle=\(\overline n\)],
a1 -- [gluon] b1,
a2 -- [gluon] b2,
};
\end{gathered}
+\cdots +
\begin{gathered}
\feynmandiagram [layered layout, large, horizontal=a to b] {
i2-- [opacity=0] b1 [particle=\( a_{2...p}\)]
-- [opacity=0] b2 [particle=\( a_{r...n-1}\)],
i1 [particle=\(\underline 1\)]
-- [fermion] a1
-- [ghost] a2
-- [fermion] a3 [particle=\(\overline n\)],
a1 -- [gluon] b1,
a2 -- [gluon] b2,
};
\end{gathered}
+\cdots . \nonumber
\end{eqnarray}
Analytically, it reads
\begin{align}\label{AS1n} 
&
A (\underline{1} ,2, \ldots, n-1, \overline{n} )   = \nonumber \\
&
\sum_{i=1}^{n-2} \sum_{\substack{(p_1,\ldots, p_i) \\  \in \, C_i(n-2)}  } \!\!\!
\bar{u}_1 \slashed{a}_{2, p_1+1} 
\frac{  (\slashed{k}_{1,p_1+1}+m ) }{(s_{1,p_1+1}-m^2)  }  
\slashed{a}_{p_1+2, p_1+p_2+1} 
...
\frac{  (\slashed{k}_{1,n-p_i-1} +m^2 ) }{  (s_{1,n-p_i-1}-m^2) } \slashed{a}_{n-p_i , n-1}  v_n, \qquad
\end{align}
with the definitions:
\begin{align}
k^\mu_{p,q} &= k^\mu_{p}+k^\mu_{p+1}+\cdots + k^\mu_{q} \, ,\nonumber\\
s_{p,q} &= s_{p\,p+1\ldots q} = k_{p,q}\cdot k_{p,q}, \nonumber \\
a^\mu_{p,q} & =a^\mu_{p\,p+1\ldots q}.
\end{align}
$C_i(n)$ is the set of  {\it i-tuples} of positive integers that add to $n$. For example,
\begin{equation}\label{Cidef}
\begin{array}{rcl}
C_1(4) &=& \left\{  (4)  \right\},   \\
C_2(4) &=& \left\{  (1,3) , \, (2,2), \, (3,1)  \right\}, \\
C_3(4) &=& \left\{  (1,1,2) , \, (1,2,1), \, (2,1,1)  \right\}, \\
C_4(4) &=& \left\{  (1,1,1,1)  \right\}.  
\end{array}
\end{equation} 
This proposal has been numerically verified for six and seven points.

\subsubsection{Four and Six-point amplitudes with quark-antiquark pairs}

For this computation, we choose the following configuration,
\begin{align}
\tilde{\bar{\Psi}}_{1\,i}=\bar{u}_1 \, \delta^{i_1}_i, \quad          
\tilde{\bar{\Psi}}_{2\,i}=\bar{u}_2 \, \delta^{i_2}_i, \quad
\tilde{\Psi}_{3}^{i}= v_3 \, \delta^{i}_{i_3},  \quad
\tilde{\Psi}_{4}^{i}= v_4 \, \delta^{i}_{i_4}.
\end{align}

We can then use either \eqref{eq:QCD-flav-amps} and \eqref{eq:QCD-amps} to obtain amplitudes with and without flavour, $\mathscr{A}_{4,2}$ and $\mathscr{A}^{\rm Flav.}_{4,2}$ respectively. They are simply expressed as
\begin{align}
& \mathscr{A}_{4,2}  =  \frac{ [-(\bar{u}_1 \gamma_\mu v_4) (\bar{u}_2 \gamma^\mu v_3)]  }{s_{23}} (T_a)^{i_{1}}_{\phantom{i_{1}}i_4} (T^a)^{i_{2}}_{\phantom{i_{2}}i_3}   +   
\frac{ [-(\bar{u}_2 \gamma_\mu v_4) (\bar{u}_1 \gamma^\mu v_3)]  }{s_{23}} (T_a)^{i_{2}}_{\phantom{i_{2}}i_4} (T^a)^{i_{1}}_{\phantom{i_{1}}i_3} 
  ,   \\  
& \mathscr{A}^{\rm Flav.}_{4,2}  = \frac{ [-(\bar{u}_1 \gamma_\mu v_4) (\bar{u}_2 \gamma^\mu v_3)]  }{s_{23}} (T_a)^{i_{1}}_{\phantom{i_{1}}i_4} (T^a)^{i_{2}}_{\phantom{i_{2}}i_3}  ,
\end{align}
which are very well known in the literature. In $ \mathscr{A}^{\rm Flav.}_{4,2}$, each pair $(1,4)$ and $(2,3)$ have a different flavour.

For the fully-flavoured scattering between six fermions, $\mathscr{A}^{\rm Flav.}_{6,3}$, we choose the following boundary conditions,
\begin{align}
\tilde{\bar{\Psi}}_{p\,i}=\bar{u}_p \, \delta^{i_p}_i, \quad p=1,2,4,\qquad\quad         
\tilde{\Psi}_{q}^{i}= v_q \, \delta^{i}_{i_q},  \quad q=3,5,6,
\end{align}
similarly to what was done in the CSM case.

In appendix \ref{Comp-Psi}, the building blocks of this amplitude are computed explicitly. After some simplifications, $\mathscr{A}^{\rm Flav.}_{6,3}$ can be written as
\begin{multline}
 \mathscr{A}^{\rm Flav.}_{6,3}   = 
\frac{ -\tilde f^{abc} (T_a)^{i_{1}}_{\phantom{i_{1}}i_6}  (T_b)^{i_{2}}_{\phantom{i_{2}}i_3}   (T_c)^{i_{4}}_{\phantom{i_{4}}i_5} \, n_0}{s_{61}s_{23} s_{45}} 
+ 
\Bigg[
\frac{(T^aT^b)^{i_{1}}_{\phantom{i_{1}}i_6} (T^a)^{i_{2}}_{\phantom{i_{2}}i_3} (T^b)^{i_{4}}_{\phantom{i_{4}}i_5} \, n_1  }{(s_{123} - m^2) s_{23} s_{45}} 
\\
+ \frac{(T^aT^b)^{i_{1}}_{\phantom{i_{1}}i_6}  (T^a)^{i_{4}}_{\phantom{i_{4}}i_5} (T^b)^{i_{2}}_{\phantom{i_{2}}i_3} \, n_2  }{(s_{145} - m^2) s_{23} s_{45}} +
{\rm cyclic}
\Big(
(61)\rightarrow (23) \rightarrow (45) 
\Big) 
\Bigg]
,
\end{multline}
with 
\begin{equation}
\begin{array}{rcl}
n_0&=&-
(\bar{u}_1 \gamma_\mu v_6) (\bar{u}_2 \gamma_\nu v_3)  (\bar{u}_4 \gamma_\rho  v_5) V^{\mu\nu\rho}_{k_6+k_1,k_2+k_3,k_4+k_5},\\
n_1&=&  - \left[ \bar{u}_1\gamma_\mu ( \slashed{k}_{123}+m ) \gamma_\nu  v_6 \right] (\bar{u}_2 \gamma^\mu v_3) (\bar{u}_4 \gamma^\nu v_5) ,\\
n_2&=& - \left[ \bar{u}_1\gamma_\mu ( \slashed{k}_{145}+m ) \gamma_\nu  v_6 \right] (\bar{u}_4 \gamma^\mu v_5) (\bar{u}_2 \gamma^\nu v_3) ,
\end{array}
\end{equation}
where the vertex $V^{\mu\nu\rho}_{p,q,k} $ is defined in \eqref{3pvertex}. Clearly, each pair $(1,6)$, $(2,3)$ and $(4,5)$  has a different flavour. This result is in agreement with the one obtained in \cite{Johansson:2015oia}.

\section{$L_{\infty}$-structure for sQCD}\label{sec:sQCD}
This section introduces the $L_{\infty}$-structure for scalar QCD and its tree level amplitude generators. With the obvious modifications, the technical aspects are completely analogous to CSM and QCD.

\subsection{The sQCD $L_\infty$ algebra}
The notation here is identical to the one used in QCD, except that we work with $\Omega^{0}(\RR^{1,3},V)$ and $\Omega^{0}(\RR^{1,3},\bar{V})$ instead of $\Omega^{0}(\RR^{1,3},S \otimes V)$ and $\Omega^{0}(\RR^{1,3},S \otimes \bar{V})$. The graded vector space underlying the scalar QCD $L_{\infty}$-algebra $L$ is 
\begin{align*}
L^{0} &= \Omega^{0}(\RR^{1,3},\su(N)), \\
L^{1} &= \Omega^{1}(\RR^{1,3},\su(N)) \oplus \Omega^{0}(\RR^{1,3},V) \oplus \Omega^{0}(\RR^{1,3}, \bar{V}), \\
L^{2} &= \Omega^{1}(\RR^{1,3},\su(N)) \oplus \Omega^{0}(\RR^{1,3},\bar{V}) \oplus \Omega^{0}(\RR^{1,3},V), \\
L^{3} &= \Omega^{0}(\RR^{1,3},\su(N)). 
\end{align*}
In the terminology of subsection \ref{sec:2.4}, elements $c \in L^{0}$ should be interpreted as infinitesimal gauge parameters, elements $A + \phi + \bar{\phi} \in L^{1}$ as triples of a gauge field, a scalar field and a conjugate scalar field, elements $A^* + \bar{\phi}^{*} + \phi^* \in L^{2}$ as triples of antifields conjugate to $A$, $\phi$ and $\bar{\phi}^{*}$, and elements $c^* \in L^{3}$ as antifields conjugate to $c$. The direct link between the BV transformations \eqref{eq:BVtransf-scalarQCD} and the higher order brackets on $L$ is given by formulas analogous to \eqref{eq:BV-Linfinity}, with $\phi$ and $\bar{\phi}$ in place of $\psi$ and $\bar{\psi}$, and where the BV transformations of the antifields $A^{*a}_{\mu}$, $\bar{\phi}^{*}_{i}$ and $\phi^{* i}$ contain the extra contributions $\frac{1}{3!}l_3(A,A,A)^{a}_{\mu}$, $\frac{1}{2} l_3(A,A,\phi + \bar{\phi})_{i}$ and $\frac{1}{2} l_3(A,A,\phi + \bar{\phi})^{i}$, respectively. The non-vanishing higher order brackets on $L$ are identified next. For  $l_{1}$
we have
\begin{equation}\label{eq:sQCDl1}
\begin{array}{rc}
l_{1}(c)_{\mu a}=-\partial_{\mu}c_{a} & \in L^{1},\\
l_{1}(A)_{\mu}^{a}=-\partial^{\nu}(\partial_{\mu}A_{\nu}^{a}-\partial_{\nu}A_{\mu}^{a}) & \in L^{2},\\
l_1(\phi + \bar{\phi})^{i} = -\square \phi^{i} - m^2 \phi^{i} & \in L^{2},\\
l_1(\phi + \bar{\phi})_{i} = -\square \bar{\phi}_{i}  - m^2 \bar{\phi}_{i}, & \in L^{2},\\
l_{1}(A^{*})_{a}=-\partial^{\mu}A_{\mu a}^{*} & \in L^{3},
\end{array}
\end{equation}
satisfying the relation \eqref{eq:l1-relation}. For $l_{2}$ we have
\begin{equation}\label{eq:sQCDl2}
\begin{array}{rc}
l_2(c_1,c_2)^{a} = \ui \tilde{f}_{bc}^{\phantom{bc}a} c_1^{b} c_2^{c} & \in L^{0},\\
l_2(c_1,A_2)^{a}_{\mu} =  \ui \tilde{f}_{bc}^{\phantom{bc}a}   c_1^{b} A_{2\mu}^{c} & \in L^{1},\\
l_2(c_1,A_2^*)^{a}_{\mu} = \ui \tilde{f}_{bc}^{\phantom{bc}a} c_1^{b} A_{2\mu}^{*c} & \in L^{2},\\
l_2(c_1,c_2^*)^{a}  =- \ui \tilde{f}_{bc}^{\phantom{bc} a}  c_1^{b} c_2^{*c} & \in L^{3},\\
l_{2}(A^{1},A^{2})_{\mu a}=\tilde{f}_{abc}\partial^{\nu}(A_{\mu b}^{1}A_{\nu c}^{2})-\tilde{f}_{abc}(\partial_{\mu}A_{\nu b}^{1}-\partial_{\nu}A_{\mu b}^{1})A_{c}^{2\nu}+(1\leftrightarrow2) & \in L^{2}\\
l_2(A_1,A_2^*)^{a} = \ui f_{bc}^{\phantom{bc}a} A_{1\mu}^{b} A_2^{*c\mu} & \in L^{3}\\
l_2(c_1,\phi_2)^{i} = -\ui c_1^{a} (T_{a})^{i}_{\phantom{i}j} \phi_2^{j} & \in L^{1},\\
l_2(c_1,\bar{\phi}_2)_{i} = \ui c_1^{a} \bar{\phi}_{2j} (T_{a})^{j}_{\phantom{j}i} & \in L^{1},\\
l_2(c_1,\phi^*_2)^{i} = - \ui c_1^{a} (T_{a})^{i}_{\phantom{i}j} \phi_2^{*j} & \in L^{2},\\
l_2(c_1,\bar{\phi}^*_2)_{i} =  \ui c_1^{a} \bar{\phi}_{2j}(T_{a})^{j}_{\phantom{j}i} & \in L^{2},\\
l_2(A_1,\phi_2)^{i} = \ui (T^{a})^{i}_{\phantom{i}j}\{ \partial_{\mu}(A^{\mu}_{1 a} \phi_2^{j}) + A^{\mu}_{1a} \partial_{\mu}\phi_2^{j}\} & \in L^{2},\\
l_2(A_1,\bar{\phi}_2)_{i} = -\ui (T^{a})^{j}_{\phantom{j}i}\{ \partial_{\mu}(A^{\mu}_{1 a}\bar{\phi}_{2j}) + A^{\mu}_{1a} \partial_{\mu}\bar{\phi}_{2j}\} & \in L^{2},\\
l_2 (\phi_1 + \bar{\phi}_1,\phi_2 + \bar{\phi}_2)^{a}_{\mu}= \ui (T^{a})^{i}_{\phantom{i}j} \{\bar{\phi}_{1i}\partial_{\mu} \phi_2^{j} - \partial_{\mu}\bar{\phi}_{1i}\phi_2^{j} + (1\leftrightarrow2)\} & \in L^{2},\\
l_2 (\phi_1 + \bar{\phi}_1, \bar{\phi}_2^{*} + \phi_2^{*})^{a} = \ui  (T^{a})^{i}_{\phantom{i}j} \{\bar{\phi}_{1i} \phi^{*j}_{2} - \bar{\phi}_{2i}\phi^{j}_{1}\} & \in L^{3},\\
\end{array}
\end{equation}
satisfying the relation \eqref{eq:l2-relation}. And for $l_{3}$
we have
\begin{equation}\label{eq:sQCDl3}
\begin{array}{rc}
l_{3}(A_{1},A_{2},A_{3})_{\mu a}=-\tilde{f}_{abe}\tilde{f}_{cde}A_{1 \nu b}A_{2 c}^{\nu}A_{3 \mu d}+\mathrm{sym}(1,2,3) & \in L^{2},\\
l_{3}(A,\phi_1 + \bar{\phi}_1,\phi_2 + \bar{\phi}_2)_{\mu a}=A_{\mu b}\bar{\phi}_{i}^{1}(T^{a}T^{b}+T^{b}T^{a})_{\hphantom{i}j}^{i}\phi^{2j}+(1\leftrightarrow2) & \in L^{2},\\
l_{3}(A_{1},A_{2},\phi)^{i}= A_{1 \mu a}A_{2 b}^{\mu}(T^{a}T^{b}+T^{b}T^{a})_{\hphantom{i}j}^{i}\phi^{j} & \in L^{2},\\
l_{3}(A_{1},A_{2},\bar{\phi})_{i}= A_{1 \mu a} A_{2 b}^{\mu}\bar{\phi}_{j}(T^{a}T^{b}+T^{b}T^{a})_{\hphantom{j}i}^{j} & \in L^{2},\\
l_{3}(\phi_1 + \bar{\phi}_1,\phi_2 + \bar{\phi}_2,\phi_3 + \bar{\phi}_3)^i=\lambda (\bar{\phi}_{1}\phi_{2})\phi_{3}^{i})+\mathrm{sym}(1,2,3) & \in L^{2}, \\
l_{3}(\phi_1 + \bar{\phi}_1,\phi_2 + \bar{\phi}_2,\phi_3 + \bar{\phi}_3)_i=\lambda (\bar{\phi}_{1}\phi_{2})\bar{\phi}_{3i}+\mathrm{sym}(1,2,3) & \in L^{2},
\end{array}
\end{equation}
satisfying the relation \eqref{eq:l3-relation}.

The underlying cochain complex of $L$ is
$$
\begin{tikzpicture}[scale=1.0,baseline=-0.1cm, inner sep=1mm,>=stealth]
\node (0) at (-0.9,0)  {$\Omega^{0}(\RR^{1,3},\su(N))$};
\node (1) at (2.68,0)  {$\Omega^{1}(\RR^{1,3},\su(N))$};
\node (2) at (8.32,0)  {$\Omega^{1}(\RR^{1,3},\su(N))$};
\node (3) at (11.9,0)  {$\Omega^{0}(\RR^{1,3},\su(N)).$};
\node (4) at (2.68,-0.6) {$\oplus$};
\node (5) at (8.32,-0.6) {$\oplus$};
\node (6) at (2.68,-1.2) {$\Omega^{0}(\RR^{1,3}, V)$};
\node (7) at (8.32,-1.2) {$\Omega^{0}(\RR^{1,3}, \bar{V})$};
\node (8) at (2.68,-1.8) {$\oplus$};
\node (9) at (8.32,-1.8) {$\oplus$};
\node (10) at (3.85,-1.8) {};
\node (11) at (7.15,-1.8) {};
\node (12) at (2.68,-2.4) {$\Omega^{0}(\RR^{1,3}, \bar{V})$};
\node (13) at (8.32,-2.4) {$\Omega^{0}(\RR^{1,3}, V)$};
\tikzset{every node/.style={fill=white}} 
\draw[->, thick] (0) to node[midway,above=0.2pt] {\scriptsize$\ud$} (1); 
\draw[->, thick] (1) to node[midway,above=0.2pt] {\scriptsize$\delta \ud$} (2); 
\draw[->, thick] (2) to node[midway,above=0.2pt] {\scriptsize$\delta$} (3); 
\draw[->, thick] (10) to node[midway,above=0.2pt] {\scriptsize$\begin{pmatrix}
0 & \square + m^2  \\
\square + m^2 & 0 
\end{pmatrix}$} (11);
\end{tikzpicture}
$$

The cyclic structure is a simple parallel to the previous ones in \eqref{eq:inner-CSM} and \eqref{eq:inner-QCD},
\begin{align}\label{eq:inner-sQCD}
\begin{split}
\langle c , c^*  \rangle &=  \int_{\RR^{1,3}} \ud^4 x \,  c^*_{a} c^{a}, \\
\langle A , A^*  \rangle &=  \int_{\RR^{1,3}} \ud^4 x \, A^{*\mu}_{a} A_{\mu}^{a} , \\
\langle \phi + \bar{\phi}  , \bar{\phi}^* + \phi^*  \rangle &= \int_{\RR^{1,3}} \ud^4 x \, (  \phi^{*i} \bar{\phi}_{i} +  \bar{\phi}^{*}_{i} \phi^{i}  ), 
\end{split}
\end{align}
where $c \in L^{0}$, $c^* \in L^{3}$, $A+ \phi + \bar{\phi} \in L^{1}$ and $A^* +  \bar{\phi}^{*} + \phi^* \in L^{2}$. 

Now it is a straightforward computation to reproduce the homotopy Maurer-Cartan action for the $L_{\infty}$-algebra $L$,
\begin{equation}
S_{\mathrm{MC}}[A + \phi + \bar{\phi}] =  \int_{\RR^{1,3}} \ud^4 x \,\left\{ -  \frac{1}{4}F^{a}_{\mu\nu} F^{\mu\nu}_a +  D_{\mu} \bar{\phi} D^{\mu} \phi - m^2 \bar{\phi} \phi \right\}.
\end{equation}
which is simply the scalar QCD action displayed in \eqref{eq:sQCDaction}.

Just as in section \ref{sec:CSM} and section \ref{sec:QCD}, we adjust the definition of the $L_{\infty}$-algebra $L$ so as to include perturbiner expansions. We will consider an infinite set $(k_i)_{i \geq 1}$ of momentum vectors in $\RR^{1,3}$, and using the notation therein, consider the associated spaces $\mathscr{E}^{r} (\RR^{1,3},\su(N))$ and $\mathscr{E}^{0}(\RR^{1,3},V)$. The graded vector space underlying the cyclic $L_{\infty}$-algebra $L$ then becomes
\begin{align*}
L^{0} &= \mathscr{E}^{0}(\RR^{1,3},\su(N)), \\
L^{1} &= \mathscr{E}^{1}(\RR^{1,3},\su(N)) \oplus \mathscr{E}^{0}(\RR^{1,3},V) \oplus \mathscr{E}^{0}(\RR^{1,3}, \bar{V}), \\
L^{2} &= \mathscr{E}^{1}(\RR^{1,3},\su(N)) \oplus \mathscr{E}^{0}(\RR^{1,3},\bar{V}) \oplus \mathscr{E}^{0}(\RR^{1,3},V), \\
L^{3} &= \mathscr{E}^{0}(\RR^{1,3},\su(N)). 
\end{align*}
The corresponding higher order brackets and cyclic inner product are given respectively by \eqref{eq:sQCDl1}, \eqref{eq:sQCDl2}, \eqref{eq:sQCDl3} and \eqref{eq:inner-sQCD}. 

\subsection{The perturbiner expansion for sQCD}
We turn next to the perturbiner expansion for scalar QCD. As indicated earlier, the approach and calculations are very similar to those we have made in \ref{sec:4.2}. 

Like before, in order to describe  the minimal $L_{\infty}$-structure on the cohomology $H^{\sbullet}(L)$ of $L$, we need to define a projection $p \colon L \to H^{\sbullet}(L)$ and a contracting homotopy $h \colon L \to L$. The cochain complex underlying $H^{\sbullet}(L)$ is
$$
\begin{tikzpicture}[scale=1.0,baseline=-0.1cm, inner sep=1mm,>=stealth]
\node (0) at (-1,0)  {$\ker (\ud)$};
\node (1) at (2,0)  {$\ker(\delta \ud)/\im(\ud)$};
\node (2) at (8,0)  {$\ker(\delta \ud)/\im(\ud)$};
\node (3) at (11,0)  {$\ker(\ud).$};
\node (4) at (2,-0.6) {$\oplus$};
\node (5) at (8,-0.6) {$\oplus$};
\node (6) at (2,-1.55) {$\ker \begin{pmatrix}
0 & -(\square + m^2)  \\
-(\square + m^2) & 0 
\end{pmatrix}$};
\node (7) at (8,-1.55) {$\coker \begin{pmatrix}
0 & -(\square + m^2)  \\
-(\square + m^2) & 0 
\end{pmatrix}$};
\tikzset{every node/.style={fill=white}} 
\draw[->, thick] (0) to node[midway,above=0.2pt] {\scriptsize$0$} (1); 
\draw[->, thick] (1) to node[midway,above=0.2pt] {\scriptsize$0$} (2); 
\draw[->, thick] (2) to node[midway,above=0.2pt] {\scriptsize$0$} (3); 
\draw[->, thick] (6) to node[midway,above=0.2pt] {\scriptsize$0$} (7);
\end{tikzpicture}
$$ 
From this, it follows that the four components of the projection $p^{(0)}$, $p^{(1)}$, $p^{(2)}$ and $p^{(3)}$ can be chosen to be the natural projections induced by this decomposition. To define the contracting homotopy, in addition to the massless Feynman propagator $\Dsf^{\mathrm{F}}$ of the previous sections, we need also the massive Feynman propagator $\Gsf^{\mathrm{F}}$ defined on the space of $0$-forms on $\RR^{1,3}$ with values in $V \oplus \bar{V}$ as
\begin{equation}
\Gsf^{\mathrm{F}} = -\frac{1}{\square + m^2}\begin{pmatrix}
0 &1  \\
1 & 0 
\end{pmatrix},
\end{equation}
or, when acting on plane waves of the form $\ue^{-\ui k \cdot x}$,
\begin{equation}
\Gsf^{\mathrm{F}} = \frac{1}{k^2 - m^2}\begin{pmatrix}
0 &1  \\
1 & 0 
\end{pmatrix}.
\end{equation}
As before, we extend both $\Dsf^{\mathrm{F}}$ and $\Gsf^{\mathrm{F}}$ so that we obtain linear operators $\Dsf^{\mathrm{F}} \colon \mathscr{E}^{r}(\RR^{1,3},\su(N)) \to  \mathscr{E}^{r}(\RR^{1,3},\su(N))$ and $\Gsf^{\mathrm{F}}\colon \mathscr{E}^{0}(\RR^{1,3},V) \oplus  \mathscr{E}^{0}(\RR^{1,3},\bar{V}) \to \mathscr{E}^{0}(\RR^{1,3},\bar{V}) \oplus \mathscr{E}^{0}(\RR^{1,3},V)$. The three non-zero components of the contracting homotopy $h$ may then be expressed as
\begin{align}\label{eq:5.6}
\begin{split}
h^{(1)} = \left(\begin{array}{cc} \Dsf^{\mathrm{F}}\circ \delta &\, 0 \end{array} \right) &\colon L^{1} \longrightarrow L^{0}, \\
h^{(2)} = \left(\begin{array}{cc} \Dsf^{\mathrm{F}} \circ P_{\mathrm{e}} & 0 \\ 0 & \Gsf^{\mathrm{F}} \end{array} \right) &\colon L^{2} \longrightarrow L^{1}, \\
h^{(3)} = \left(\begin{array}{c} \ud \circ \Dsf^{\mathrm{F}}  \\ 0 \end{array} \right) &\colon L^{3} \longrightarrow L^{2}.
\end{split}
\end{align}
Writing down the components explicitly, we have
\begin{align}\label{eq:sQCDhomotopy}
\begin{split}
h^{(1)}(A)^{a} &= - \frac{\partial_{\mu}}{\square} A^{a \mu},  \\
h^{(2)}(A^*)^{a}_{\mu} &= \frac{1}{\square}\left( \eta_{\mu\nu} - \frac{\partial_{\mu} \partial_{\nu}}{\square}\right) A^{*a \nu}, \\
 h^{(2)}(\bar{\phi}^{*} + \phi^*)^{i} &= -\frac{1}{\square + m^2} \phi^{* i}, \\
  h^{(2)}(\bar{\phi}^{*} + \phi^*)_{i} &= -\frac{1}{\square + m^2} \bar{\phi}^{*}_{i}, \\
h^{(3)}(c^{*})^{a}_{\mu} &=  - \frac{\partial_{\mu}}{\square} c^{* a},
\end{split}
\end{align}
where $A \in L^{1}$, $A^* + \bar{\phi}^{*} + \phi^* \in L^{2}$ and $c^* \in L^{3}$. As usual, it should be mentioned that the formulas for the $L_{\infty}$-quasi-isomorphism and the higher brackets for the minimal $L_{\infty}$-structure on $H^{\sbullet}(L)$ are derived under the assumption that $h^{(1)}(A + \phi + \bar{\phi}) = 0$ for any $A + \phi + \bar{\phi} \in L^1$. Owing to \eqref{eq:sQCDhomotopy}, this implies that the first summands $A \in \mathscr{E}^{1}(\RR^{1,3},\su(N))$ satisfy the Lorentz gauge condition $\partial_{\mu} A^{a\mu} = 0$. 

We now have the ingredients necessary to formulate the perturbiner expansion for scalar QCD. First of all, we pick a Maurer-Cartan element $A ' + \phi' + \bar{\phi}'\in H^{1}(L) = \left( \ker(\delta \ud)/\im(\ud)\right) \oplus \ker \left( \begin{smallmatrix}
0 & -(\square + m^2)  \\
-(\square + m^2) & 0 
\end{smallmatrix} \right)$, and write
\begin{align}\label{eqn:sQCDplanewave}
\begin{split}
A'^{a \mu} &= \sum_{p \geq 1} \mathcal{A}_{p}^{a\mu} \, \ue^{-\ui k_p \cdot x}, \\
\phi'^{i} &= \sum_{p \geq 1} \Phi^{i}_{p} \, \ue^{-\ui k_p \cdot x}, \\
\bar{\phi}'_{i} &= \sum_{p \geq 1} \bar{\Phi}_{pi} \, \ue^{-\ui k_p \cdot x},
\end{split}
\end{align}
assuming the physical input described around equation \eqref{eq:grassmann-polarization}. Then we define the perturbiner expansion to be the Maurer-Cartan element $A + \phi + \bar{\phi}$ in $L$ given by the formula
\begin{equation}
A +\phi + \bar{\phi}= \sum_{n \geq 1} \frac{1}{n!} f_{n} (A' + \phi' + \bar{\phi}',\dots,A' + \phi' + \bar{\phi}'). 
\end{equation}
The task is to work out the components $A \in \mathscr{E}^{1}(\RR^{1,3},\su(N))$ and $\phi \in \mathscr{E}^{0}(\RR^{1,3},V)$ and $\bar{\phi} \in \mathscr{E}^{0}(\RR^{1,3},\bar{V})$. By the same argument used to derive \eqref{eq:QCDperturbiner-quasi}, we start with
\begin{align}\label{eq:sQCDperturbiner-quasi}
\begin{split}
A^{a \mu} &= \sum_{n \geq 1} \sum_{p + 2q = n}  \frac{1}{p! (2q)!}  f_{n}(A' , \dots, A', \phi' + \bar{\phi}', \dots, \phi' + \bar{\phi}')^{a \mu}, \\
\phi^{i} & = \sum_{n \geq 1} \sum_{p + 2q +1 = n}  \frac{1}{p! (2q + 1)!}  f_{n}(A' , \dots, A', \phi' + \bar{\phi}', \dots, \phi' + \bar{\phi}')^{i}, \\
\bar{\phi}_{i} & = \sum_{n \geq 1} \sum_{p + 2q +1 = n}  \frac{1}{p! (2q + 1)!}  f_{n}(A' , \dots, A', \phi' + \bar{\phi}', \dots, \phi' + \bar{\phi}')_{i},
\end{split}
\end{align}
where the expressions for $f_{n}(A' , \dots, A', \phi' + \bar{\phi}', \dots, \phi' + \bar{\phi}')^{a \mu}$, $f_{n}(A' , \dots, A',\phi' + \bar{\phi}', \dots, \phi' + \bar{\phi}')^{i}$ and $f_{n}(A' , \dots, A',\phi' + \bar{\phi}', \dots, \phi' + \bar{\phi}')_{i}$ are identical to those of \eqref{eq:QCDfnvector}, \eqref{eq:QCDfnfundamental} and \eqref{eq:QCDfnantifundamental} with $\psi' + \bar{\psi}'$ replaced by $\phi' + \bar{\phi}'$. Next, we need to solve the ensuing recursion relations and insert the result into \eqref{eq:sQCDperturbiner-quasi}. This calculation proceeds exactly as before, using mathematical induction. Thus we obtain the color-dressed perturbiners,
\begin{align}\label{eq:5.11}
\begin{split}
A^{\mu a} &= \sum_{n \geq 1} \sum_{P \in \mathcal{OW}_{n}} \mathcal{A}^{a \mu}_{P} \, \ue^{-\ui k_{P} \cdot x} = \sum_{p} \mathcal{A}^{a\mu}_{p} \, \ue^{-\ui k_{p} \cdot x} + \sum_{p < q} \mathcal{A}^{a \mu}_{pq} \, \ue^{-\ui k_{pq} \cdot x} +  \cdots, \\
\phi^{i} &= \sum_{n \geq 1} \sum_{P \in \mathcal{OW}_{n}} \Phi^{i}_{P} \ue^{-\ui k_{P} \cdot x} = \sum_{p} \Phi^{i}_{p} \, \ue^{-\ui k_{p} \cdot x} + \sum_{p < q} \Phi^{i}_{pq} \, \ue^{-\ui k_{pq} \cdot x}  + \cdots, \\
\bar{\phi}_{i} &= \sum_{n \geq 1} \sum_{P \in \mathcal{OW}_{n}} \bar{\Phi}_{P i} \ue^{-\ui k_{P} \cdot x} = \sum_{p} \bar{\Phi}_{p i} \, \ue^{-\ui k_{p} \cdot x} + \sum_{p < q} \bar{\Phi}_{pq i} \, \ue^{-\ui k_{pq} \cdot x}  + \cdots,
\end{split}
\end{align}
where the Berends-Giele currents $\mathcal{A}^{a \mu}_{P}$, $\Phi^{i}_{P}$ and $\bar{\Phi}_{Pi}$ are determined from the recursion relations and finally expressed as
\begin{align}\label{eq:5.12}
\begin{split}
\mathcal{A}^{a \mu}_{P} &=  \frac{1}{s_{P}}  \mathcal{J}^{a \mu}_{P} +   \frac{\ui}{s_{P}} \sum_{P = Q \cup R} \big\{ -\ui\tilde{f}^{\phantom{bc}a}_{bc} (k_{Q} \cdot \mathcal{A}^{b}_{R}) \mathcal{A}^{c \mu}_{Q} + \mathcal{A}^{b}_{Q \nu} \mathcal{F}^{c \nu \mu}_{R}\big\} , \\
\mathcal{J}^{a \mu}_{P} &=  \sum_{P = Q \cup R} \{ (k^{\mu}_{R}-k^{\mu}_{Q})\bar{\Phi}_{Qi} (T^{a})^{i}_{\phantom{i}j} \Phi^{j}_{R} +\mathcal{A}^{\mu}_{Q b} \sum_{R = S \cup T} \bar{\Phi}_{S i} (T^{a}T^{b}+T^{b}T^{a})^{i}_{\phantom{i}j} \Phi^{j}_{T}\}, \\
\mathcal{F}^{a \mu \nu}_{P} &=\ui k^{\mu}_{P} \mathcal{A}^{a \nu}_{P} -\ui k^{\nu}_{P} \mathcal{A}^{a \mu}_{P} +  \ui \tilde{f}^{\phantom{bc}a}_{bc} \sum_{P = Q \cup R}  \mathcal{A}^{b \mu}_{Q} \mathcal{A}^{c \nu}_{R},\\
\Phi^{i}_P &= - \frac{1}{s_{P} - m^2} \sum_{P = Q \cup R} \{ 2(k_{R} \cdot \mathcal{A}^{a}_{Q}) (T_{a})^{i}_{\phantom{i}j} \Phi^{j}_{R} + (T_{a}T_{b})^{i}_{\phantom{i}j} \Phi^{j}_{Q} \sum_{R = S \cup T} \mathcal{A}^{a}_{S} \cdot \mathcal{A}^{b}_{T}\}, \\
\bar{\Phi}_{Pi} &= \frac{1}{s_{P} - m^2} \sum_{P = Q \cup R} \{ 2(k_{R} \cdot \mathcal{A}^{a}_{Q}) \bar{\Phi}_{R j} (T_{a})^{j}_{\phantom{j}i}  - \bar{\Phi}_{Q j} (T_{b}T_{a})^{j}_{\phantom{j}i} \sum_{R = S \cup T} \mathcal{A}^{a}_{S} \cdot \mathcal{A}^{b}_{T} \}.
\end{split} 
\end{align}
The quantities $\mathcal{J}^{a \mu}_{P}$ and $\mathcal{F}^{a \mu \nu}_{P}$ correspond to the expansion coefficients of the matter current $J^{a \mu}$ and the field strength $F^{a\mu\nu}$. So, once again, we find that the recursion relations for the perturbiner coefficients are predetermined by the recursion relations for the $L_{\infty}$-quasi-isomorphism from $H^{\sbullet}(L)$ to $L$.

\subsection{Tree-level scattering amplitudes for sQCD}
We can now discuss the tree-level scattering amplitudes for scalar QCD. The approach is entirely analogous to that of QCD in the previous section. 

The first step is to insert the plane wave superpositions in \eqref{eqn:sQCDplanewave} into the homotopy Maurer-Cartan action in the cohomology, $S'_{\mathrm{MC}}$. Since the construction is analogous to QCD, we will just present the result. It is given by
\begin{multline}
\label{eq:MC-sQCD}
S'_{\mathrm{MC}} = (2 \pi)^4 \sum_{n \geq 1}\frac{1}{(n+1)}\sum_{p \geq 1} \sum_{\substack{P \in \mathcal{OW}_{n} \\ \hspace{-1pt}P = Q \cup R}} \delta(k_{pP}) \Big\{ \tilde{f}_{abc} \big( (k_{Q} \cdot \mathcal{A}^{b}_{R})(\mathcal{A}^{c}_{Q}\cdot\mathcal{A}_{p}^{a}) + \mathcal{A}^{b}_{Q \nu} \mathcal{A}_{p\mu}^{a}\mathcal{F}^{\mu \nu c}_{R}\big)\\
+ ((k_{R}-k_{Q})\cdot\mathcal{A}_{ap})\bar{\Phi}_{Qi} (T^{a})^{i}_{\phantom{i}j} \Phi^j_{R}
+ (\mathcal{A}_{bQ}\cdot\mathcal{A}_{ap})\sum_{R=S\cup T}\bar{\Phi}_{Si}(T^{b}T^{a}+T^{a}T^{b})^{i}_{\phantom{i}j}\Phi^j_{T} \\
+ 2(k_{R} \cdot \mathcal{A}^{a}_{Q})\bar{\Phi}_{pi} (T_{a})^{i}_{\phantom{i}j} \Phi^j_{R} 
- 2(k_{R} \cdot \mathcal{A}^{a}_{Q}) \bar{\Phi}_{Ri}(T_{a})^{i}_{\phantom{i}j}\Phi^j_p   \\
+[\bar{\Phi}_{pi}(T_{a}T_{b})^{i}_{\phantom{i}j} \Phi^j_{Q}+ \bar{\Phi}_{Qi}(T_{b}T_{a})^{i}_{\phantom{i}j}\Phi^j_p] \sum_{R = S \cup T} \mathcal{A}^{a}_{S} \cdot \mathcal{A}^{b}_{T}\Big\}.
\end{multline}

As before, we can remove the extra coefficients by taking the diagonal sum and imposing momentum conservation. This leave us with 
\begin{multline}
\label{eq:MC-sQCD-diag}
S'^{\mathrm{sQCD}}_{\mathrm{MC}} = \sum_{n \geq 1}\sum_{\substack{P \in \mathcal{OW}_{n} \\ \hspace{-1pt}P = Q \cup R}} \Big\{ \tilde{f}_{abc} \big( (k_{Q} \cdot \mathcal{A}^{b}_{R})(\mathcal{A}^{c}_{Q}\cdot\mathcal{A}_{p}^{a}) + \mathcal{A}^{b}_{Q \nu} \mathcal{A}_{p\mu}^{a}\mathcal{F}^{\mu \nu c}_{R}\big)\\
+ ((k_{R}-k_{Q})\cdot\mathcal{A}_{ap})\bar{\Phi}_{Qi} (T^{a})^{i}_{\phantom{i}j} \Phi^j_{R}
+ (\mathcal{A}_{bQ}\cdot\mathcal{A}_{ap})\sum_{R=S\cup T}\bar{\Phi}_{Si}(T^{b}T^{a}+T^{a}T^{b})^{i}_{\phantom{i}j}\Phi^j_{T} \\
+ 2(k_{R} \cdot \mathcal{A}^{a}_{Q})\bar{\Phi}_{pi} (T_{a})^{i}_{\phantom{i}j} \Phi^j_{R} 
- 2(k_{R} \cdot \mathcal{A}^{a}_{Q}) \bar{\Phi}_{Ri}(T_{a})^{i}_{\phantom{i}j}\Phi^j_p   \\
+[\bar{\Phi}_{pi}(T_{a}T_{b})^{i}_{\phantom{i}j} \Phi^j_{Q}+ \bar{\Phi}_{Qi}(T_{b}T_{a})^{i}_{\phantom{i}j}\Phi^j_p] \sum_{R = S \cup T} \mathcal{A}^{a}_{S} \cdot \mathcal{A}^{b}_{T}\Big\}.
\end{multline}

Now we introduce the boundary conditions operators for sQCD, with the same operator for gluons that we have in \eqref{eq:glu-ferm-oper} plus the ones for the scalars, given by
\ba\label{eq:glu-scal-oper} 
\tilde{\bar{\Phi}}_{qi}\frac{\partial}{\partial\bar{\Phi}_{qi}},\quad\quad\tilde{\Phi}_{r}^i\frac{\partial}{\partial\Phi_r^i}.
\ea
The boundary conditions for the scalars will be $\tilde{\bar\Phi}_{p\,i}=\delta^{i_p}_i$, $\tilde{\Phi}_{q}^{i}=\delta^{i}_{i_q}$. For a general $m$-point amplitude with $m-2k$ gluons and $k$ scalar pairs we have the following expression 
\ba\label{eq:sQCD-amps}
\mathscr{A}_{m,k} = \prod_{p\in \mathcal{S}^1_{m-2k}}\Big(\tilde{\mathcal{A}}_{p}^{\mu a}\frac{\partial}{\partial\mathcal{A}_{p}^{\mu a}}\Big)\prod_{q\in \mathcal{S}^2_{k}}\Big(\tilde{\bar{\Phi}}_{qi}\frac{\partial}{\partial\bar{\Phi}_{qi}}\Big)\prod_{r\in \mathcal{S}^3_{k}}\Big(\tilde{\Phi}_{r}^j\frac{\partial}{\partial\Phi_r^j}\Big)S'^{\mathrm{diag}}_{\mathrm{MC}}\Bigg|_{\mathcal{A},\Phi,\bar{\Phi}=0}
\ea 
where the $\mathcal{S}^i_r$ are again non-intersecting sets of size $r$ where the particles take the labels. The case with no scalars falls again into Yang-Mills amplitudes the same way it happened in QCD. We can jump directly to the fully-flavoured case, where only the last term in \eqref{eq:MC-sQCD-diag} contributes in the Dyck localisation for the amplitudes we deal with here.  

The fully-flavoured scattering amplitudes generator for sQCD after Dyck localisation reads
\begin{multline}\label{eq:flav-sQCD-gener}
\mathbf{G}_{sQCD}=\sum_{n \geq 1} \sum_{\substack{P \in \mathcal{OW}_{n-1} \\ \hspace{-8pt}P\in \mathrm{Dyck}}} \Big\{ 2(k_{1} \cdot \mathcal{A}^{a}_{P}) \bar\Phi_{1 j}(T_{a})^{j}_{\phantom{j}k}\Phi^k_{n+1}- \bar\Phi_{1 j} (T_{b}T_{a})^{j}_{\phantom{j}k}\Phi_{n+1}^k \sum_{P = Q \cup R} (\mathcal{A}^{a}_{Q} \cdot \mathcal{A}^{b}_{R})+\\
+\sum_{\substack{P = Q \cup R \\ \hspace{-2pt}Q,R\in \mathrm{Dyck}}} \frac{1}{(s_{1Q} - m^2)}\big[2(k_{1} \cdot \mathcal{A}^{a}_{Q}) \bar{\Phi}_{1i} (T_{a})^{i}_{\phantom{i}j}  - \bar{\Phi}_{1i} (T_{b}T_{a})^{i}_{\phantom{i}j} \sum_{Q = S \cup T} (\mathcal{A}^{a}_{S} \cdot \mathcal{A}^{b}_{T})\big] \times \\
\times \big[ 2(k_{1Q} \cdot \mathcal{A}^{c}_{R})(T_{c})^{j}_{\phantom{j}k}\Phi^k_{n+1}
- (T_{d}T_{c})^{j}_{\phantom{j}k}\Phi_{n+1}^k \sum_{R = S \cup T} (\mathcal{A}^{c}_{S} \cdot \mathcal{A}^{d}_{T})\big]\Big\},
\end{multline}
where the Dyck localisation is analogous to the QCD case. Here, for example, the multiparticle fields ${\cal A}^\mu_{M a}$ are Dyck localised. Additionally, it is useful to point out that in the internal deconcatenations not all the words have Dyck character (just their union), unless explicitly stated.

Mimicking the expression \eqref{eq:QCD-flav-amps} in QCD, we are now ready for the calculation of fully-flavoured scattering amplitudes in scalar QCD, given by
\begin{multline}\label{eq:sQCD-flav-amps}
\mathscr{A}^{\rm Flav.}_{m,k} = \Big(\tilde{\bar{\Phi}}_{1i}\frac{\partial}{\partial\bar{\Phi}_{1i}}\Big) \Big(\tilde{\Phi}_{m}^j\frac{\partial}{\partial\Phi_m^j}\Big) \times \\
\times \prod_{\substack{q=2 }}^{k}\Big(\tilde{\bar{\Phi}}_{2q-2,i}\frac{\partial}{\partial\bar{\Phi}_{2q-2,i}}\Big)\Big(\tilde{\Phi}_{2q-1}^j\frac{\partial}{\partial\Phi_{2q-1}^j}\Big)
\prod_{p = 2k}^{m-1}\Big(\tilde{\mathcal{A}}_{p}^{\mu a}\frac{\partial}{\partial\mathcal{A}_{p}^{\mu a}}\Big)\mathbf{G}_{sQCD} \Bigg|_{\mathcal{A},\Phi,\bar{\Phi}=0}.
\end{multline}

Now we will present some examples in order to illustrate the use of the amplitudes generator. 

\subsection{Examples}\label{ExampleScalar}

Such as it was made in the QCD section \ref{ExampleFermion},  first we consider the scattering between two scalars and two and three gluons. These two amplitudes will give us the tools to obtain a general and compact formula to go beyond than five-point.

Later, we are going to study amplitudes that just include scalars particles (with and without flavors).

\subsubsection{Four-point (two scalars and two gluons)}\label{2saclar2gluon}

In order to compute the scattering of two scalar fields and two gluons, we use the following set up,
\begin{eqnarray}
&& \tilde{\bar\Phi}_{1\,i}=\delta^{i_1}_i, \qquad \tilde{\Phi}_{4}^{i}=\delta^{i}_{i_4} , \nonumber \\ 
&& \tilde{\cal A}^\mu_{2\,a}= \epsilon_2^\mu \, \delta_{a}^{a_2}, \qquad \tilde{\cal A}^\mu_{3\,a}= \epsilon_3^\mu \, \delta_{a}^{a_3}  ,
\end{eqnarray}
where the total amplitude, $\mathscr{A}_{4,1}$, is given by \eqref{eq:sQCD-amps}. Thus, it is needed to calculate the multiparticle field $ \bar{{\Phi}}_{123\, i}$, then, 
\begin{align}
(s_{123}-m^2) \, {\bar \Phi}_{123\, i} &= -  {\bar \Phi}_{1\, j}  \, (T^a T^b)^{j}_{\phantom{j}i} 
\left[ 
( {\cal A}_{2\, a} \cdot  {\cal A}_{3\, b} ) + ( {\cal A}_{2\, b} \cdot  {\cal A}_{3\, a} )
\right] 
- 2  ( {\cal A}_{2\, a} \cdot k_{13}) {\bar \Phi}_{13\, j}  (T^a)^{j}_{\phantom{j}i} 
\qquad\qquad
\nonumber
\\
&
\,
- 2  ( {\cal A}_{23\, a} \cdot k_{1}) {\bar \Phi}_{1\, j}  (T^a)^{j}_{\phantom{j}i} 
- 2  ( {\cal A}_{3\, a} \cdot k_{12}) {\bar \Phi}_{12\,j} (T^a)^{j}_{\phantom{j}i} \, ,
\qquad 
\end{align}
with
\begin{align}
{\cal A}_{23\, a}^\mu & =   \tilde f_{aa_2a_3}\,   a_{23}^\mu ,  \quad
 {\bar \Phi}_{12\,j}   =
- \frac{   2 ( a_{2} \cdot k_{1}) (T^{a_2})^{i_1}_{\phantom{i_1}j}  }{ s_{12} - m^2} , \quad
 {\bar \Phi}_{13\, j}   =
-  \frac{   2 ( a_{3} \cdot k_{1}) (T^{a_3})^{i_1}_{\phantom{i_1}j}  }{ s_{13} - m^2} , 
\end{align}
and where the current $a^\mu_M$ was defined in \eqref{amudef2}. Therefore, $\mathscr{A}_{4,1} $ becomes
\begin{align}
\mathscr{A}_{4,1}   =&  (T^{a_2}T^{a_3})^{i_1}_{\phantom{i_1}i_4} \left[   \frac{ (a_2\cdot ( k_1-k_{34}) ) (a_3\cdot ( k_{12}-k_{4}) ) }{ s_{12} - m^2 }  -(a_2 \cdot a_3 )        \right]  \nonumber \\
&
+(T^{a_3}T^{a_2})^{i_1}_{\phantom{i_1}i_4} \left[   \frac{ (a_3\cdot ( k_1-k_{24}) ) (a_2\cdot ( k_{13}-k_{4}) ) }{ s_{13} - m^2 }  -(a_3 \cdot a_2 )        \right]  \nonumber\\
&
+\tilde f_{a_2a_3 c}(T^{c})^{i_1}_{\phantom{i_1}i_4} ( a_{23} \cdot (k_1-k_4) ) .
\end{align}
Now,  using the algebra,  $[T^{a_2},T^{a_3}]=\tilde f_{a_2a_3c} T^c$, it is straightforward to arrive at,
\begin{align}
\mathscr{A}_{4,1} & =  (T^{a_2}T^{a_3})^{i_1}_{\phantom{i_1}i_4} \,  A (\overline{1} ,2, 3, \underline{4} )  + (2 \leftrightarrow 3) ,
\end{align}
being $A (\overline{1} ,2, 3, \underline{4} )$ the color-ordered amplitude,
\begin{align}\label{4p-ordered}
A (\overline{1} ,2, 3, \underline{4} ) & =   \frac{ (a_2\cdot ( k_1-k_{34}) ) (a_3\cdot ( k_{12}-k_{4}) ) }{ s_{12} - m^2 }  -(a_2 \cdot a_3 )      +   ( a_{23} \cdot (k_1-k_4)) .
\end{align}

\subsubsection{Five-point (two scalars and three gluons)}

Here, we focus on the interaction between two scalar fields and  three gluons. 
The boundary condition we choose are the following
\begin{eqnarray}
&& \tilde{\bar\Phi}_{1\,i}=\delta^{i_1}_i, \qquad \tilde{\Phi}_{5}^{i}=\delta^{i}_{i_5} , \nonumber \\ 
&& \tilde{\cal A}^\mu_{2\,a}= \epsilon_2^\mu \, \delta_{a}^{a_2}, \qquad \tilde{\cal A}^\mu_{3\,a}= \epsilon_3^\mu \, \delta_{a}^{a_3} ,  \qquad \tilde{\cal A}^\mu_{4\,a}= \epsilon_4^\mu \, \delta_{a}^{a_4} .
\end{eqnarray}
The total amplitude, $\mathscr{A}_{5,1}$, may be computed  from \eqref{eq:sQCD-amps}. So, 
such as it was done above in section \ref{2saclar2gluon}, after some simple manipulations,
it is not hard to show that $\mathscr{A}_{5,1} $ arrive at
\begin{align}
\mathscr{A}_{5,1} & = \sum_{\rho\in S_3} (T^{a_{\rho(2)}}T^{a_{\rho(3)}} T^{a_{\rho(4)}}    )^{i_1}_{\phantom{i_1}i_5} \,  A (\overline{1} ,\rho(2), \rho(3), \rho(4), \underline{5} )   ,
\end{align}
where the ordered amplitude, $A (\overline{1} ,2, 3,4, \underline{5} ) $, is read as
\begin{align}\label{5p-ordered}
&
A (\overline{1} ,2, 3,4, \underline{5} )  =   \nonumber\\
&
\frac{ [a_2\cdot ( k_1-k_{35}) ] [a_3\cdot ( k_{12}-k_{45}) ] [a_4\cdot ( k_{13}-k_{5}) ] }{ (s_{12} - m^2) (s_{123} - m^2)   } -
 \frac{(a_2\cdot a_3) [a_4\cdot ( k_{13}-k_{5}) ] }{ (s_{123} - m^2)   } +
\nonumber \\
& 
 \frac{ [a_2\cdot ( k_1-k_{35}) ] [(-1)(a_3\cdot a_4) ]  }{ (s_{12} - m^2)   } +
 \frac{ [a_{23}\cdot ( k_1-k_{45}) ]  [a_4\cdot ( k_{13}-k_{5}) ]  }{  (s_{123} - m^2)  } 
 - 
  (a_{23}\cdot a_4) + \nonumber \\
  &
 \frac{ [a_2\cdot ( k_1-k_{35}) ] [a_{34}\cdot ( k_{12}-k_{5}) ]  }{ (s_{12} - m^2)   } -(a_2\cdot a_{34})  
 + [a_{234}\cdot ( k_{1}-k_{5}) ] ,
\end{align} 
with $a^\mu_M$ defined in \eqref{amudef2}.

\subsubsection{Generalisation (two scalars and $n-2$ gluons)}

Notice that from the color-stripped Feynman rules for scalar QCD
\begin{align}\label{FeynmanScalars}
\!\!\!\!\!\!\!\!\!\!\!\!\!\!\!\!\!\!\!
\begin{gathered}
\feynmandiagram [layered layout, large, horizontal=a to b] {
i2 -- [opacity=0] b1 [particle=\(  \mu \)]
-- [opacity=0] b2 
-- [opacity=0] b3 
[particle=\( \nu \)],
i1 
-- [opacity=0]  a1 [particle=\(k_I\)]
-- [scalar] a2
-- [scalar] a3 
[particle=\(k_J\)]
--  [opacity=0]  a4 [particle=\(\)],
a2-- [gluon] b1,
a2-- [gluon] b3,
};
\end{gathered}
\!\!\!\!\!\!\!\!\!\!\!\!\!
=- \, \eta^{\mu\nu} ,\quad
\begin{gathered}
\feynmandiagram [layered layout, large, horizontal=a to b] {
i2-- [opacity=0] b1 [particle=\( \mu \)]
-- [opacity=0] b2 [particle=\( \)],
i1 [particle=\(k_I\)]
-- [scalar] a1
-- [scalar] a2 [particle=\( k_J\)],
a1 -- [gluon] b1,
};
\end{gathered}
\!\!= (k_I-k_J)^\mu,  
\quad
\begin{gathered}
\feynmandiagram [layered layout, large, horizontal=a to b] {
i1 [particle=\(k_I\)]
-- [scalar] a1
-- [scalar] a2 [particle=\((-k_I)\)],
};
\end{gathered}
= \frac{1}{k_I^2-m^2},
\end{align}
where $I,J$ are multiparticle labels,
the ordered amplitude, $A (\overline{1} ,2, 3, \underline{4} )$, can be rewritten as
\begin{equation}\label{A1234}
A (\overline{1} ,2, 3, \underline{4} ) =
\begin{gathered}
\feynmandiagram [layered layout, large, horizontal=a to b] {
i2-- [opacity=0] b1 [particle=\( a_2\)]
-- [opacity=0] b2 [particle=\( a_3\)],
i1 [particle=\(\overline 1\)]
-- [scalar] a1
-- [scalar] a2
-- [scalar] a3 [particle=\(\underline 4\)],
a1 -- [gluon] b1,
a2 -- [gluon] b2,
};
\end{gathered}
+
\begin{gathered}
\feynmandiagram [layered layout, large, horizontal=a to b] {
i2-- [opacity=0] b1 [particle=\( a_{23}\)]
-- [opacity=0] b2 [particle=\( \)],
i1 [particle=\(\overline 1\)]
-- [scalar] a1
-- [scalar] a2 [particle=\(\underline 4\)],
a1 -- [gluon] b1,
};
\end{gathered}
+
\!\!\!\!\!\!\!\!\!\!
\begin{gathered}
\feynmandiagram [layered layout, large, horizontal=a to b] {
i2 -- [opacity=0] b1 [particle=\( a_2\)]
-- [opacity=0] b2 
-- [opacity=0] b3 
[particle=\( a_3\)],
i1 
-- [opacity=0]  a1 [particle=\(\overline 1\)]
-- [scalar] a2
-- [scalar] a3 
[particle=\(\underline 4\)]
--  [opacity=0]  a4 [particle=\(\)],
a2-- [gluon] b1,
a2-- [gluon] b3,
};
\end{gathered} .
\end{equation}

Analogously, the result obtained in \eqref{5p-ordered}  for $A (\overline{1} ,2, 3,4, \underline{5} )$ gives a similar  diagrammatic structure
\begin{align}\label{A12345}
&
A (\overline{1} ,2, 3, 4, \underline{5} ) = \nonumber\\
&
\begin{gathered}
\feynmandiagram [layered layout, large, horizontal=a to b] {
i2-- [opacity=0] b1 [particle=\( a_2\)]
-- [opacity=0] b2 [particle=\( a_3\)]
-- [opacity=0] b3 [particle=\( a_4\)],
i1 [particle=\(\overline 1\)]
-- [scalar] a1
-- [scalar] a2
-- [scalar] a3 
-- [scalar] a4
[particle=\(\underline 5\)],
a1 -- [gluon] b1,
a2 -- [gluon] b2,
a3 -- [gluon] b3,
};
\end{gathered}
+
\begin{gathered}
\feynmandiagram [layered layout, large, horizontal=a to b] {
i2-- [opacity=0] b1 [particle=\( a_{23}\)]
-- [opacity=0] b2 [particle=\( a_4\)],
i1 [particle=\(\overline 1\)]
-- [scalar] a1
-- [scalar] a2
-- [scalar] a3 [particle=\(\underline 5\)],
a1 -- [gluon] b1,
a2 -- [gluon] b2,
};
\end{gathered}
+
\begin{gathered}
\feynmandiagram [layered layout, large, horizontal=a to b] {
i2-- [opacity=0] b1 [particle=\( a_{2}\)]
-- [opacity=0] b2 [particle=\( a_{34}\)],
i1 [particle=\(\overline 1\)]
-- [scalar] a1
-- [scalar] a2
-- [scalar] a3 [particle=\(\underline 5\)],
a1 -- [gluon] b1,
a2 -- [gluon] b2,
};
\end{gathered}
+
\begin{gathered}
\feynmandiagram [layered layout, large, horizontal=a to b] {
i2-- [opacity=0] b1 [particle=\( a_{234}\)]
-- [opacity=0] b2 [particle=\( \)],
i1 [particle=\(\overline 1\)]
-- [scalar] a1
-- [scalar] a2 [particle=\(\underline 5\)],
a1 -- [gluon] b1,
};
\end{gathered}
\nonumber
\\
&
+
\!\!\!\!\!\!\!\!
\begin{gathered}
\feynmandiagram [layered layout, large, horizontal=a to b] {
i2 -- [opacity=0] b1 [particle=\( a_2\)]
-- [opacity=0] b2 
-- [opacity=0] b3 
[particle=\( a_3\)]
-- [opacity=0] b4 
[particle=\( a_4\)],
i1 
-- [opacity=0]  a1 [particle=\(\overline 1\)]
-- [scalar] a2
-- [scalar] a3 
--  [scalar]  a4 
--  [scalar]  a5
[particle=\( \underline 5 \)],
a2-- [gluon] b1,
a2-- [gluon] b3,
a4-- [gluon] b4,
};
\end{gathered}
+
\begin{gathered}
\feynmandiagram [layered layout, large, horizontal=a to b] {
i2 -- [opacity=0] b1 [particle=\( a_2\)]
-- [opacity=0] b2 
[particle=\( a_3\)]
-- [opacity=0] b3 
-- [opacity=0] b4 
[particle=\( a_4\)],
i1 [particle=\(\overline 1\)]
-- [scalar]  a1 
-- [scalar] a2
-- [scalar] a3 
--  [scalar]  a4 [particle=\( \underline 5 \)]
--  [opacity=0]  a5
[particle=\( \)],
a1-- [gluon] b1,
a3-- [gluon] b2,
a3-- [gluon] b4,
};
\end{gathered}
\!\!\!\!\!\!\!\!
+
\!\!\!\!\!
\begin{gathered}
\feynmandiagram [layered layout, large, horizontal=a to b] {
i2 -- [opacity=0] b1 [particle=\( a_{23}\)]
-- [opacity=0] b2 
-- [opacity=0] b3 
[particle=\( a_{4}\)],
i1 
-- [opacity=0]  a1 [particle=\(\overline 1\)]
-- [scalar] a2
-- [scalar] a3 
[particle=\(\underline 5\)]
--  [opacity=0]  a4 [particle=\(\)],
a2-- [gluon] b1,
a2-- [gluon] b3,
};
\end{gathered}
\nonumber
\\
&
+
\!\!\!\!\!\!\!\!\!
\begin{gathered}
\feynmandiagram [layered layout, large, horizontal=a to b] {
i2 -- [opacity=0] b1 [particle=\( a_{2}\)]
-- [opacity=0] b2 
-- [opacity=0] b3 
[particle=\( a_{34}\)],
i1 
-- [opacity=0]  a1 [particle=\(\overline 1\)]
-- [scalar] a2
-- [scalar] a3 
[particle=\(\underline 5\)]
--  [opacity=0]  a4 [particle=\(\)],
a2-- [gluon] b1,
a2-- [gluon] b3,
};
\end{gathered}
.
\end{align}

Thus, by the result obtained in \eqref{A1234} and \eqref{A12345} (for $A(1,2,3,4)$ and $A(1,2,3,4,5)$, respectively), it is straightforward to spot a pattern, that being
\begin{align}\label{eq:AnsQCD}
&
A (\overline{1} ,2, \ldots,n-1, \underline{n} ) = \nonumber\\
&
\begin{gathered}
\feynmandiagram [layered layout, large, horizontal=a to b] {
i2-- [opacity=0] b1 [particle=\( a_2\)]
-- [opacity=0] b2 [particle=\( a_3\)]
-- [opacity=0] b3 [particle=\( a_{n-1}\)],
i1 [particle=\(\overline 1\)]
-- [scalar] a1
-- [scalar] a2
-- [ghost] a3 
-- [scalar] a4
[particle=\(\underline n\)],
a1 -- [gluon] b1,
a2 -- [gluon] b2,
a3 -- [gluon] b3,
};
\end{gathered}
+
\begin{gathered}
\feynmandiagram [layered layout, large, horizontal=a to b] {
i2-- [opacity=0] b1 [particle=\( a_{23}\)]
-- [opacity=0] b2 [particle=\( a_{n-1}\)],
i1 [particle=\(\overline 1\)]
-- [scalar] a1
-- [ghost] a2
-- [scalar] a3 [particle=\(\underline n\)],
a1 -- [gluon] b1,
a2 -- [gluon] b2,
};
\end{gathered}
+\cdots +
\begin{gathered}
\feynmandiagram [layered layout, large, horizontal=a to b] {
i2-- [opacity=0] b1 [particle=\( a_{2...p}\)]
-- [opacity=0] b2 
-- [opacity=0] b3 [particle=\( a_{r...n-1}\)],
i1 [particle=\(\overline 1\)]
-- [scalar] a1
-- [ghost] a2
-- [ghost] a3 
-- [scalar] a4
[particle=\(\underline n\)],
a1 -- [gluon] b1,
a3 -- [gluon] b3,
};
\end{gathered}
 \nonumber\\
&
+
\cdots
+
\begin{gathered}
\feynmandiagram [layered layout, large, horizontal=a to b] {
i2-- [opacity=0] b1 [particle=\( a_{23...n-1}\)]
-- [opacity=0] b2 [particle=\( \)],
i1 [particle=\(\overline 1\)]
-- [scalar] a1
-- [scalar] a2 [particle=\(\underline n\)],
a1 -- [gluon] b1,
};
\end{gathered}
\,\,\,
+
\!\!\!\!\!\!
\begin{gathered}
\feynmandiagram [layered layout, large, horizontal=a to b] {
i2 -- [opacity=0] b1 [particle=\( a_2\)]
-- [opacity=0] b2 
-- [opacity=0] b3 
[particle=\( a_3\)]
-- [opacity=0] b4 
[particle=\( a_{n-1}\)],
i1 
-- [opacity=0]  a1 [particle=\(\overline 1\)]
-- [scalar] a2
-- [ghost] a3 
--  [ghost]  a4 
--  [scalar]  a5
[particle=\( \underline n \)],
a2-- [gluon] b1,
a2-- [gluon] b3,
a4-- [gluon] b4,
};
\end{gathered}
\!
+
\cdots
+
\nonumber
\\
&
\begin{gathered}
\feynmandiagram [layered layout, large, horizontal=a to b] {
i2 -- [opacity=0] b1 
[particle=\( a_{2...p}\)]
-- [opacity=0] b2 
-- [opacity=0] b3 
[particle=\( a_{i...i+r}\)]
-- [opacity=0] b4 
-- [opacity=0] b5 
[particle=\( a_{i+r+1...l }\)]
-- [opacity=0] b6 
-- [opacity=0] b7 
[particle=\( a_{q...q+t}\)]
-- [opacity=0] b8
-- [opacity=0] b9 
[particle=\( a_{j...j+h}\)]
-- [opacity=0] b10 
-- [opacity=0] b11 
[particle=\( a_{j+h+1...w}\)]
-- [opacity=0] b12
-- [opacity=0] b13
[particle=\( a_{s...n-1}\)] 
,
i1 [particle=\(\overline 1\)]
-- [scalar] a1 
-- [ghost] a2
-- [ghost] a3 
-- [ghost] a4
-- [ghost] a5
-- [ghost] a6
-- [ghost] a7 
-- [ghost] a8
-- [ghost] a9
-- [ghost] a10  
--  [ghost]  a11 
--  [ghost]  a12
--  [ghost]  a13  
--  [scalar]  a14
[particle=\( \underline n \)]
[particle=\( \)],
a1-- [gluon] b1,
a4-- [gluon] b3,
a4-- [gluon] b5,
a7-- [gluon] b7,
a10-- [gluon] b9,
a10-- [gluon] b11,
a13-- [gluon] b13,
};
\end{gathered}+\cdots  \nonumber
\\
&
+
\!\!\!\!\!\!\!\!\!
\begin{gathered}
\feynmandiagram [layered layout, large, horizontal=a to b] {
i2 -- [opacity=0] b1 [particle=\( a_{2}\)]
-- [opacity=0] b2 
-- [opacity=0] b3 
[particle=\( a_{34...n-1}\)],
i1 
-- [opacity=0]  a1 [particle=\(\overline 1\)]
-- [scalar] a2
-- [scalar] a3 
[particle=\(\underline n \)]
--  [opacity=0]  a4 [particle=\(\)],
a2-- [gluon] b1,
a2-- [gluon] b3,
};
\end{gathered} .
\end{align}

Such as it was done in \eqref{AS1n} for QCD, we are able to propose an analytic expression for $A (\overline{1} ,2, \ldots, n-1, \underline{n} )$, which has following form
\begin{align}\label{eq:Gen-scalar} 
&
A (\overline{1} ,2, \ldots, n-1, \underline{n} )   = 
\nonumber\\
&
\sum_{\substack{ i=2 \\ i \in 2\mathbb{N}  }}^{n-2}   \sum_{\substack{(p_1,\ldots, p_i) \\  \in \, C_i(n-2)}  }  
\left\{
(a_{2, p_1+1} \cdot \Delta\cdot a_{p_1+2, p_{12}+1} )\,
\frac{  1}{(s_{1,p_{12}+1}-m^2)  } 
( a_{p_{12}+2,p_{1,3}+1} \cdot \Delta \cdot a_{p_{1,3}+2,p_{1,4}+1} )
\right.
\nonumber\\
&
\frac{  1       }{  (s_{1,p_{1,4}+1}-m^2)   }  
\,
\times
\cdots
\times
\,
\frac{  1       }{  (s_{1,p_{1,i-2}+1}-m^2)   }  
( a_{p_{1,i-2}+2,p_{1,i-1}+1} \cdot \Delta \cdot a_{n-p_{i},n-1} ) 
\,\,
+
\nonumber\\
&
(a_{2, p_1+1} \cdot \bar{k}_{1|p_1+2,n}  ) 
\frac{  1}{(s_{1,p_1+1}-m^2)  } 
( a_{p_{1}+2,p_{12}+1} \cdot \Delta \cdot a_{p_{12}+2,p_{1,3}+1} )
\,
\frac{  1       }{  (s_{1,p_{1,3}+1}-m^2)   }  
\cdots
\nonumber\\
&
\left.
\frac{  1       }{  (s_{1,p_{1,i-3}+1}-m^2)   }  
( a_{p_{1,i-3}+2,p_{1,i-2}+1} \cdot \Delta \cdot a_{p_{1,i-2}+2,p_{1,i-1}+1} )
\frac{  1    }{  (s_{1,n-p_i-1}-m^2)   }  (a_{n-p_i, n-1} \cdot  \bar{k}_{1,n-p_i-1|n}  )  
\right\}
\nonumber\\
&
+\!\!
\sum_{\substack{ i=1 \\ i \in 2\mathbb{N} +1  }}^{n-2}   \sum_{\substack{(p_1,\ldots, p_i) \\  \in \, C_i(n-2)}  }  
\left\{
(a_{2, p_1+1} \cdot \bar{k}_{1|p_1+2,n}  ) 
\frac{  1 }{(s_{1,p_1+1}-m^2)  }   (a_{p_{1}+2, p_{12}+1} \cdot \Delta \cdot 
a_{p_{12}+2, p_{1,3}+1}
 ) 
 \frac{  1       }{  (s_{1,p_{1,3}+1}-m^2)   }  
\right.
\nonumber\\
&
( a_{p_{1,3}+2,p_{1,4}+1} \cdot \Delta \cdot a_{p_{1,4}+2,p_{1,5}+1} )
\cdots
\frac{  1       }{  (s_{1,p_{1,i-2}+1}-m^2)   }  
( a_{p_{1,i-2}+2,p_{1,i-1}+1} \cdot \Delta \cdot a_{n-p_{i},n-1} )
\,\,
+
\nonumber\\
&
(a_{2, p_1+1} \cdot \Delta\cdot a_{p_1+2, p_{12}+1} )\,
\frac{  1}{(s_{1,p_{12}+1}-m^2)  } 
( a_{p_{12}+2,p_{1,3}+1} \cdot \Delta \cdot a_{p_{1,3}+2,p_{1,4}+1} )
\cdots
\frac{  1       }{  (s_{1,p_{1,i-3}+1}-m^2)   }  
\nonumber\\
&
\left.
( a_{p_{1,i-3}+2,p_{1,i-2}+1} \cdot \Delta \cdot a_{p_{1,i-2}+2,p_{1,i-1}+1} )
\frac{   1    }{  (s_{1,n-p_i-1}-m^2)   } (a_{n-p_i, n-1} \cdot  \bar{k}_{1,n-p_i-1|n}  )  
\right\}
\nonumber\\
&
-
\sum_{ i=1 }^{n-2}   \sum_{\substack{(p_1,\ldots, p_i) \\  \in \, C_i(n-2)}  }  
\left\{
(a_{2, p_1+1} \cdot \bar{k}_{1|p_1+2,n}  )
\,
\frac{  1}{(s_{1,p_1+1}-m^2)  } \,  (a_{p_1+2, p_{12}+1} \cdot \bar{k}_{1,p_1+1|p_{12}+2,n} ) \times
\right.
\nonumber\\
&
\left.
\frac{  1       }{  (s_{1,p_{12}+1}-m^2)   }   
\cdots
\frac{  1       }{  (s_{1,n-p_i-1}-m^2)   }   
\,
 (a_{n-p_i, n-1} \cdot  \bar{k}_{1,n-p_i-1|n}  ) 
\right\} \, ,
 \nonumber\\
\end{align}
where we have defined
\begin{align}\label{DefSQCD}
&
p_{1, r} = p_1+p_2+\cdots + p_r \nonumber \\
&
\bar{k}_{p,q|r,s} = k_{p,q} - k_{r,s}, \nonumber\\
&
(a_{p ,p+r} \cdot \Delta \cdot a_{p+r+1,p+r+1+t}) \nonumber \\ 
&
= a_{p ,p+r}^\mu \left(
-\eta^{\mu\nu} + \frac{\bar{k}^\mu_{1,p-1 | p+r+1,n} \, \bar{k}^\nu_{1,p+r | p+r+1+t+1,n}   }{s_{1,p+r} -m^2}
\right) 
a_{p+r+1,p+r+1+t}^\nu \, ,\, 
\end{align}
and with $C_i(n-2)$ given in \eqref{Cidef}. For instance, when $n=4$, the previous formula becomes
\begin{equation}\label{A1234S2}
A (\overline{1} ,2, 3, \underline{4} )  = (a_2\cdot \Delta \cdot a_3)  + (a_{23} \cdot \bar{k}_{1|4})
\end{equation}
with
\begin{equation}
 (a_2\cdot \Delta \cdot a_3) = a_2^\mu \, \Big(  - \eta^{\mu\nu} + \frac{  \bar{k}_{1|34}^\mu \, \bar{k}_{12|4}^\nu }{s_{12}-m^2}          \Big) \, a_3^\nu,
\end{equation}
from \eqref{DefSQCD}. It is trivial to check that \eqref{4p-ordered}  and \eqref{A1234S2} are equal.

Finally, we have verified this formula up to seven-point (numerically).

\subsubsection{Four-point (only scalars)}

Our second example is four scalar fields with and without flavours. By choosing the boundary conditions
\begin{eqnarray}\label{bc-4s}
&& \tilde{\bar{\Phi}}_{1,i}=\delta^{i_1}_i, \qquad \tilde{\bar{\Phi}}_{2,i}=\delta^{i_2}_i, \nonumber\\ 
&& \tilde{\Phi}_{3}^{i}=\delta^{i}_{i_3}, \qquad \tilde{\Phi}_{4}^{i}=\delta^{i}_{i_4} ,
\end{eqnarray}
the amplitudes, $\mathscr{A}_{4,2} $ and $\mathscr{A}_{4,2}^{\rm Flav.}$, are given by the equations  \eqref{eq:sQCD-amps} and \eqref{eq:sQCD-flav-amps}, respectively.

From the Perturbiners, $\bar{\Phi}_{123, i}$ and $\bar{\Phi}^{\rm Dyck}_{123, i}$ (Dyck means localisation on Dyck words\footnote{See appendix \ref{Comp-Phi}.}, such as it was explained in section \ref{DyckLocalization}),
and the current ${\cal A}_{M a}^\mu$
\begin{eqnarray}\label{A-1323}
{\cal A}_{13 \, a}^\mu  = \frac{ {\cal J}_{13\, a}^\mu }{ s_{13} } = \frac{  (k_1-k_3)^\mu (T^a)^{i_1}_{\phantom{i_1}i_3} }{ s_{13} }, \qquad
{\cal A}_{23 \, a}^\mu  = \frac{ {\cal J}_{23\, a}^\mu }{ s_{23} } = \frac{  (k_2-k_3)^\mu (T^a)^{i_2}_{\phantom{i_2}i_3} }{ s_{23} }, 
\end{eqnarray}
one arrives at following expressions for the amplitudes
\begin{eqnarray}
\mathscr{A}_{4,2} 
&=& 
(T^a)^{i_1}_{\phantom{i_1}i_4} (T^a)^{i_2}_{\phantom{i_2}i_3} \, A(\overline{1},\underline{4} , \overline{2},\underline{3} )  +
(T^a)^{i_1}_{\phantom{i_1}i_3} (T^a)^{i_2}_{\phantom{i_2}i_4} \, A(\overline{1},\underline{3} , \overline{2},\underline{4} )  , \\
\mathscr{A}_{4,2}^{\rm Flav.}  
&=&
(T^a)^{i_1}_{\phantom{i_1}i_4} (T^a)^{i_2}_{\phantom{i_2}i_3} \, A(\overline{1},\underline{4} , \overline{2},\underline{3} ),
\end{eqnarray}
with
\begin{eqnarray}
 A(\overline{1},\underline{4} , \overline{2},\underline{3} )  
= 
\frac{  (k_1-k_4)\cdot ( k_{2} - k_{3} )  }{s_{23}}
 , 
\qquad
 A(\overline{1},\underline{3} , \overline{2},\underline{4} )  
= 
\frac{  (k_2-k_4)\cdot ( k_{1} - k_{3} )  }{s_{13}} 
  .
\end{eqnarray}

\subsubsection{Six-point (only scalars)}

In this section, we consider the six-point amplitude totally flavoured (three different flavours). 
For this computation, we choose the boundary conditions,
\begin{eqnarray}\label{bc-6s}
&& \tilde{\bar{\Phi}}_{1,i}=\delta^{i_1}_i, \qquad \tilde{\bar{\Phi}}_{2,i}=\delta^{i_2}_i, \qquad \tilde{\bar{\Phi}}_{4,i}=\delta^{i_4}_i \nonumber\\ 
&& \tilde{\Phi}_{3}^{i}=\delta^{i}_{i_3}, \qquad \tilde\Phi_{5}^{i}=\delta^{i}_{i_5} , \qquad \tilde\Phi_{6}^{i}=\delta^{i}_{i_6} \, ,
\end{eqnarray}
and the amplitude, $\mathscr{A}_{6,3}^{\rm Flav.}$, is given by \eqref{eq:sQCD-flav-amps}.  In order to perform this calculation, we must carry out the current,  $\bar{\Phi}^{\rm Dyck}_{12345, i}$, {\it i.e.} 
\begin{align}\label{}
&
-(s_{12345} - m^{2}) \, 
\bar{\Phi}^{\rm Dyck}_{12345, i} = 2 ( k_1\cdot {\cal A}^{\text{Dyck}}_{2345, a}) \bar{\Phi}_{1,j}  (T^a)^{j}_{\phantom{j}i} \nonumber \\
&
+ \bar{\Phi}_{1,j} (T^a T^b)^{j}_{\phantom{j}i} \left[ ({\cal A}_{23, a} \cdot {\cal A}_{45, b} ) + ({\cal A}_{45, a} \cdot {\cal A}_{23, b} )  \right] \qquad \nonumber\\
& + 2 (k_{123} \cdot {\cal A}_{45, a}) \bar{\Phi}^{\rm Dyck}_{123, j} (T^a)^{j}_{\phantom{j}i} + 2 (k_{145} \cdot {\cal A}_{23, a}) \bar{\Phi}^{\rm Dyck}_{145, j} (T^a)^{j}_{\phantom{j}i}  \, . \qquad
\end{align}
In appendix \ref{Comp-Phi}, we have detailed this computation, therefore, by using the result obtained in \eqref{Phi12345}, the amplitude $\mathscr{A}_{6,3}^{\rm Flav.} $ turns into
\begin{multline}
\mathscr{A}_{6,3}^{\rm Flav.} =
\frac{ \tilde f^{abc} (T_c)^{i_{1}}_{\phantom{i_{1}}i_6}  (T_a)^{i_{2}}_{\phantom{i_{2}}i_3}   (T_b)^{i_{4}}_{\phantom{i_{4}}i_5} \, n_0}{s_{61}s_{23} s_{45}} 
+ \Bigg[ \frac{(T^aT^b)^{i_{1}}_{\phantom{i_{1}}i_6} (T^a)^{i_{2}}_{\phantom{i_{2}}i_3} (T^b)^{i_{4}}_{\phantom{i_{4}}i_5} \, n_1  }{(s_{123} - m^2) s_{23} s_{45}} \\
+ \frac{(T^bT^a)^{i_{1}}_{\phantom{i_{1}}i_6}  (T^a)^{i_{2}}_{\phantom{i_{2}}i_3} (T^b)^{i_{4}}_{\phantom{i_{4}}i_5} \, n_2  }{(s_{145} - m^2) s_{23} s_{45}} +
{\rm cyclic}
\left(
\begin{matrix}
(61)\rightarrow (23) \rightarrow (45) 
\end{matrix}
\right)
\Bigg] ,
\end{multline}
with
\begin{equation}
\begin{array}{rcl}
n_0&=&(k_6-k_1)_\mu (k_2-k_3)_\nu (k_4-k_5)_\rho V^{\mu\nu\rho}_{k_6+k_1,k_2+k_3,k_4+k_5} , \\
n_1&=& - 4 (k_1\cdot (k_2-k_3)) (k_6\cdot (k_4-k_5)) -  (s_{123} - m^2) (k_2-k_3)\cdot (k_4-k_5) , \\
n_2&=& - 4 (k_1\cdot (k_4-k_5)) (k_6\cdot (k_2-k_3)) -(s_{145} - m^2) (k_2-k_3)\cdot (k_4-k_5).
\end{array}
\end{equation}
The vertex $V^{\mu\nu\rho}_{p,q,k} $ has been defined in \eqref{3pvertex}. Again, notice that each  pair $(1,6)$, $(2,3)$ and $(4,5)$ has a different flavour.

Finally, we can check that $n_0$, $n_1$ and $n_2$ satisfy the colour-kinematics relation \cite{Bern:2008qj}
\begin{equation}
(T^aT^b)^{i_{1}}_{\phantom{i_{1}}i_6}    - (T^bT^a)^{i_{1}}_{\phantom{i_{1}}i_6}  =\tilde  f^{abc} (T_c)^{i_{1}}_{\phantom{i_{1}}i_6} \,\,\,  \rightarrow  \,\,\,
n_1-n_2=n_0. \quad
\end{equation}
After some  simple manipulations,
it is straightforward to write down $\mathscr{A}_{6,3}^{\rm Flav.}$ in terms  of the Johansson-Ochirov color base   \cite{Johansson:2015oia}.

\section{Summary and discussion}\label{sec:concl}

In this section we will summarise our results and point out the applications, generalisations and future developments.

\subsection{Main results}

In this work we used algebraic tools inspired by string field theory for the underlying structure of the BV formalism and exploited its connection with the perturbiner expansion. We computed the multiparticle solutions (Berends-Giele currents) of gauge theories coupled to matter, including recursive formulas to compute their tree level amplitudes. We focused here in three special theories: Chern-Simons-Matter (CSM), Quantum Chromodynamics (QCD), and scalar Quantum Chromodynamics (sQCD), all of them single-flavoured with $SU(N)$ gauge group.

\subsection{Applications}

The first main outcome of our work is a set of covariant and compact formulae for color-stripped amplitudes for the scattering of one massive quark-antiquark (or scalar-antiscalar) pair with $(n-2)$-gluons with arbitrary polarisations. In the DDM base \cite{DelDuca:1999rs}, these amplitudes can be expressed as
\begin{align}
\mathscr{A}_{n,1} & = \sum_{\rho\in S_{n-2}} (T^{a_{\rho(2)}}T^{a_{\rho(3)}} \cdots T^{a_{\rho(n-1)}})^{i_1}_{\phantom{i_1}i_n} \,  A (\overline{1} ,\rho(2), \rho(3), \ldots , \rho(n-1), \underline{n} )   ,
\end{align}
where $A (\overline{1} ,2, 3, \ldots ,n-1, \underline{n} )$ denotes the color-stripped amplitude (primitive). In QCD, they are simply expressed as
\begin{align}
A (\overline{1} ,2, \ldots,n-1, \underline{n} ) 
=
\sum_{i=1}^{n-2} \sum_{\substack{(p_1,\ldots, p_i) \\  \in \, C_i(n-2)}  } 
\begin{gathered}
\feynmandiagram [layered layout, large, horizontal=a to b] {
i2-- [opacity=0] b1 [particle=\( a_{p_1}\)]
-- [opacity=0] b2 [particle=\( a_{p_2}\)]
-- [opacity=0] b3 [particle=\( a_{p_i}\)],
i1 [particle=\(\overline 1\)]
-- [fermion] a1
-- [fermion] a2
-- [ghost] a3 
-- [fermion] a4
[particle=\(\underline n\)],
a1 -- [gluon] b1,
a2 -- [gluon] b2,
a3 -- [gluon] b3,
};
\end{gathered}
\,\,\, ,
\end{align}
which we presented in equation \eqref{AS1n},  using the Feynman rules in \eqref{FeynmanSpinors}. For scalar QCD this  ordered amplitude has a similar form, but more diagrams must be included since there is a quartic interaction between scalars and gluons (see equations  \eqref{eq:AnsQCD} and \eqref{eq:Gen-scalar}).

From the BCFW-shifting method, Ochirov in \cite{Ochirov:2018uyq} and Forde, Kosower, Badger, Glover, Khoze, Svrcek in \cite{Forde:2005ue,Badger:2005zh}, obtained a general formula for the same type of amplitudes (massive fermions and scalars, respectively) but where all gluons have identical helicity, {\it e.g.} $A (\overline{1} ,2^+, 3^+, \ldots ,(n-1)^+, \underline{n} )$. At four and five-point, it is straightforward to check that our results are in total agreement with their results. However, we would like to understand how those formulae match at the $n$-point case. For the scalar case, Ahmadiniaz, Bastianelli, and Corradini in \cite{Ahmadiniaz:2015xoa} used the worldline formalism to compute the gluon-irreducible contribution to the tree level amplitude of $(n-2)$ gluons and two scalars, which agrees to the corresponding terms in equations \eqref{eq:AnsQCD} and \eqref{eq:Gen-scalar}. 

The second main outcome of our work is related to Dyck word localizations and their connection to flavoured amplitudes. In \cite{Johansson:2015oia},  by considering the Melia base of primitive amplitudes in QCD (see \cite{Melia:2013bta}),  Johansson and Ochirov observed that the $n$-point scattering of $k=n/2$ quark-antiquark pairs with different flavours can be written as
\begin{equation}
\mathscr{A}^{\rm Flav.}_{n,n/2} = \sum_{\sigma\in {\rm Dyck}_{k-1}} C(\underline{1},\bar{2},\sigma) \, A (\underline{1} ,\bar{2}, \sigma ) ,
\end{equation}
where $C(\underline{1},\bar{2},\sigma)$ is the Johansson-Ochirov color base. Here, we showed that these amplitudes can be obtained directly from single-flavoured theories if we use a Dyck localization on the Berends-Giele currents. For example, in QCD (and similarly for CSM and scalar QCD), with boundary conditions
\begin{equation}\label{}
\tilde{\bar{\Psi}}_{qi}=\bar{u}_q\delta_i^{i_q}, \quad q=1,2,4,...,n-2,
\qquad\tilde{\Psi}_r^i=v_q\delta^i_{i_r} , \quad r=3,5,...,n-1,n
\end{equation}
it is straightforward to check that the amplitude generator \eqref{eq:QCD-flav-amps} provides the fully-flavoured amplitude $\mathscr{A}^{\rm Flav.}_{n,n/2}$, where the quark-antiquark pairs, $(1,n)$, $(2,3)$ ... $(n-2,n-1)$, have a different flavours, but with the same mass.

\subsection{Generalisation: different masses for flavours}

In order to describe different flavours in a given theory, we usually have to add a number of $N_f$ copies of the matter kinetic term to the action, each with a different mass. The distinct flavours and masses are then introduced when drawing different Feynman diagrams for a certain amplitude. Here can shortcut this procedure by looking at the contracting homotopy for the matter fields, without directly modifying the action.

Looking at \eqref{eq:3.6}, we see that the contracting homotopy for fermions keeps track of the particles' momenta in the fermionic multi-particle solutions, but only one mass is taken into account, as can be seen in the fermionic perturbiners \eqref{eq:pert-CSM} and \eqref{eq:QCDperturbinercoefficients}. In \eqref{eq:3.12} and \eqref{eq:QCDperturbiner-quasi} we can see that fermionic perturbiner contains a number of $2n+1$ fermionic single particle elements and it is precisely that $+1$ ``unpaired" fermion the one that carries the mass after the contraction. A slight modification in the fermionic part of \eqref{eq:3.6} that takes into account the different masses is
\ba\label{eq:mass-homot}
\widetilde{\Ssf}^{\mathrm{F}} =\frac{1}{k^2-m_f^2}\left(\begin{array}{cc} 0 & -(\slashed{k}+m_f)) \\ -(\slashed{k}-m_f) & 0 \end{array} \right).
\ea 
Free of deconcatenations, this fermion is easily identified in the Dyck language localisation for fully-flavoured amplitudes. In this case, we can use \eqref{eq:mass-homot} to modify the fermionic Berends-Giele currents to
\begin{subequations}\label{ferm-pert-masses}
\begin{multline}
\Psi^i_{pQ}= -\left(\frac{\slashed{k}_{pQ} + m_p}{s_{pQ} - m_p^2}\right)\slashed{\mathcal{A}}_{Q}^{a} (T_{a})^{i}_{\phantom{i}j}\Psi^j_p \\
+\left(\frac{\slashed{k}_{pQ} + m_p}{s_{pQ} - m_p^2}\right) \sum_{\substack{\hspace{-5pt}Q = R \cup S \\ \hspace{-2pt}R,S\in \mathrm{Dyck}}} \slashed{\mathcal{A}}_{R}^{a} (T_{a})^{i}_{\phantom{i}j}\left(\frac{\slashed{k}_{pS} + m_p}{s_{pS} - m_p^2}\right)(T_{b})^{j}_{\phantom{j}k} \slashed{\mathcal{A}}_{S}^{b}\Psi^k_p,
\end{multline}
\begin{multline}
\bar{\Psi}_{pQi}= \bar{\Psi}_{pj}\slashed{\mathcal{A}}_{Q}^{a} (T_{a})^{k}_{\phantom{k}i} \left(\frac{\slashed{k}_{pQ} - m_p}{s_{pQ} - m_p^2}\right) \\
 +\sum_{\substack{\hspace{-5pt}Q = R \cup S \\ \hspace{-2pt}R,S\in \mathrm{Dyck}}}\bar{\Psi}_{pj}(T_b)^{j}_{\phantom{j}k}\slashed{\mathcal{A}}_{R}^{b}\left(\frac{\slashed{k}_{pR} - m_p}{s_{pR} - m_p^2}\right)\slashed{\mathcal{A}}_{S}^{a} (T_{a})^{k}_{\phantom{k}i} \left(\frac{\slashed{k}_{pQ} - m_p}{s_{pQ} - m_p^2}\right),
\end{multline}
\end{subequations}
for QCD and similarly for CSM. Then, the generator for fully-flavoured amplitudes takes the form
\begin{equation}
\mathbf{G}_{CSM}=\sum_{n \geq 1} \sum_{\substack{Q \in \mathcal{OW}_{2n} \\ \hspace{-8pt}Q\in \mathrm{Dyck}}}\Big\{ \bar{\Psi}_{1i}\slashed{\mathcal{A}}_{Q}^{a} (T_{a})^{i}_{\phantom{i}j} \Psi_{2n+2}^j + \sum_{\substack{Q = R \cup S \\ \hspace{-8pt}R,S\in \mathrm{Dyck}}} \bar{\Psi}_{1i}(T_b)^{i}_{\phantom{i}j}\slashed{\mathcal{A}}_{R}^{b}\left(\frac{\slashed{k}_{1R} - m_1}{s_{1R} - m_1^2}\right)\slashed{\mathcal{A}}_{S}^{a} (T_{a})^{j}_{\phantom{j}k} \Psi_{2n+2}^k  \Big\}.
\end{equation}
This construction is easily extended to the theory with scalars.

\subsection{Future perspectives}

There are some immediate directions to turn following the results presented here. In a work in progress we are analysing the kinematic numerators obtained for CSM with different flavours. Since the theory is trivalent, it is a perfect candidate for colour-kinematics and double-copy constructions of 3D gravity-matter amplitudes. If, as suggested in \cite{Johansson:2019dnu}, 3D axiodilaton gravity coupled to matter comes from the double copy of QCD$_3$, we expect an interesting topological gravity theory coming from the double-copy of CSM. The Dyck localisation procedure could also be extended to supersymmetric theories and its effect on the colour-stripped amplitudes might shed some light  on a possible algebraic construction of the Johansson-Ochirov basis for different theories. Finally, another natural step is to understand multiparticle solutions at loops using strong homotopy algebras, motivated by the recent results in \cite{Jurco:2019yfd}.


\acknowledgments
We are thankful to Oliver Scholetterer for valuable comments in the final version of the draft. H.G. is supported by the Danish National Research
Foundation (DNRF91) as well as the Carlsberg Foundation. H.G. acknowledges 
partial support from University Santiago de Cali (USC). R.L.J. would like to thank the Czech Science Foundation - GA\v{C}R for financial support under the grant 19-06342Y.  A.Q.V. thanks the support provided by COLCIENCIAS through grant number  FP44842-013-2018 of the Fondo Nacional de Financiamiento para la Ciencia, la Tecnolog\'ia y la Inovaci\'on.


\appendix

\section{Computing $\bar{\Psi}^{\rm Dyck}_{12345,i}$ for QCD} \label{Comp-Psi}

In this section, we will outline how to calculate the current $\bar{\Psi}_{12345, i}$ on the support of the Dyck constraint, namely  $\bar{\Psi}^{\rm Dyck}_{12345,i}=\bar{\Psi}^{\rm Dyck}_{(1(23)(45),i}$.  Since here we focus on Dyck words then, the upper text ``${\rm Dyck}$" in the Perturbiners is no longer necessary. 

Let us consider $ N $ as an odd-length word, this implies there is only one open parenthesis (otherwise, the current vanishes trivially), such as in $\bar{\Psi}_{(1(23)(45),i}$. By the rules described in section \ref{DyckLocalization}, the letter with the open parenthesis must be fixed (factorised), for example $N\rightarrow (p \, M$, where $M$ is a Dyck word.  Therefore, the current $\bar{\Psi}_{N, i} $ turns into\footnote{It is useful to remind that for the QCD examples, we are taking the transformation, $k^\mu_p\,\rightarrow\, -k^\mu_p$, in order to compare our results with the ones obtained in \cite{Johansson:2015oia}.}
\begin{align}\label{QCD-Dyck}
-\bar{\Psi}_{(p\,M, i} (\slashed{k}_{pM}-m) = 
\bar{\Psi}_{p, j} \, \slashed{\mathcal{A}}_{M}^{a} (T_{a})^{j}_{\phantom{j}i}  + \!\!\!  \sum_{\substack{ M = Q \cup R \\ Q,R\in \mathrm{Dyck}}}  \left[  \bar{\Psi}_{p, l} \,\slashed{\mathcal{A}}_{Q}^{b} (T_b)^{l}_{\phantom{l}j}\frac{\slashed{k}_{1Q} + m}{s_{1Q} - m^2} \right] \slashed{\mathcal{A}}_{R}^{a} (T_{a})^{j}_{\phantom{j}i}  \,\, ,
\end{align}
where the multiparticle fields, ${\cal A}^\mu_{Y a}$, respect Dyck's condition, ${\cal A}^\mu_{Y a}= {\cal A}^{\mu , \rm Dyck}_{Y a}$. Clearly,  a similar expansion can be written for the current $\Psi^{i\, \rm Dyck}_{(p\,M} $.

Now, from the boundary conditions,
\begin{align}
\bar{\Psi}_{p\,i}=\bar{u}_p \, \delta^{i_p}_i, \quad p=1,2,4,\qquad\quad         
\Psi_{q}^{i}= v_q \, \delta^{i}_{i_q},  \quad q=3,5,6,
\end{align}
and \eqref{QCD-Dyck}, 
the current $\bar{\Psi}_{12345, i} $ becomes
\begin{align}\label{PsiDyck}
-\bar{\Psi}_{(1(23)(45), i} (\slashed{k}_{12345}-m)= \left[\bar{\Psi}_{1, j} \, \slashed{\mathcal{A}}_{2345\, a}  + \bar{\Psi}_{(1(23), j} \, \slashed{\mathcal{A}}_{45\, a}  + \bar{\Psi}_{(1(45), j} \, \slashed{\mathcal{A}}_{23\, a} \right] (T^{a})^{j}_{\phantom{j}i} \, .
\end{align}
Notice that $\bar{\Psi}_{(1(23), j} $ and $\bar{\Psi}_{(1(45), j} $ mean the square bracket in \eqref{QCD-Dyck}, {\it i.e.}
\begin{align}
&
\bar{\Psi}_{(1(23), j} \, \slashed{\mathcal{A}}_{45\, a} (T^{a})^{j}_{\phantom{j}i} =
\frac{ - \left[ \bar{u}_1\gamma_\mu ( \slashed{k}_{123}+m ) \gamma_\nu  \right] (\bar{u}_2 \gamma^\mu v_3) (\bar{u}_4 \gamma^\nu v_5)   }{(s_{123}-m^2) \, s_{23} \, s_{45}} \,  (T^a T^b)^{i_1}_{\phantom{i_1}i} (T^a)^{i_{2}}_{\phantom{i_{2}}i_3} (T^b)^{i_{4}}_{\phantom{i_{4}}i_5}, \quad \nonumber \\
&
\bar{\Psi}_{(1(45), j} \, \slashed{\mathcal{A}}_{23\, a} (T^{a})^{j}_{\phantom{j}i} =
\frac{ - \left[ \bar{u}_1\gamma_\mu ( \slashed{k}_{145}+m ) \gamma_\nu  \right] (\bar{u}_4 \gamma^\mu v_5) (\bar{u}_2 \gamma^\nu v_3)   }{(s_{145}-m^2) \, s_{23} \, s_{45}} \,  (T^a T^b)^{i_1}_{\phantom{i_1}i} (T^a)^{i_{4}}_{\phantom{i_{4}}i_5} (T^b)^{i_{2}}_{\phantom{i_{2}}i_3}, \quad
\end{align}
where
\begin{align}
{\cal A}^{\mu}_{23\, a} = \frac{{\cal J}^\mu_{23\, a}  }{s_{23}} = \frac{ (\bar{u}_2 \gamma^\mu v_3) }{s_{23}} \, (T_a)^{i_{2}}_{\phantom{i_{2}}i_3}, \qquad 
{\cal A}^{\mu}_{45\, a} = \frac{{\cal J}^\mu_{45\, a}  }{s_{45}} = \frac{(\bar{u}_4 \gamma^\mu v_5) }{s_{45}} \, (T_a)^{i_{4}}_{\phantom{i_{4}}i_5}\, .
\end{align}

Subsequently, the contribution $ {\cal A}^{\mu}_{2345, a}$ is given by
\begin{align}\label{A2345Dqcd}
{\cal A}^{\mu}_{2345, a} &= \frac{ {\cal J}_{2345, a}^{\mu }  }{s_{2345}}  + \nonumber\\
&
\frac{i\tilde f^{abc}}{s_{2345}}
\left\{
i {\cal A}_{23 b}^\mu (k_{23} \cdot {\cal A}_{45 c} ) - ({\cal A}_{23 b} \cdot {\cal F}_{45 c} )^\mu 
+
i {\cal A}_{45 b}^\mu (k_{45} \cdot {\cal A}_{23 c} ) - ({\cal A}_{45 b} \cdot {\cal F}_{23 c} )^\mu 
\right\},
\end{align}
where the only non-vanishing allowed contributions in the second line come from the de-concatenations, 
\begin{equation}
(23)(45)=R \cup S \,\rightarrow \, (23) \cup (45) ; \, (45) \cup (23) .  
\end{equation}
Notice that deconcatenations such as, for instance $(2(4 \cup 3)5)$ and $(2 \cup 3)(45)$, vanish trivially, while, $3)(4 \cup (25)$ or $3)5) \cup (2(4$, are not allowed. Under support of the boundary conditions, it is straightforward to see the curvatures, ${\cal F}_{23 c}^{\nu\mu}$ and ${\cal F}_{45 c}^{\nu\mu}$, are given by
\begin{align}
{\cal F}_{23 c}^{\nu\mu} = i k_{23}^\nu {\cal A}_{23 c}^\mu - i k_{23}^\mu {\cal A}_{23 c}^\nu \, , \qquad 
{\cal F}_{45 c}^{\nu\mu} = i k_{45}^\nu {\cal A}_{45 c}^\mu - i k_{45}^\mu {\cal A}_{45 c}^\nu, 
\end{align}
therefore, the second line in \eqref{A2345Dqcd}  turns into
\begin{align}\label{}
&
\bar{\Psi}_{1, j} \, \gamma_\mu\, \left[
{\cal A}^{\mu}_{2345, a} - \frac{  {\cal J}_{2345, a}^{\mu }  }{s_{61}}  
\right] \, (T^a)^{j}_{\phantom{j}i} =
\nonumber\\
&
 - \frac{ \tilde f^{abc} (T_a)^{i_1}_{\phantom{i_1}i}  (T_b)^{i_{2}}_{\phantom{i_{2}}i_3}   (T_c)^{i_{4}}_{\phantom{i_{4}}i_5} }{s_{61}s_{23} s_{45}} \times
\left\{-
(\bar{u}_1 \gamma_\mu \,) (\bar{u}_2 \gamma_\nu v_3)  (\bar{u}_4 \gamma_\rho  v_5) V^{\mu\nu\rho}_{k_6+k_1,k_2+k_3,k_4+k_5}
\right\},
\end{align}
where we have used momentum conservation, $k_1+k_2+\cdots + k_6=0$, and the equations of motions
\begin{equation}
\bar{u}_p \, \slashed{k_p} =m \,\bar{u}_p, \qquad  \slashed{k_p}\, v_p =-m \,v_p\, .
\end{equation}
Additionally, we have introduced the vertex
\begin{equation}\label{3pvertex}
V^{\mu\nu\rho}_{p,q,k} = \eta^{\mu\nu} (p-q)^\rho + \eta^{\rho\mu} (k-p)^\nu +  \eta^{\nu\rho} (q-k)^\mu  .
\end{equation} 

Finally, we now compute the current, $ {\cal J}_{(23)(45), a}^{\mu} $. It is straightforward to see that 
all possible non-vanishing deconcatenations are 
\begin{equation}
(23)(45)=R \cup S \,\rightarrow \,  (23)(4 \cup 5) ; \, (4 \cup (23)5) ; 
\, (2(45) \cup 3) ; \, (2 \cup 3)(45) ,    \nonumber 
\end{equation}
thus, one has
\begin{align}
&
 {\cal J}_{2345, a}^{\mu } =  \nonumber \\
& 
 \bar{\Psi}_{(23)(4,i} \gamma^\mu\, (T_a)^{i}_{\phantom{i}j} \Psi_{5)}^j + \bar{\Psi}_{(2(45),i} \gamma^\mu\, (T_a)^{i}_{\phantom{i}j} \Psi_{3)}^j + \bar{\Psi}_{(2,i} \gamma^\mu\, (T_a)^{i}_{\phantom{i}j} \Psi^{j}_{3)(45)} + \bar{\Psi}_{(4,i} \gamma^\mu\, (T_a)^{i}_{\phantom{i}j} \Psi^{j}_{(23)5)}. \nonumber \\
\end{align}
Therefore, from \eqref{QCD-Dyck},  it is not hard to compute the above contributions so, these land at
\begin{align}
&
\bar{\Psi}_{(23)(4,i} \gamma^\mu\, (T_a)^{i}_{\phantom{i}j} \Psi_{5)}^j = \frac{ (T^bT^a)^{i_{4}}_{\phantom{i_{4}}i_5} (T^b)^{i_{2}}_{\phantom{i_{2}}i_3}  }{(s_{423} - m^2 ) \, s_{23}} \times \left\{
-\left[ \bar{u}_4 \gamma_\nu (\slashed{k}_{423}+m ) \gamma_\mu v_5 \right]  (\bar{u}_2 \gamma^\nu v_3)
\right\} ,\nonumber \\
&
\bar{\Psi}_{(2(45),i} \gamma^\mu\, (T_a)^{i}_{\phantom{i}j} \Psi_{3)}^j = \frac{ (T^bT^a)^{i_{2}}_{\phantom{i_{2}}i_3} (T^b)^{i_{4}}_{\phantom{i_{4}}i_5}  }{(s_{245} - m^2 ) \, s_{45}} \times \left\{
-\left[ \bar{u}_2 \gamma_\nu (\slashed{k}_{245}+m ) \gamma_\mu v_3 \right]  (\bar{u}_4 \gamma^\nu v_5)
\right\} ,\nonumber \\
&
\bar{\Psi}_{(2,i} \gamma^\mu\, (T_a)^{i}_{\phantom{i}j} \Psi_{3)(45)}^{j} = \frac{ (T^aT^b)^{i_{2}}_{\phantom{i_{2}}i_3} (T^b)^{i_{4}}_{\phantom{i_{4}}i_5}  }{(s_{261} - m^2 ) \, s_{45}} \times \left\{
-\left[ \bar{u}_2 \gamma_\mu (\slashed{k}_{261}+m ) \gamma_\nu v_3 \right]  (\bar{u}_4 \gamma^\nu v_5)
\right\} ,\nonumber \\
&
\bar{\Psi}_{(4,i} \gamma^\mu\, (T_a)^{i}_{\phantom{i}j} \Psi_{(23)5)}^{j} = \frac{ (T^aT^b)^{i_{4}}_{\phantom{i_{4}}i_5} (T^b)^{i_{2}}_{\phantom{i_{2}}i_3}  }{(s_{461} - m^2 ) \, s_{23}} \times \left\{
-\left[ \bar{u}_4 \gamma_\mu (\slashed{k}_{461}+m ) \gamma_\nu v_5 \right]  (\bar{u}_2 \gamma^\nu v_3)
\right\} . \nonumber \\
\end{align}
Putting all results together we arrive at
\begin{align}\label{Psi12345}
&
-\bar{\Psi}_{(12345, i} (\slashed{k}_{12345}-m) \Psi_6^i= 
\frac{ -\tilde f^{abc} (T_a)^{i_{1}}_{\phantom{i_{1}}i_6}  (T_b)^{i_{2}}_{\phantom{i_{2}}i_3}   (T_c)^{i_{4}}_{\phantom{i_{4}}i_5} \, n_0}{s_{61}s_{23} s_{45}} 
+ 
\left[
\frac{(T^aT^b)^{i_{1}}_{\phantom{i_{1}}i_6} (T^a)^{i_{2}}_{\phantom{i_{2}}i_3} (T^b)^{i_{4}}_{\phantom{i_{4}}i_5} \, n_1  }{(s_{123} - m^2) s_{23} s_{45}} 
\right.
\nonumber \\
&
\left.
+ \frac{(T^aT^b)^{i_{1}}_{\phantom{i_{1}}i_6}  (T^a)^{i_{4}}_{\phantom{i_{4}}i_5} (T^b)^{i_{2}}_{\phantom{i_{2}}i_3} \, n_2  }{(s_{145} - m^2) s_{23} s_{45}} +
{\rm cyclic}
\Big(
(61)\rightarrow (23) \rightarrow (45) 
\Big) 
\right]
,
\end{align}
where 
\begin{align}
&
n_0=-
(\bar{u}_1 \gamma_\mu v_6) (\bar{u}_2 \gamma_\nu v_3)  (\bar{u}_4 \gamma_\rho  v_5) V^{\mu\nu\rho}_{k_6+k_1,k_2+k_3,k_4+k_5} \, ,\nonumber\\
&
n_1=  - \left[ \bar{u}_1\gamma_\mu ( \slashed{k}_{123}+m ) \gamma_\nu  v_6 \right] (\bar{u}_2 \gamma^\mu v_3) (\bar{u}_4 \gamma^\nu v_5) \, ,\nonumber
  \\
&
n_2= - \left[ \bar{u}_1\gamma_\mu ( \slashed{k}_{145}+m ) \gamma_\nu  v_6 \right] (\bar{u}_4 \gamma^\mu v_5) (\bar{u}_2 \gamma^\nu v_3) . \nonumber \\
\end{align}
Notice that in \eqref{Psi12345} we have introduced $\Psi_6^i$ on the left hand side.

\section{Computing $\bar{\Phi}^{\rm Dyck}_{12345,i}$ for sQCD } \label{Comp-Phi}

In this section, we  calculate the current $\bar{\Phi}_{12345, i}$ on the support of the Dyck condition,  $\bar{\Phi}^{\rm Dyck}_{12345,i}=\bar{\Phi}^{\rm Dyck}_{(1(23)(45),i}$. Notice that the upper text ``${\rm Dyck}$" in the Perturbiners is no longer necessary, since we are only focusing on Dyck words. 

Such as in the above section, let us consider $ N $ as an odd-length word. By the rules in section \ref{DyckLocalization}, the letter with the open parenthesis must be fixed, {\it i.e.} $N\rightarrow (p \, M$, where $M$ is a Dyck word.  Therefore, the current $\bar{\Phi}_{N, i} $ turns into
\begin{align}\label{scalar-Dyck}
&
-(s_{pM}-m^{2}) \, \bar{\Phi}_{(p\,M, i} = 
\left\{ 2(k_{p}\cdot\mathcal{A}_{M}^{a}) \bar{\Phi}_{p, j}  (T_{a})_{\hphantom{j}i}^{j}  
+
\bar{\Phi}_{p, j}  (T^{b}T^{a})_{\hphantom{j}i}^{j} \sum_{M =P \cup Q} (\mathcal{A}_{Pa}\cdot\mathcal{A}_{Qb}) 
\right\}
\nonumber \\
&
+\!\!\!\! \sum_{\substack{ \hspace{-5pt} M = R \cup S \\ \hspace{-2pt}R,S\in \mathrm{Dyck}}}
\left\{ 2(k_{pR}\cdot\mathcal{A}_{S}^{a})  \left[ - \frac{ 2 (k_p\cdot {\cal A}_R^c ) \bar{\Phi}_{p, l}  (T_c)^{l}_{\phantom{l}j}  + \bar{\Phi}_{p, l}(T^cT^d)^{l}_{\phantom{l}j} \sum_{R = U\cup V}  (\mathcal{A}_{Ud}\cdot\mathcal{A}_{Vc})  }{(s_{pR} - {m}^2 )} \right]  (T_{a})_{\hphantom{j}i}^{j} + \right. \nonumber\\
&
\left.
 \left[ - \frac{ 2 (k_p\cdot {\cal A}_R^c ) \bar{\Phi}_{p, l}  (T_c)^{l}_{\phantom{l}j}  + \bar{\Phi}_{p, l}(T^cT^d)^{l}_{\phantom{l}j} \sum_{R= U \cup V}  (\mathcal{A}_{Ud}\cdot\mathcal{A}_{Vc})  }{(s_{pR} - {m}^2 )} \right]  (T^{b}T^{a})_{\hphantom{j}i}^{j} \sum_{S =P \cup Q} (\mathcal{A}_{Pa}\cdot\mathcal{A}_{Qb}) \right\} ,
\end{align}
where the multiparticle fields, ${\cal A}^\mu_{Y a}$, respect Dyck's condition, ${\cal A}^\mu_{Y a}= {\cal A}^{\mu , \rm Dyck}_{Y a}$. In a similar way, we have the current ${\Phi}^{i\, \rm Dyck}_{(p\,M} $.
Additionally, it is useful to remind that
the ordered deconcatenation in the sums, $\sum_{M =P \cup Q} $,  $\sum_{R = U\cup V}  $ and $\sum_{S =P \cup Q}$, are such that the union is Dyck.

Now, from the boundary conditions,
\begin{eqnarray}\label{}
\bar{\Phi}_{1,i}=\delta^{i_1}_i, \quad \bar{\Phi}_{2,i}=\delta^{i_2}_i, \quad \bar{\Phi}_{4,i}=\delta^{i_4}_i, \quad \Phi_{3}^{i}=\delta^{i}_{i_3}, \quad \Phi_{5}^{i}=\delta^{i}_{i_5} ,
\end{eqnarray}
and using the equation \eqref{scalar-Dyck}, 
the current $\bar{\Phi}_{(1(23)(45), i} $ becomes
\begin{align}\label{phiDyck}
&
-(s_{12345}-m^{2}) \, 
\bar{\Phi}_{(1(23)(45), i} = 2 ( k_1\cdot {\cal A}_{2345, a}) \bar{\Phi}_{1,j}  (T^a)^{j}_{\phantom{j}i} \nonumber \\
&
+ \bar{\Phi}_{1,j} (T^a T^b)^{j}_{\phantom{j}i} \left[ ({\cal A}_{23, a} \cdot {\cal A}_{45, b} ) + ({\cal A}_{45, a} \cdot {\cal A}_{23, b} )  \right] \qquad \nonumber\\
& + 2 (k_{123} \cdot {\cal A}_{45, a}) \bar{\Phi}_{(1(23), j} (T^a)^{j}_{\phantom{j}i} + 2 (k_{145} \cdot {\cal A}_{23, a}) \bar{\Phi}_{(1(45), j} (T^a)^{j}_{\phantom{j}i}  ,  \qquad
\end{align}
where we have denoted as $\bar{\Phi}_{123, j} $ and $\bar{\Phi}_{145, j} $ the square brackets in \eqref{scalar-Dyck}, 
\begin{align}
\bar{\Phi}_{(1(23), j} = - \frac{2(k_1\cdot {\cal A}_{23 c} ) \bar{\Phi}_{1,l} (T^c)^{l}_{\phantom{l}j}  }{(s_{123} - m^2)} , \qquad
\bar{\Phi}_{(1(45), j} = - \frac{2(k_1\cdot {\cal A}_{45 c} ) \bar{\Phi}_{1,l} (T^c)^{l}_{\phantom{l}j}  }{(s_{145} - m^2)} .\qquad
\end{align}
The contributions ${\cal A}_{23 c}$ and ${\cal A}_{45 c}$ were already computed in \eqref{A-1323}, so,
\begin{eqnarray}\label{A-2345}
{\cal A}_{23 \, c}^\mu  = \frac{ {\cal J}_{23\, c}^\mu }{ s_{23} } = \frac{  (k_2-k_3)^\mu (T^c)^{i_2}_{\phantom{i_2}i_3} }{ s_{23} }, \qquad
{\cal A}_{45 \, c}^\mu  = \frac{ {\cal J}_{45\, c}^\mu }{ s_{45} } = \frac{  (k_4-k_5)^\mu (T^c)^{i_4}_{\phantom{i_4}i_5} }{ s_{45} }. 
\end{eqnarray}
On the other hand, from the perturbiner expansion of ${\cal A}^\mu_{Y, a}$, the contribution $ {\cal A}^{\mu}_{2345, a}$ 
\begin{align}\label{A2345}
{\cal A}^{\mu}_{2345, a} &= \frac{ {\cal J}_{2345, a}^{\mu }  }{s_{2345}}  + \nonumber\\
&
\frac{i\tilde f^{abc}}{s_{2345}}
\left\{
i {\cal A}_{23 b}^\mu (k_{23} \cdot {\cal A}_{45 c} ) - ({\cal A}_{23 b} \cdot {\cal F}_{45 c} )^\mu 
+
i {\cal A}_{45 b}^\mu (k_{45} \cdot {\cal A}_{23 c} ) - ({\cal A}_{45 b} \cdot {\cal F}_{23 c} )^\mu 
\right\},
\end{align}
where the only non-vanishing allowed contributions in the second line come from the deconcatenations, 
\begin{equation}
(23)(45)=R \cup S \,\rightarrow \, (23) \cup (45) ; \, (45) \cup (23) ,  
\end{equation}
such as it was explained in the previous section.
Under support of the boundary conditions, the curvatures, ${\cal F}_{23 c}^{\nu\mu}$ and ${\cal F}_{45 c}^{\nu\mu}$, are given by
\begin{align}
{\cal F}_{23 c}^{\nu\mu} = i k_{23}^\nu {\cal A}_{23 c}^\mu - i k_{23}^\mu {\cal A}_{23 c}^\nu \, , \qquad 
{\cal F}_{45 c}^{\nu\mu} = i k_{45}^\nu {\cal A}_{45 c}^\mu - i k_{45}^\mu {\cal A}_{45 c}^\nu, 
\end{align}
therefore, the second line in \eqref{A2345}  can be written as
\begin{align}\label{}
&
{\cal A}^{\mu}_{2345, a} - \frac{  {\cal J}_{2345, a}^{\mu }  }{s_{2345}}  = 
 - \frac{ \tilde f^{abc} (T_a)^{i_1}_{\phantom{i_1}i}  (T_b)^{i_{2}}_{\phantom{i_{2}}i_3}   (T_c)^{i_{4}}_{\phantom{i_{4}}i_5} }{s_{61}s_{23} s_{45}} \times
\nonumber\\
&
\left\{
(k_6-k_1)_\mu (k_2-k_3)_\nu (k_4-k_5)_\rho V^{\mu\nu\rho}_{k_6+k_1,k_2+k_3,k_4+k_5}
+ (s_{23}-s_{45})(k_2-k_3)\cdot(k_4-k_5)
\right\},
\end{align}
where we have used momentum conservation, $k_1+k_2+\cdots + k_6=0$, and the 
vertex $V^{\mu\nu\rho}_{p,q,k} $ was defined in \eqref{3pvertex}.

The last step is to compute the current $ {\cal J}_{2345, a}^{\mu } $. For this case, the non-vanishing deconcatenations are 
\begin{equation}
(23)(45)=R \cup S \,\rightarrow \, (23) \cup (45) ; \, (45) \cup (23);\, (23)(4 \cup 5) ; \, (4 \cup (23)5) ; 
\, (2(45) \cup 3) ; \, (2 \cup 3)(45) ,    \nonumber 
\end{equation}
then one has, 
\begin{align}\label{Jcurrent}
&
 {\cal J}_{2345, a}^{\mu } =  \nonumber \\
& 
 {\cal A}_{23}^{b\, \mu} \, \Theta_{45,ba} + {\cal A}_{45}^{b\, \mu} \, \Theta_{23,ba}
 +(k_{234} - k_5)^\mu \bar{\Phi}_{(23)(4, i} (T^a)^{i}_{\phantom{i}j} \Phi^j_{5)}
 +(k_{4} - k_{235})^\mu \bar{\Phi}_{(4, i} (T^a)^{i}_{\phantom{i}j} \Phi^{j}_{(23)5)} \nonumber \\
 &
  +(k_{245} - k_3)^\mu \bar{\Phi}_{(2(45), i} (T^a)^{i}_{\phantom{i}j} \Phi^j_{3)}
    +(k_{2} - k_{345})^\mu \bar{\Phi}_{(2, i} (T^a)^{i}_{\phantom{i}j} \Phi^{j}_{3)(45)} ,\nonumber \\
\end{align}
with
\begin{align}
\Theta_{23,ba} = (T_bT_a)^{i_{2}}_{\phantom{i_{2}}i_3} + (T_aT_b)^{i_{2}}_{\phantom{i_{2}}i_3} , \qquad
\Theta_{45,ba} = (T_bT_a)^{i_{4}}_{\phantom{i_{4}}i_5} + (T_aT_b)^{i_{4}}_{\phantom{i_{4}}i_5} \, .
\end{align}
Finally, from \eqref{scalar-Dyck}, the currents $\bar{\Phi}_{Y, i}$ and ${\Phi}^{i}_{Y}$ in \eqref{Jcurrent} become,
\begin{align}
&
\bar{\Phi}_{(23)(4, i} = -\frac{ 2(k_4\cdot {\cal A}_{23}^a ) \bar{\Phi}_{4,j} (T^a)^{j}_{\phantom{j}i}  }{ (s_{234} - m^2) } , \qquad
\bar{\Phi}_{(2(45), i} = -\frac{ 2(k_2\cdot {\cal A}_{45}^a ) \bar{\Phi}_{2,j} (T^a)^{j}_{\phantom{j}i}  }{ (s_{245} - m^2) } , \\
&
{\Phi}^{j}_{(23)5)} = \frac{ 2(k_5\cdot {\cal A}_{23}^a )  (T^a)^{j}_{\phantom{j}i} {\Phi}_{5}^{i} }{ (s_{235} - m^2) } , \qquad
{\Phi}^{j}_{3)(45)} = \frac{ 2(k_3\cdot {\cal A}_{45}^a )  (T^a)^{j}_{\phantom{j}i} {\Phi}_{3}^i  }{ (s_{345} - m^2) } .
\end{align}
Therefore, by putting everything together one arrives at
\begin{align}\label{Phi12345}
&
(s_{12345}-m^{2}) \, 
\bar{\Phi}_{(1(23)(45), i} = 
\frac{ \tilde f^{abc} (T_a)^{i_1}_{\phantom{i_1}i}  (T_b)^{i_{2}}_{\phantom{i_{2}}i_3}   (T_c)^{i_{4}}_{\phantom{i_{4}}i_5} \, n_0}{s_{61}s_{23} s_{45}} 
+ 
\left[
\frac{(T^aT^b)^{i_1}_{i} (T^a)^{i_{2}}_{\phantom{i_{2}}i_3} (T^b)^{i_{4}}_{\phantom{i_{4}}i_5} \, n_1  }{(s_{123} - m^2) s_{23} s_{45}} 
\right.
\nonumber \\
&
\left.
+ \frac{(T^aT^b)^{i_1}_{i}  (T^a)^{i_{4}}_{\phantom{i_{4}}i_5} (T^b)^{i_{2}}_{\phantom{i_{2}}i_3} \, n_2  }{(s_{145} - m^2) s_{23} s_{45}} +
{\rm cyclic}
\left(
\begin{matrix}
(61)\rightarrow (23) \rightarrow (45) \\
(i \, i_1)\rightarrow (i_2 i_3) \rightarrow (i_4i_5)
\end{matrix}
\right) 
\right]
,
\end{align}
where 
\begin{align}
&
n_0=(k_6-k_1)_\mu (k_2-k_3)_\nu (k_4-k_5)_\rho V^{\mu\nu\rho}_{k_6+k_1,k_2+k_3,k_4+k_5} \nonumber\\
&
n_1= - 4 \left\{
(k_1\cdot (k_2-k_3)) (k_6\cdot (k_4-k_5)) + \frac{ (s_{123} - m^2) (k_2-k_3)\cdot (k_4-k_5) }{4}
\right\} \nonumber \\
&
n_2= - 4 \left\{
(k_1\cdot (k_4-k_5)) (k_6\cdot (k_2-k_3)) + \frac{ (s_{145} - m^2) (k_2-k_3)\cdot (k_4-k_5) }{4}
\right\} \nonumber .\\
\end{align}
Here, we have used the momentum conservation, $k_1+\cdots +k_6=0$, the algebra, $[T^a, T^b]=\tilde f^{abc} \,T^c$, and the on-shell condition, $k_i^2=m^2$.


\bibliographystyle{JHEP}

\providecommand{\href}[2]{#2}\begingroup\raggedright\endgroup

\end{document}